\pdfoutput=1
\documentclass[12pt]{article}
\usepackage{caption}
\usepackage{amsmath,amsfonts,amsthm,amssymb}
\usepackage{subfigure}
\usepackage{setspace}
\usepackage{Tabbing}
\usepackage{lastpage}
\usepackage{extramarks}
\usepackage{chngpage}
\usepackage[usenames,dvipsnames]{color}
\usepackage{graphicx,float,wrapfig}
\usepackage{lineno}
\usepackage{float}
\def\br{\begin{eqnarray}}
\def\er{\end{eqnarray}}
\def\be{\begin{equation}}
\def\ee{\end{equation}}
\def\({\left(}
\def\){\right)}

\def\rlx{\relax\leavevmode}
\def\IR{\rlx\hbox{\rm I\kern-.18em R}}
\def\vp{\varphi}

\def\IZ{\rlx\hbox{\sf Z\kern-.4em Z}}
\def\IR{\rlx\hbox{\rm I\kern-.18em R}}
\def\IC{\rlx\hbox{\,$\inbar\kern-.3em{\rm C}$}}
\def\one{\hbox{{1}\kern-.25em\hbox{l}}}

\begin{document}

\begin{titlepage}
\vspace*{-1cm}

\vskip 2cm

\vspace{.2in}
\begin{center}
{\large\bf Further comments on BPS systems}
\end{center}

\vspace{.5cm}

\begin{center}

P. Klimas~$^{\star}$ and W.J. Zakrzewski~$^{\dagger}$

\small
\par \vskip .2in \noindent

 \par \vskip .2in \noindent
$^{(\star)}$Departamento de F\'isica,\\Universidade Federal de Santa Catarina,\\
 Trindade, CEP 88040-900, Florian\'opolis-SC, Brazil\\
email: pawel.klimas@ufsc.br

\small


\par \vskip .2in \noindent
$^{(\dagger)}$~Department of Mathematical Sciences,\\
 University of Durham, Durham DH1 3LE, U.K.\\
email: W.J.Zakrzewski@durham.ac.uk

\normalsize
\end{center}


\begin{abstract}

We look at BPS systems involving two interacting Sine-Gordon like fields both when one of them has a kink solution and the second one either a kink or an antikink solution.
The interaction between the two fields is controlled by a parameter
 $\lambda$ which has to satisfy $\vert \lambda\vert< 2$. We then take 
these solitonic static solutions (with solitons well localised) and construct from them systems involving two solitons in each field (kinks and antikinks) and then  use them as initial conditions for their evolution in Lorentz covariant versions of such models. This way we study their 
interactions and compare them with similar interactions
involving only one Sine-Gordon field.

In particular, we look at the behaviour of two static kinks in each field (which for one field repel each other) and of a system involving kinks and anti-kinks (which for one field attract each other) 
and look how their behaviour depends on the strength
of the interaction $\lambda$ between the two fields. 


Our simulations have led us to look again at the static BPS solutions of systems involving more fields. We have found that such ostensibly `static' BPS solutions can exhibit small motions due to the excitation of their zero modes. These exitations arise from small unevoidable numerical errors (the overall translation is cancelled by the conservation of momentum) but as systems of two or more fields have more that one zero mode such motions can be generated and are extremely small. The energy of our systems has been conserved to within $10^{-5}\%$.

\end{abstract} 
\end{titlepage}


\section{Introduction}
\label{sec:intro}
\setcounter{equation}{0}

In our previous paper \cite{first} we considered a $(1+1)$-dimensional Minkowski space-time theory involving two coupled real scalar fields $\varphi_a$, $a=1,2$, defined by the Lagrangian 
\be
{\cal L}= \frac{1}{2}\left[ (\partial_{\mu}\varphi_1)^2 + 
(\partial_{\mu}\varphi_2)^2 - \lambda \partial_{\mu}\varphi_1\,\partial^{\mu}\varphi_2\right] -V(\varphi_1,  \varphi_2),
\label{lagrangiannewtwo}
\ee
in which $V(\varphi_1,\varphi_2)$ was being given by 
\be
 V=\frac{1}{2}\(F_1^2+F_2^2-\lambda F_1F_2\)\,= \frac{2}{4-\lambda^2}\left[M^2\(\vp_1\)+N^2\(\vp_2\)+\lambda M\(\vp_1\)\,N\(\vp_2\)\right].
\label{poten}
\ee

The functions $M$ and $N$ of fields were chosen to be the generalisations of the Sine-Gordon terms; so we took
\be
M(\varphi_1)\,=\,4\sin(\varphi_1),\qquad  N(\varphi_2)\,=\,4\epsilon\sin(\varphi_2),
\label{topo2d}
\ee
where $\epsilon$ could be $\pm1$. 
As we discussed it in \cite{first} the model satisfies the BPS conditions which give us 
the BPS equations which take the form:
\begin{align} 
\partial_x \varphi_1\,&=\,\frac{4}{4- \lambda^2}
\Big(4\sin(\varphi_1)+2\lambda \epsilon \sin(\varphi_2)\Big),\label{aa1}\\
\partial_x \varphi_2\,&=\,
\frac{4}{4-\lambda^2}\Big(2\lambda \sin(\varphi_1)\,+
4\epsilon \sin(\varphi_2)\Big)\label{aa2}.
\end{align}
As any BPS model this model also has a pre-potential $U$ which, in this case, takes the form
\[
U=-4\big(\cos(\varphi_1)+\epsilon \cos(\varphi_2)\big).
\]

\begin{figure}[h!]
\centering
  {\includegraphics[width=1.0\textwidth,height=0.43\textwidth]{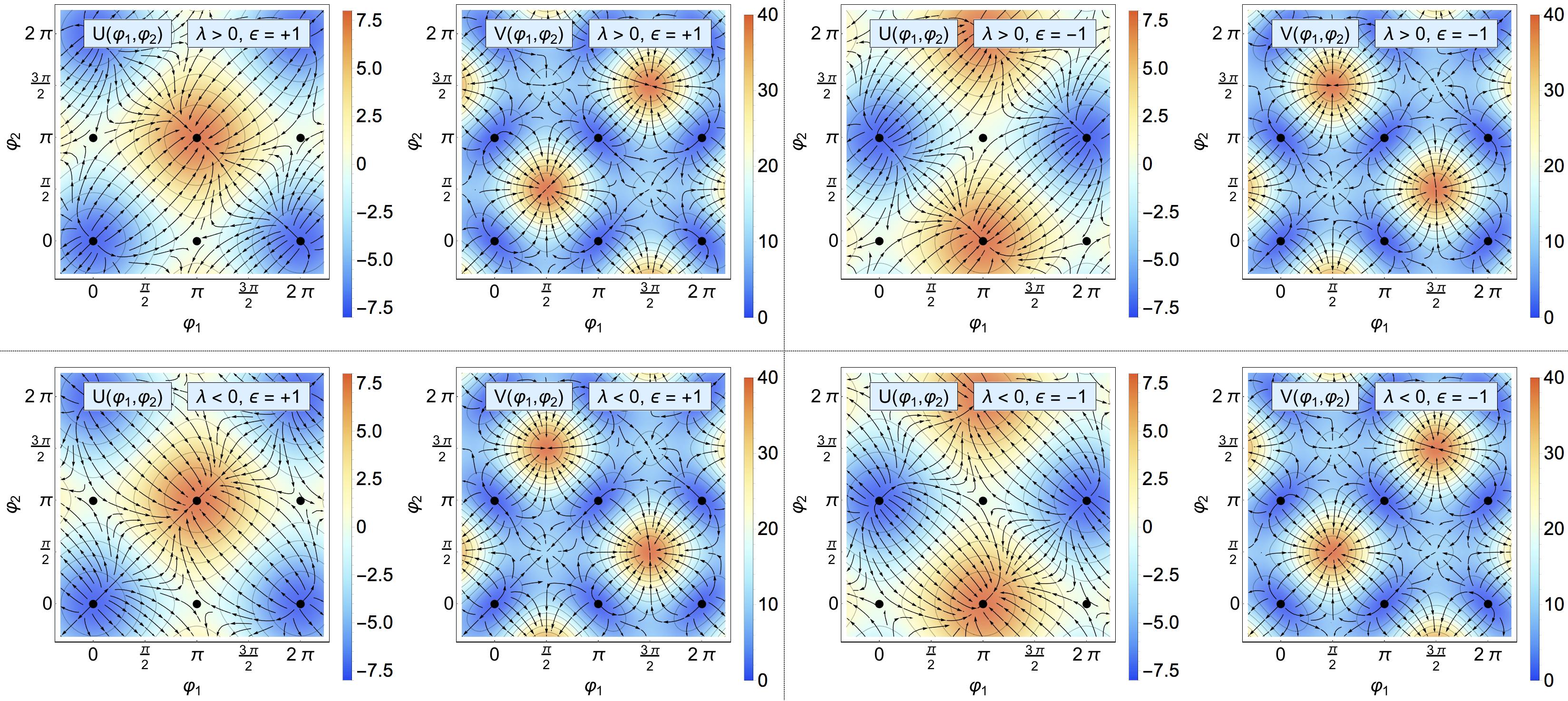}}
   \caption{Pre-potential $U$ together with its modified gradient flow and potential $V$ together with its gradient flow in models with different signs of $\lambda$ and $\epsilon$. Here $|\lambda|=1.2$.}
  \label{fig:grid}
\end{figure}

Two models which differ only by $\epsilon$ have different potentials and pre-potentials.  In Fig.\ref{fig:grid} we show plots of pre-potential $U$ and potential $V$ for four cases corresponding to choices of signs $\epsilon=\pm1$ and ${\rm sgn}(\lambda)=\pm1$. We have also plotted the lines of the gradient flow $\vec\nabla V$ and of the modified gradient flow 
\[
 \vec \nabla_{\gamma} U=\frac{4}{4-\lambda^2}\Big(2\frac{\partial U}{\partial\varphi_1}+\lambda\frac{\partial U}{\partial\varphi_2}, \lambda\frac{\partial U}{\partial\varphi_1}+2\frac{\partial U}{\partial\varphi_2}\Big)
\]
which, as mentioned in our previous papers, takes an important role in study of BPS solutions. The character of extrema of the pre-potential depends on the value of $\epsilon$. Although maxima of the potential $V$ depend on $\epsilon$ the minima (and so  vacua of the model) are invariant under the sign of $\epsilon$.

The expression for the energy of the solutions of these BPS equations is given by 
\be 
E\,=\,-4\big[\cos(\varphi_1)\,+\,\epsilon \cos(\varphi_2)\big]_{-\infty}^{\infty}
\label{energ}
\ee
and is clearly independent of $\lambda$. 
Moreover, when $\epsilon=1$ each field has a kink-like solution and when $\epsilon=-1$ we have a model with a kink and an anti-kink solution. The fields of such solutions had to be determined numerically.
They were very similar to the familiar one kink (or one anti-kink) of the well known Sine-Gordon model
except that they also had small ``bumps" in their shapes at the position of the kink (or anti-kink)
of the other field. The height of these ``bumps" depended on the value of $\lambda$.

In \cite{first} we discussed various properties of such solutions. As the solitons are very well localised
we looked at configurations in which the solitons of each field were well separated and then we studied 
their behaviour when we sent these structures towards each other.  And, of course, we have looked at the dependence
of the obtained results on $\lambda$. 
The results were very interesting and we have described them in detail in \cite{first}.
However, there are many more interesting questions that one can ask about such systems.
One class of problems involves interactions between two solitons in one field. We know, that if the consider 
a simple one field Sine-Gordon model its kinks repel and a system involving a kink and an anti-kink
attract and they evolve into a breather (emitting some radiation in the process). This has been seen in processes
involving two static solitons (originally placed not too far away from each other, nor too close either).
Of course, we can also send the initial structures towards each other.
As the kinks solutions are well localised this can be easily studied.
The question then arises what happens, how these properties change, if we look at such problems in our 
two field model. These are the problems we analyse in this paper.

\section{Interactions between two solitons in each field}

First we looked at the interaction of two solitons in each field. To do this we have considered  a system of two solitons placed at
rest or sent to each other with a constant velocity. Then we considered systems involving solitons and antisolitons.
Their evolution is described by  their equations of motion, which in our case, are given by
\be
(\partial_t^2-\partial_x^2)\vp_1\,+\,\frac{4}{4-\lambda^2}\left(\frac{\partial V}{\partial{\vp_1}}\,+\, \frac{\lambda}{2}\frac{\partial V}{\partial{\vp_2}}\right)\,=\,0,
\label{eqm}
\ee
and a similar equation for $\vp_2$ with $\vp_1$ and $\vp_2$ interchanged in the expression above. 

\subsection{Two static solitons}
To consider two static solitons we sew together two one soliton fields (one from $0$ to $\pi$ and another from $\pi$ to $2\pi$).
This was done for each field $\varphi_i$. Then we performed simulations starting with each soliton at rest ({\it i.e.} with the initial
time derivative of each field being set to zero).
Of course, as each field is strongly localized, the behaviour of such a system depended on the initial distance between the two
solitons in each field. For larger distances, there was no motion. So we repeated the evolution with the solitons being
initially placed closer and closer to each other. The resultant behaviour depended on the values of $\lambda$.
This was not surprising as the energy of the initial configuration depended on $\lambda$ and it increased a little with the reduction of the distance between the solitons.
\begin{figure}[h!]
\centering
  {\includegraphics[width=0.6\textwidth,height=0.3\textwidth]{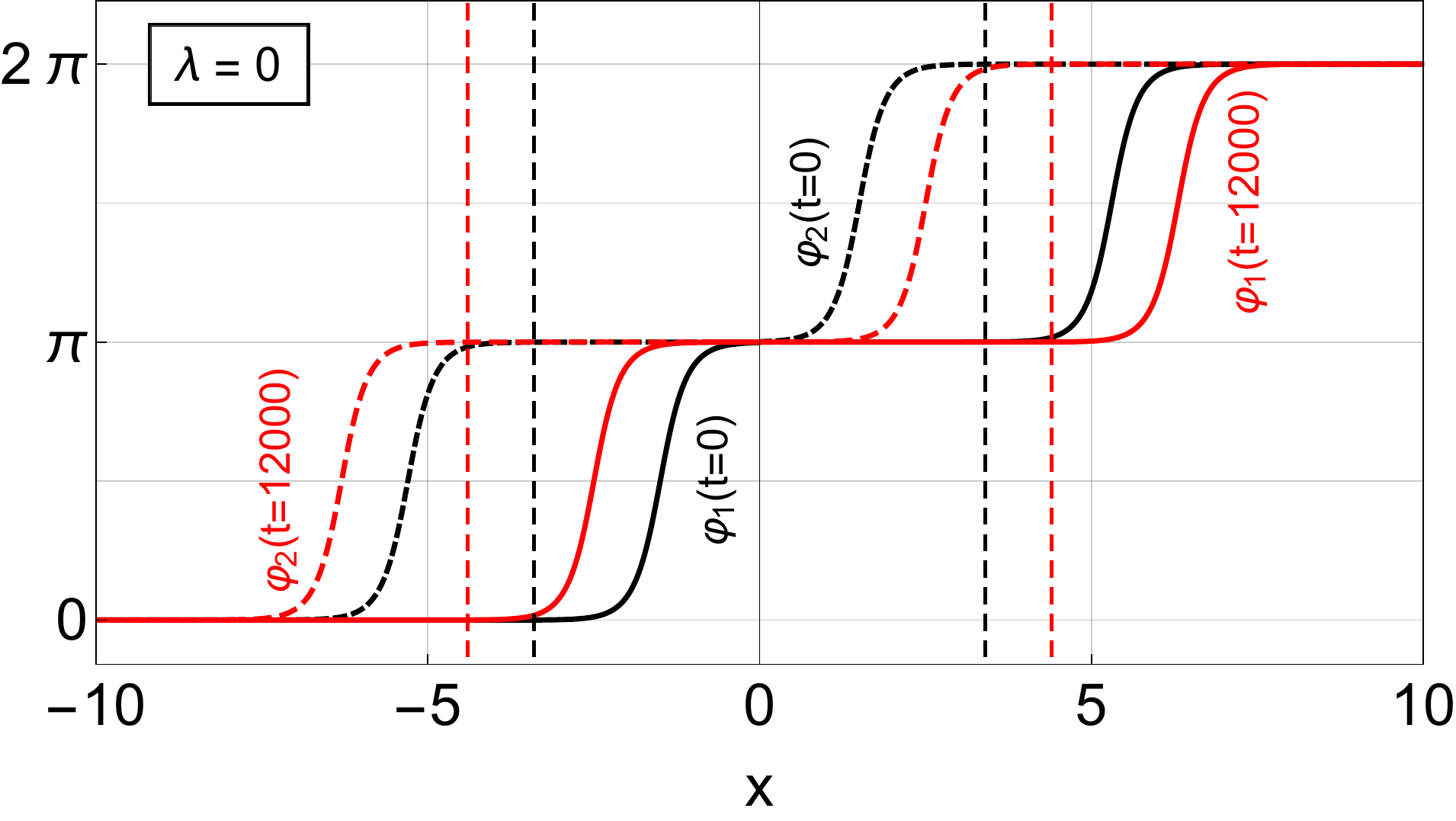}}
   \caption{The case for $\lambda=0$. Fields $\vp_1$ and $\vp_2$ at $t=0$ and at $t=12000$. The vertical lines correspond to $x=\pm 3.40$ and $x=\pm 4.40$. }
  \label{fig:8sol}
\end{figure}
 \begin{figure}[h!]
\centering
 {\includegraphics[width=0.6\textwidth,height=0.3\textwidth]{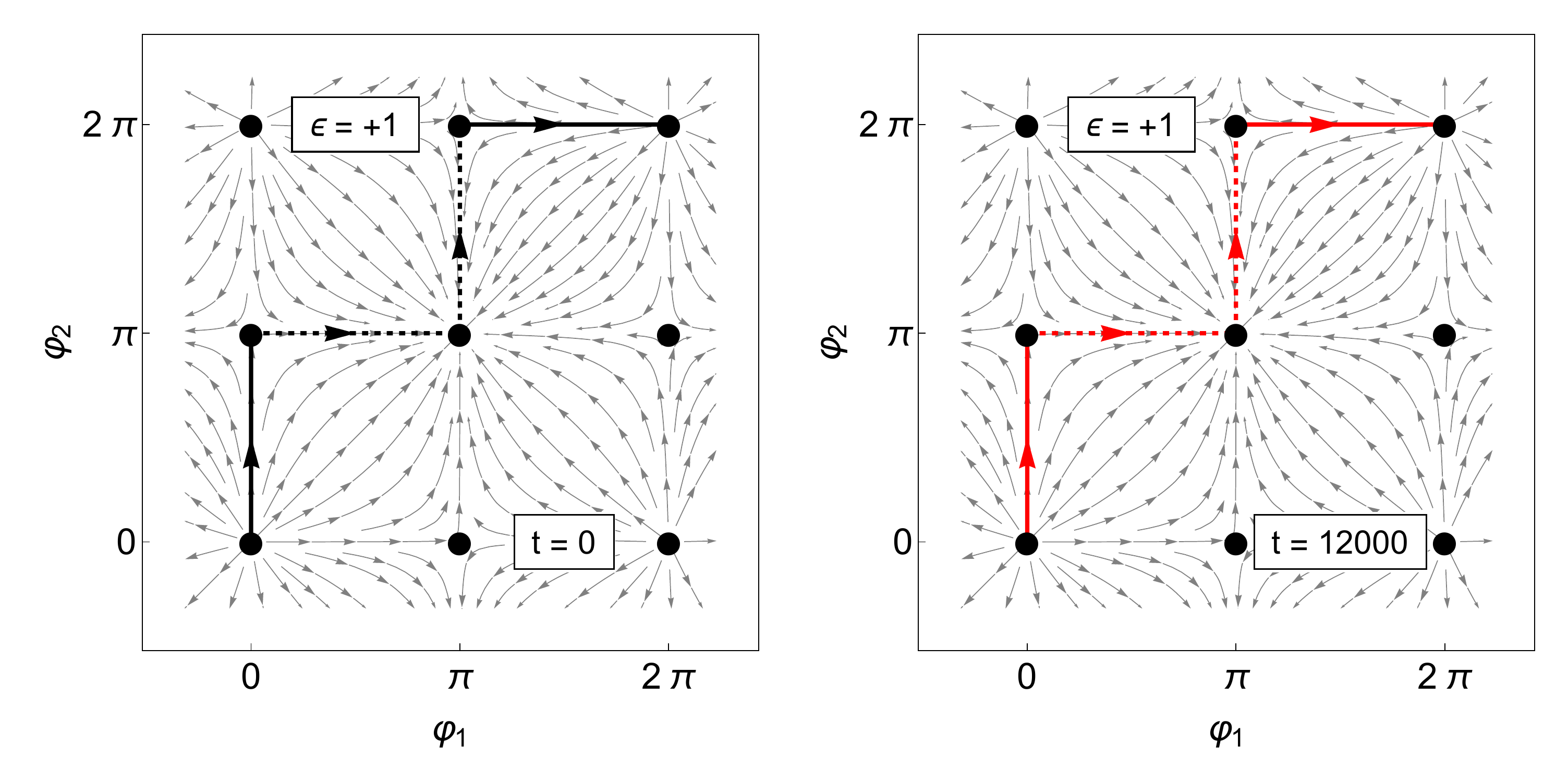}}                   
   \caption{The case for $\lambda=0$. Plot of the solution in space of fields on background provided by flows with $\epsilon=1$.}
  \label{fig:8solb}
\end{figure}

For $\lambda=0$, due to this increase of energy the solitons repelled each other (this is well known from the Sine-Gordon model
and for $\lambda=0$ the two fields $\varphi_i$ are decoupled and so the system of two independent Sine-Gordon fields  behaved as expected). 
\begin{figure}[h!]
  \centering
 {\includegraphics[width=0.8\textwidth,height=0.5\textwidth]{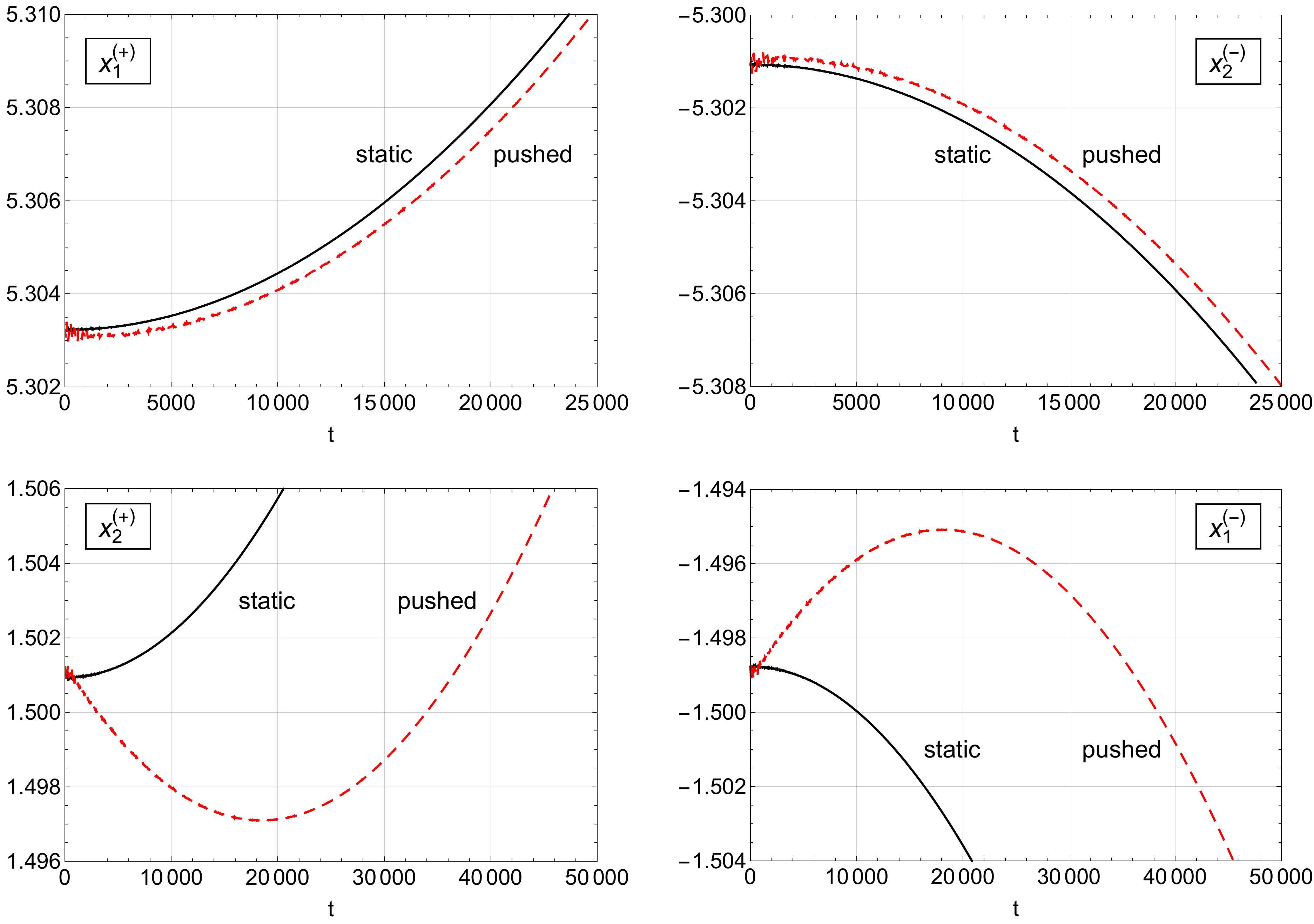}}                                
   \caption{The case $\lambda=0$. Trajectories $x_1(t)$ and $x_2(t)$ of solitons in field $\vp_1$ and $\vp_2$. They are defined as $\vp_a(x^{(-)}_a)=\frac{\pi}{2}$ and $\vp_a(x^{(+)}_a)=\frac{3\pi}{2}$ where $a=1,2$.}
  \label{fig:88sol}
\end{figure}
In Fig.\ref{fig:8sol} we present the plots of the $\varphi_1$ and $\varphi_2$ fields, when they are initially, ({\it i.e.} at $t=0$), placed sufficiently close to each other. 
We run this simulation for quite a while and the solitons started to move away from each other. The  configuration of the fields at $t=12000$ is shown in Fig.\ref{fig:8sol}. We see, from these plots, that the motion is very slow (hence our ``final'' plots of the fields are not that different from the initial ones) but we clearly see that the solitons located initially at the negative values of $x$ are moving to the left and those for positive values are moving to the right - {\it i.e.} away from each other. 
Looking at the evolution of the fields in the  $(\vp_1,\vp_2)$ space, shown in Fig.\ref{fig:8solb}, we do not see any significant difference between the curve  at $t=12000$ and the initial one. We have split the curve into three segments separated by the points where the curve crosses the line $\vp_1+\vp_2=\pi$ for first time and where it crosses  the line $\vp_1+\vp_2=3\pi$ for the last time. As for other values of $\lambda$ the behaviour of the curve close to points $(0,\pi)$ and $(\pi,2\pi)$ is quite complicated, in what fololows,  we shall use this, to some extend somewhat arbitrary, choice as a method of splitting multi-kinks into single kinks. The values of variable $x$ corresponding to these points are marked by vertical lines in Fig.\ref{fig:8sol}. 

Next we have made the solitons to move slightly towards each other; in this case they slowed down and turned around, as to be expected. In Fig.\ref{fig:88sol} we present the plots of the trajectories of solitons (dashed lines) when two of them, initially, were sent towards the other one. The solid lines correspond to the trajectories of solitons initially at rest analysed in the previous example.

For other values of $\lambda$ the results were slightly different.
\begin{figure}[h!]
  \centering
 {\includegraphics[width=0.75\textwidth,height=0.4\textwidth]{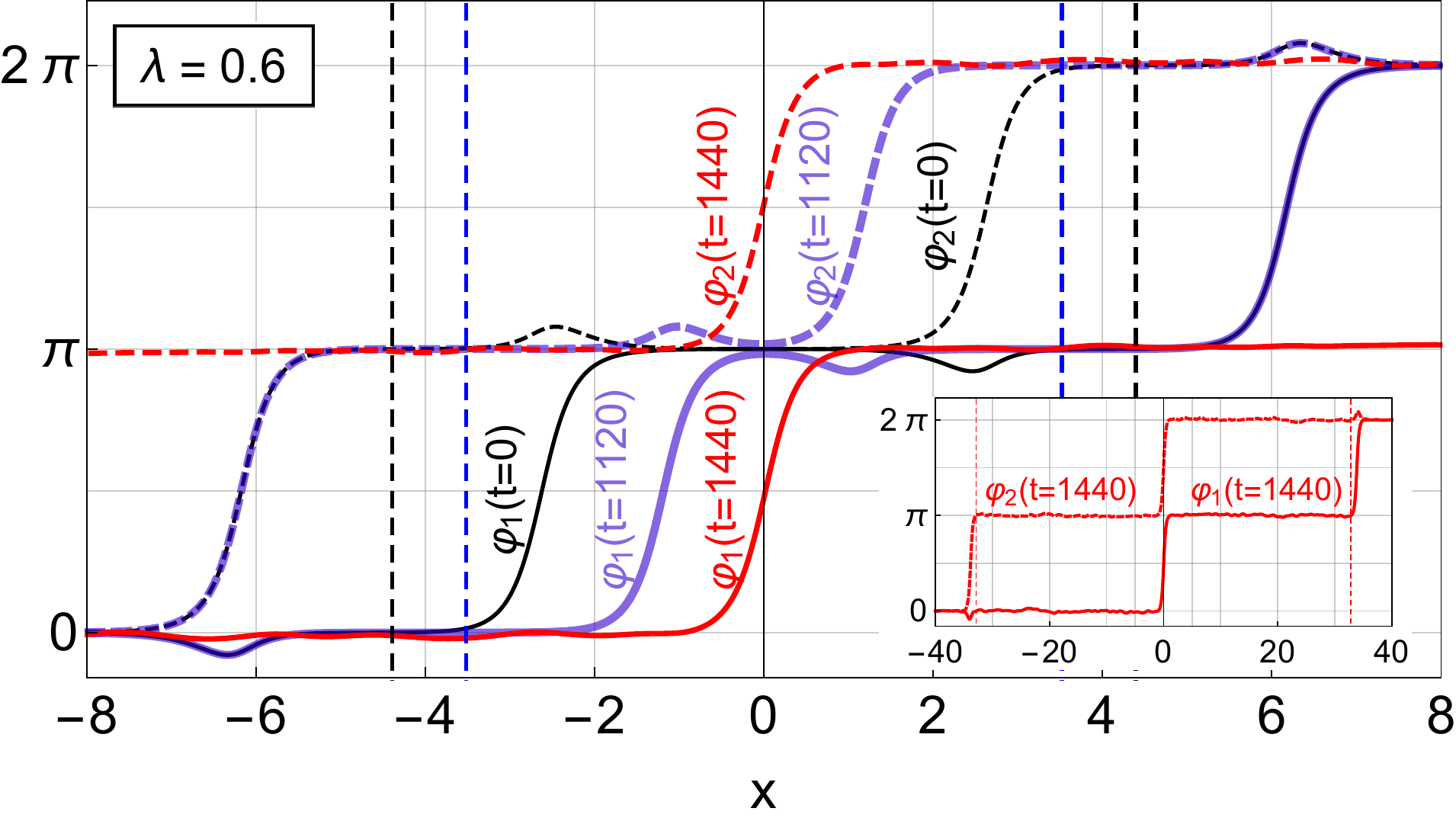}}   
   \caption{Evolution of the static fields corresponding to $\lambda=0.6$ at $t=0$, $t=1120$ and  $t=1440$. The vertical lines correspond to $x=\pm4.39$, $x=\pm3.52$ and $x=\pm 32.85$. }
  \label{fig:9sol}
\end{figure}
For relatively small values of $\lambda$ (at least up to $\lambda\sim 0.6$) we
saw the behaviour very much like for $\lambda=0$; later it has changed.
Thus for $\lambda=0.6$ the 4 solitons (in two fields) came together, then they interacted and produced a bound state at the symmetric point ($x=0$ in our case) with two external solitons moving off to the boundaries. 
\begin{figure}[h!]
  \centering
 {\includegraphics[width=0.95\textwidth,height=0.3\textwidth]{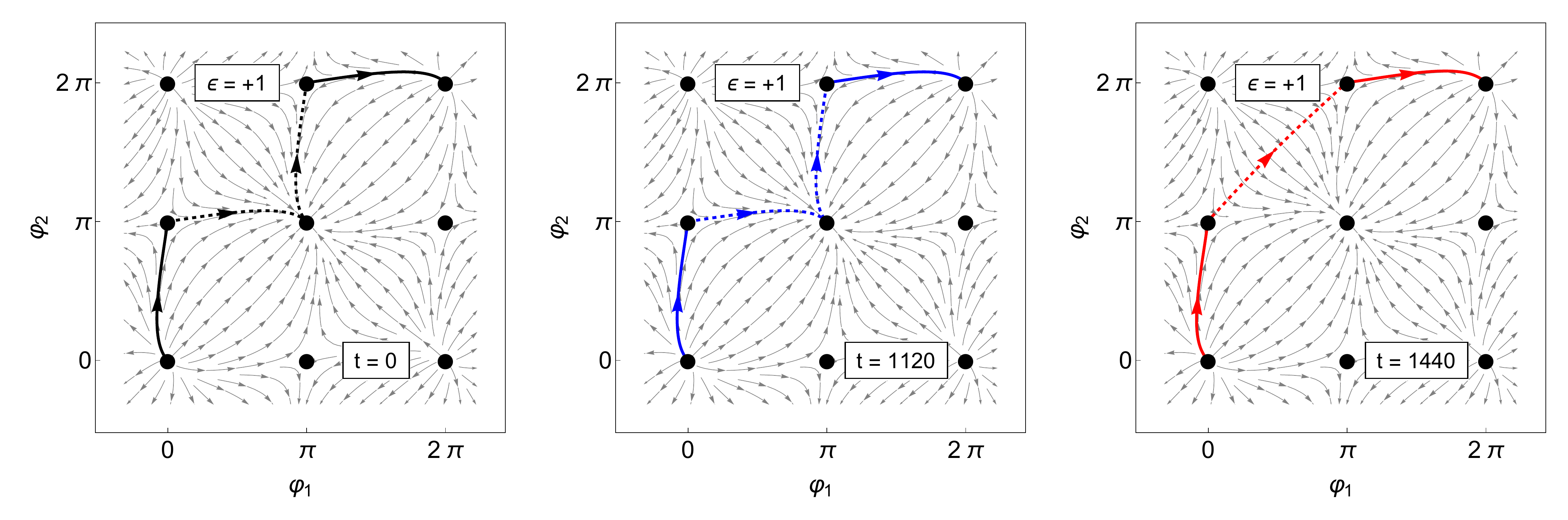}}   
   \caption{The $\lambda=0.6$  case. Evolution in space of fields on background provided by flows with $\epsilon=1$.}
  \label{fig:9solb}
\end{figure}
In Fig.\ref{fig:9sol} we present the plots of $\varphi_1$ and $\varphi_2$  seen in such a simulation, for three instants of time, namely: $t=0$, $t=1120$ and $t=1440$. 
Looking at the evolution of the system in the space of fields shown in Fig.\ref{fig:9solb} we observe that, differently from the case $\lambda=0$, the curve changes its shape a lot during the evolution of the system. When central kinks have different positions the curve passes close to the point $(\pi,\pi)$ (following approximately the modified gradient flow for $\epsilon=+1$) but when central kinks are very close to each other, the corresponding part of the curve does not follow this flow anymore.  The central part of the curve in the $(\vp_1,\vp_2)$ space became orthogonal to the flow $\epsilon=+1$ whereas  the initial and the final parts of the curve, that describe the behaviour of external kinks which escape to infinity, follow the modified gradient flow very closely.

\begin{figure}[h!]
  \centering
 {\includegraphics[width=0.75\textwidth,height=0.4\textwidth]{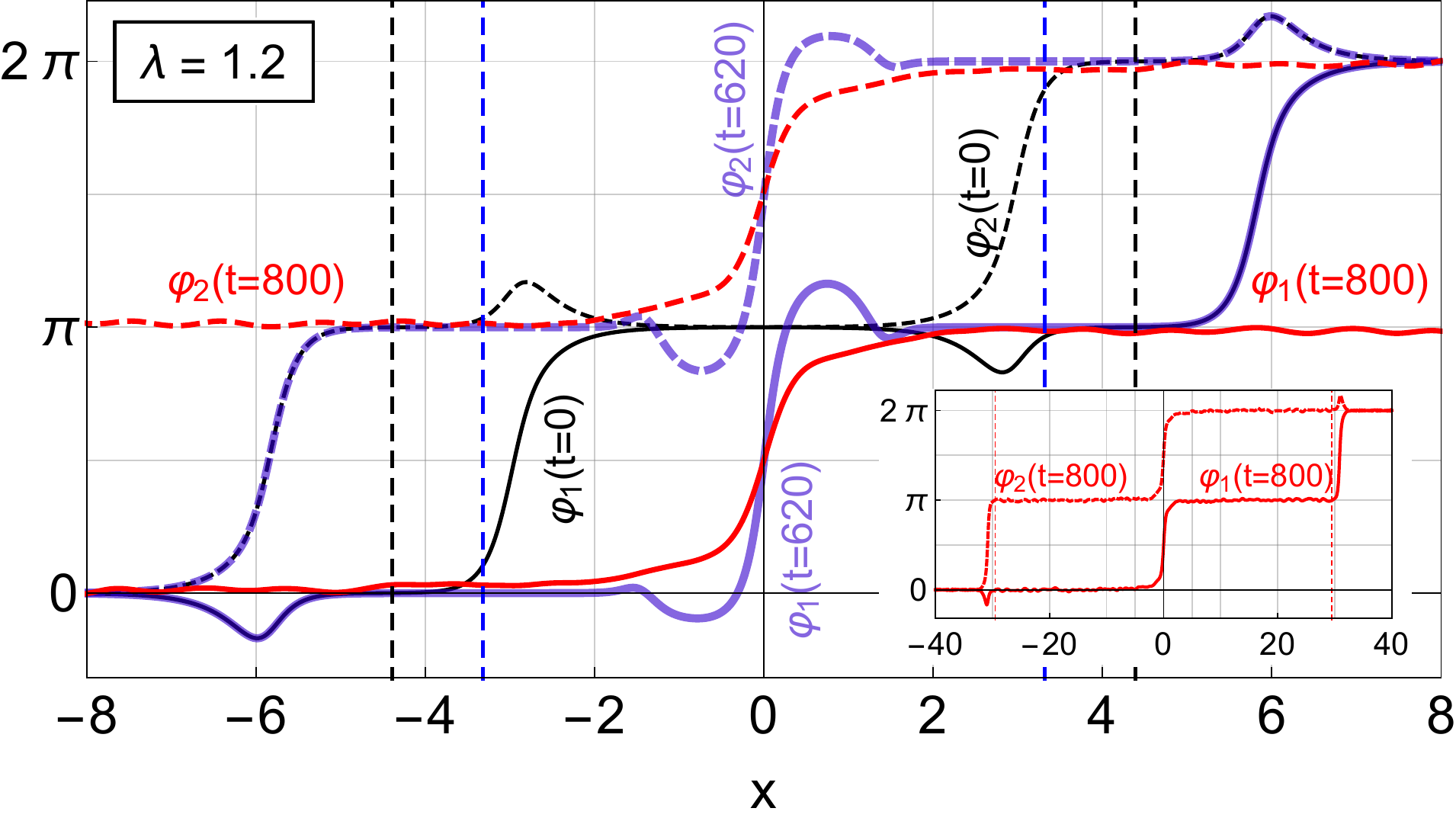}}   
   \caption{Evolution of the static fields corresponding to $\lambda=1.2$ at $t=0$,   $t=620$ and  $t=800$. The vertical lines correspond to  $x=\pm4.39$,  $x=\pm2.65$ and $x=\pm 29.51$ }
  \label{fig:10sol}
\end{figure}

We have also looked at larger values of $\lambda$.  For $\lambda=0.8$ the simulation still showed results essentially similar to those for $\lambda=0.6$. For larger values of $\lambda$ the evolution looked somewhat 
different.
In Fig.\ref{fig:10sol} we present the plots of the simulation for $\lambda=1.2$ at three instants of time, namely: $t=0$, $t=620$ and $t=800$. In this case, initially the simulation followed the path of the $\lambda=0.6$ one {\it i.e.}  a kink in one field and a ``bump''  in the other field  moved in the direction of larger values of $|x|$.  At the same time, two central kinks (bumps) were attracted towards $x=0$, where they have formed a metastable bound state. During this process lots of energy was emitted and these waves of energy affected what would happen next.
\begin{figure}[h!]
  \centering
 {\includegraphics[width=0.95\textwidth,height=0.3\textwidth]{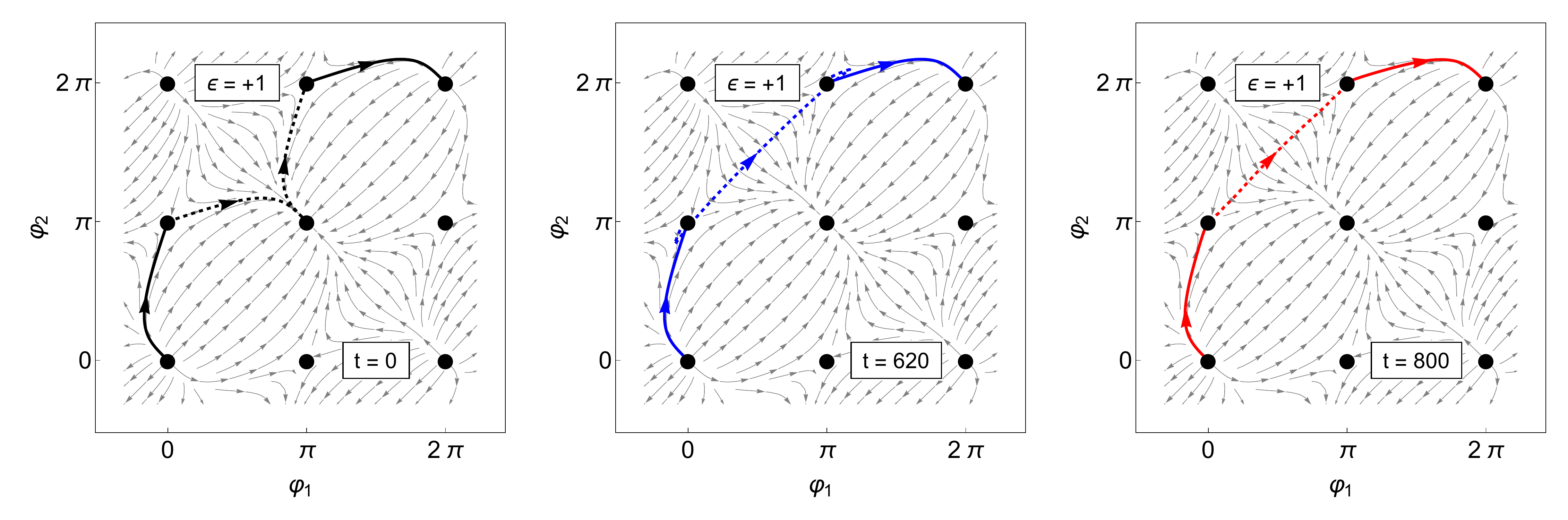}}   
   \caption{The $\lambda=1.2$  case. Evolution in space of fields at $t=0$,   $t=620$ and  $t=800$ on background provided by flow with $\epsilon=1$.}
  \label{fig:10solb}
\end{figure}
\begin{figure}[h!]
  \centering
 {\includegraphics[width=0.45\textwidth,height=0.4\textwidth]{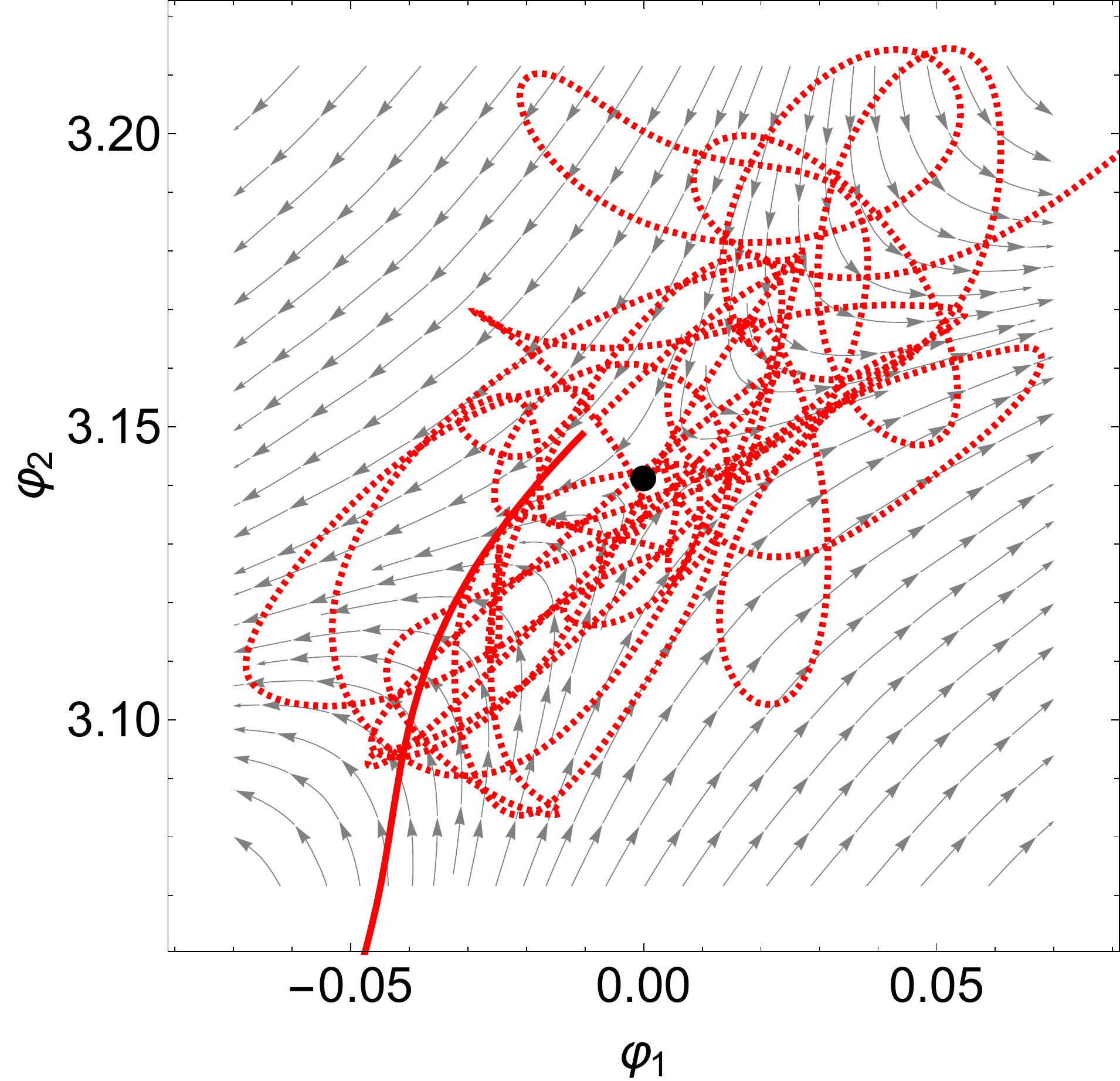}} 
   \caption{ Trajectory in a vicinity of the point $(0,\pi)$ at $t=800$.}
  \label{fig:background}
\end{figure}
Thus, four initial peaks of the energy density have evolved at this stage into three peaks and some waves of radiation energy in the system.
We have looked at the curve representing solutions in the space of fields. In Fig.\ref{fig:10solb} we present its form in the background provided by the modified gradient flow for $\epsilon=1$. The bound state corresponds to a straight line segment connecting points $(0,\pi)$ and $(\pi,2\pi)$. The straight segments of the curve are orthogonal to the gradient flow. In the vicinity of these points the linearity is lost and the curve becomes very complicated.  In Fig.\ref{fig:background} we show a blow up of the region in the vicinity of the point $(0,\pi)$. This is very much as in the example with $\lambda=0.6$.

\begin{figure}[h!]
 \centering
  \subfigure []{\includegraphics[width=0.55\textwidth,height=0.35\textwidth]{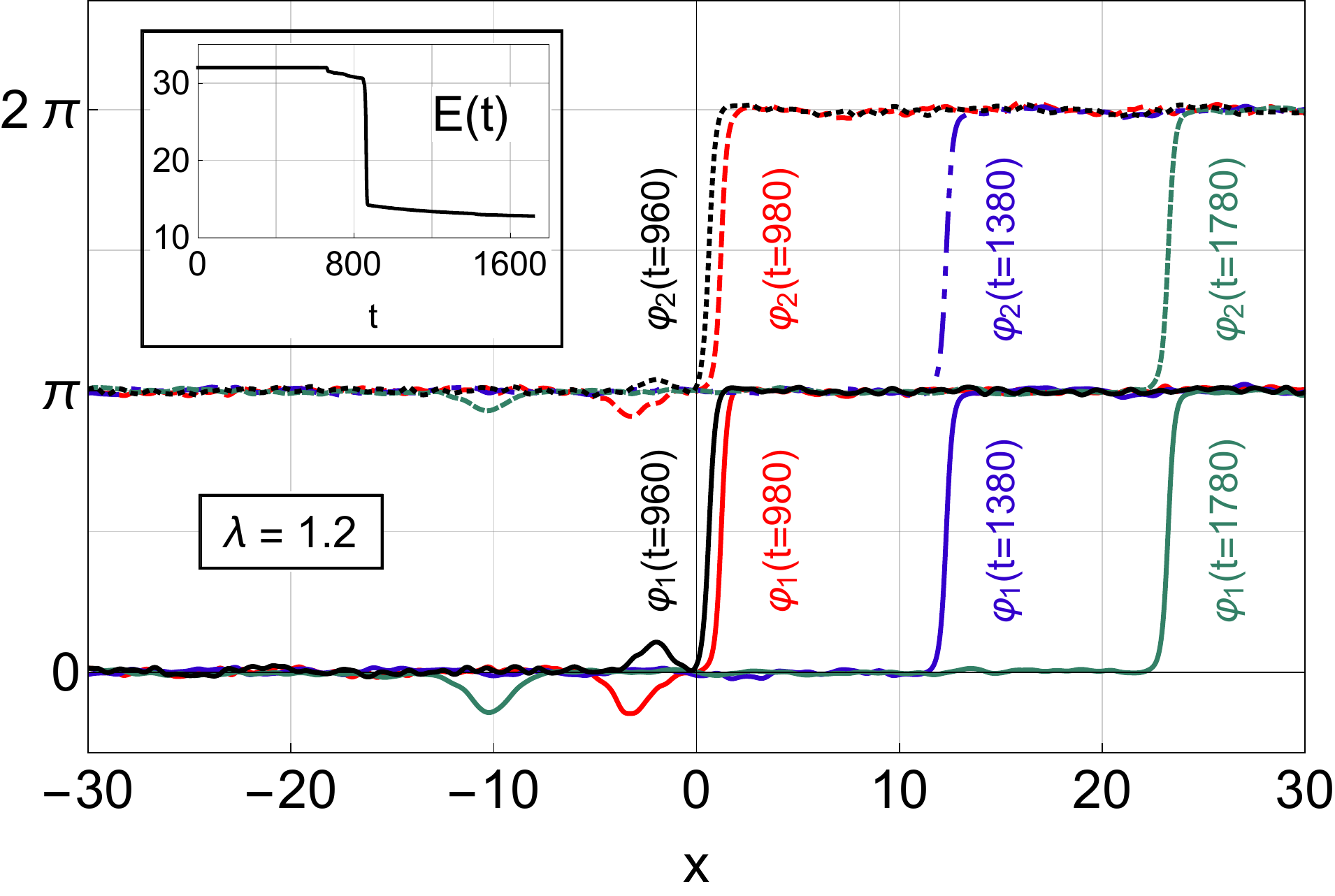}}
  \hskip0.5cm
 \subfigure[]{\includegraphics[width=0.4\textwidth,height=0.35\textwidth]{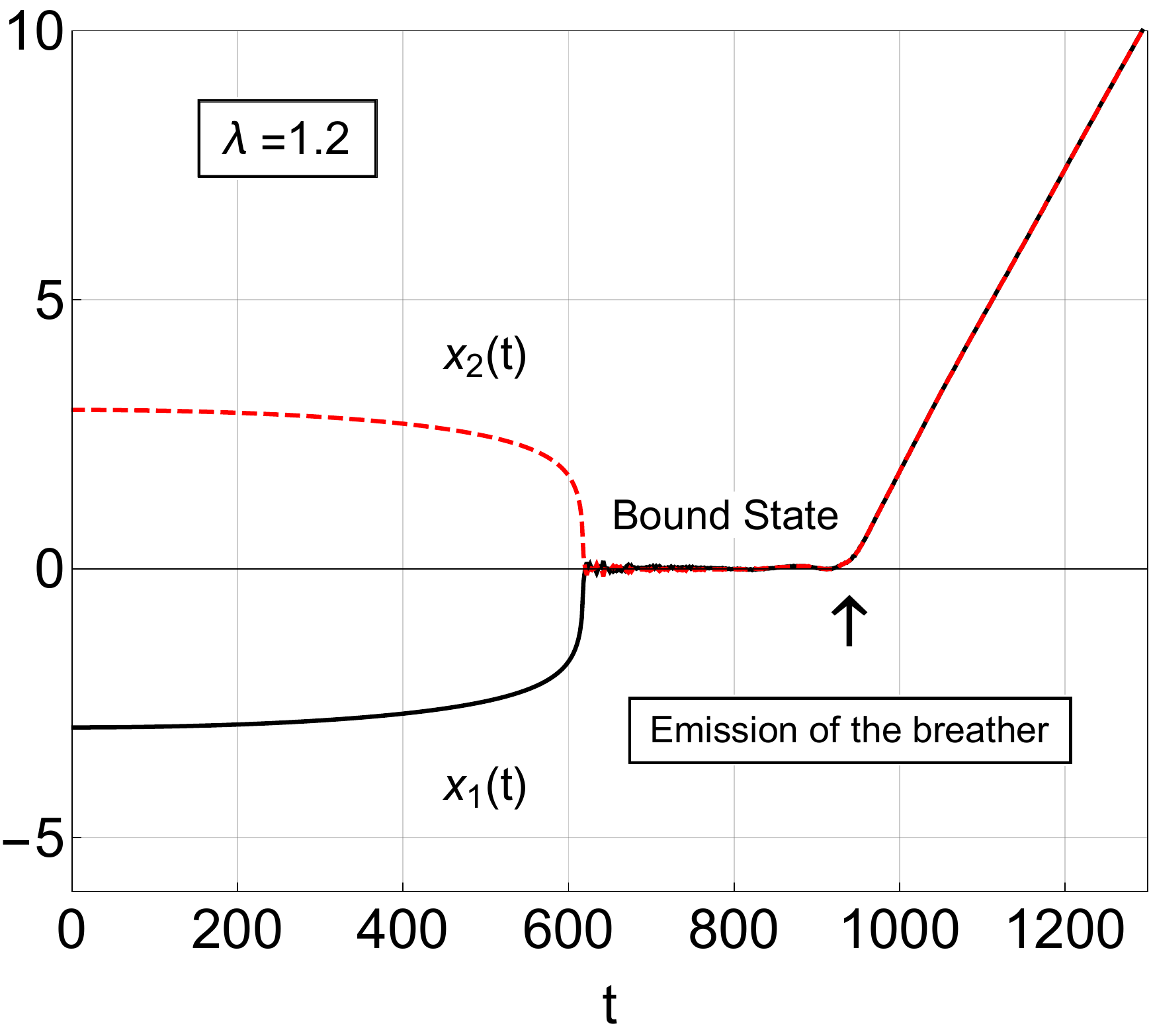}}
   \caption{a) Further evolution of the static fields corresponding to $\lambda=1.2$ and the energy of numerical  solution on the segment $x\in[-40,40]$, b) Trajectories of the final soliton $x_1(t)$ and $x_2(t)$; where $\vp_1(x_1)=\frac{\pi}{2}$ and $\vp_2(x_2)=\frac{3\pi}{2}$.} 
  \label{fig:10sol2}
\end{figure}

At larger values of time we have found that for all values of $\lambda$,  the ``outer'' kinks would have eventually reached the boundaries $x=\pm40$ where they were absorbed. This absorption results in an abrupt decrease of the energy of the system in the box $x\in[-40,40]$ and  is clearly seen in the subfigure Fig.\ref{fig:10sol2}(a) (we plot there the energy in the box). So what was left at this stage involved the middle energy peak and the waves of energy from the previous stages of the evolution.

From then on the system evolved via the interaction between the middle peak of energy and the radiation. Looking at this interaction in more detail we have found that
 the main difference between the present case and the cases discussed before was the later  behaviour of the middle energy peak. 
 Namely, in the present case it has turned out that the middle peak was not stable and it emitted a ``breather-like'' structure and started moving in the opposite direction. The speed of the ``breather-like'' object was significantly larger than the initial speed of the kinks.
In figure Fig.\ref{fig:10sol2}(a) we present few snapshots of fields  $\varphi_1$ and $\varphi_2$ in the next stage of the evolution {\it i.e.} after the emission of a breather-like object.  The breather reflected from the boundary (this is a purely numerical effect as, in an infinite system, it would have moved away to `infinity'). The breather could also have got absorbed 
at the boundary (this depends on the breather's phase when it reaches the boundary) and, in our simulation, it bounced back and returned towards the moving soliton.
The emission of the breather led to the motion of the general  bound state structure in the opposite direction to the original motion of the emitted breather. 
 This behaviour is clearly seen  as the motion of the bound state of the two kinks in direction of the positive axis $x$. In Fig.\ref{fig:10sol2}(b) we present plots of the trajectories $x_1(t)$ and $x_2(t)$ of the central kinks. They represent the motion of the points $\varphi_1=\frac{\pi}{2}$ and $\varphi_2=\frac{3\pi}{2}$. 

\begin{figure}[h!]
  \centering
 \includegraphics[width=0.65\textwidth,height=0.4\textwidth]{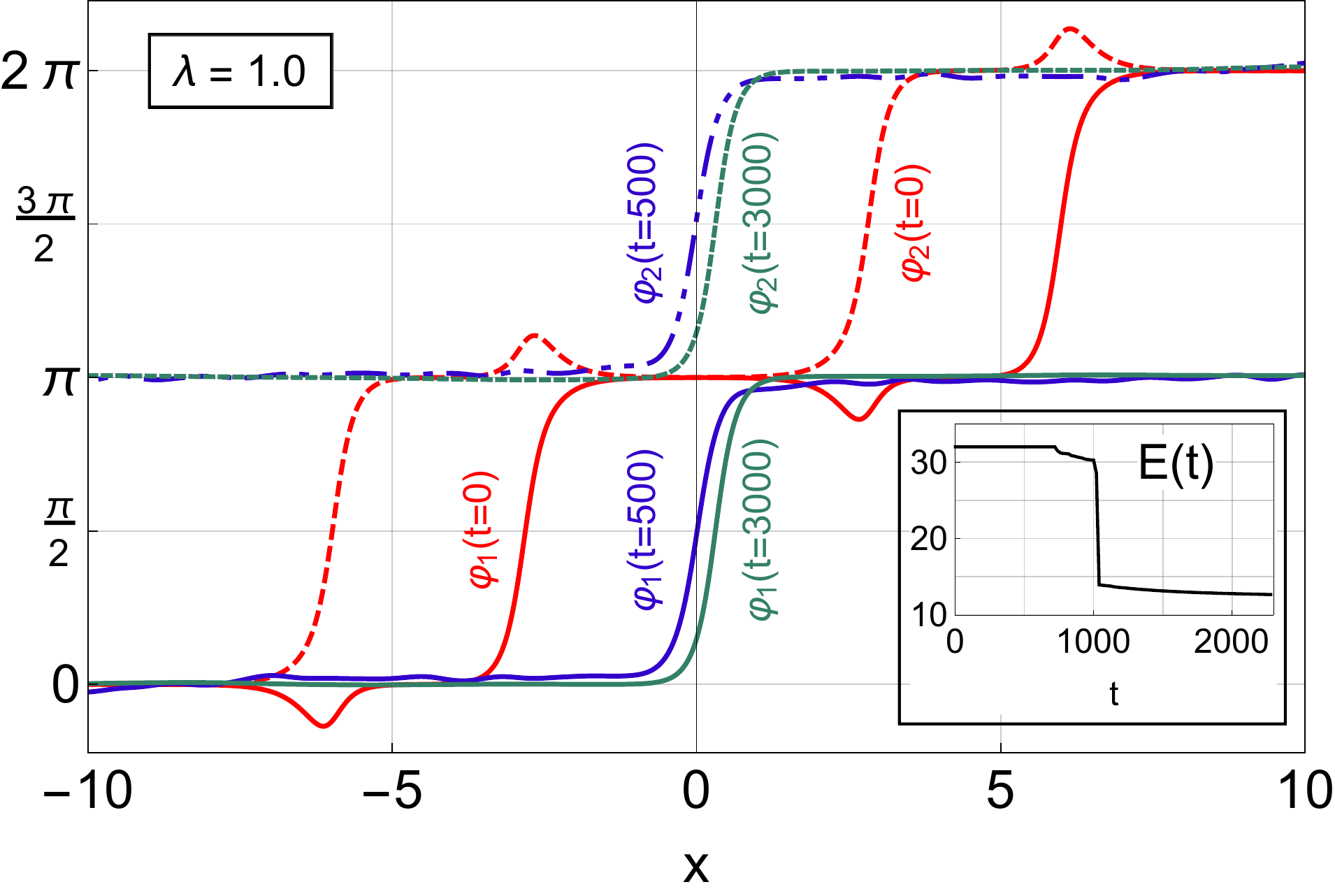}                
   \caption{Fields $\varphi_1$ and $\varphi_2$ for $\lambda=1.0$ at $t=0$, $t=500$ and $t=3000$. In the insertion - the time dependence of the energy seen in this simulation on $x\in [-20,20]$.}
  \label{fig:10new}
\end{figure}

The observed behaviour is quite generic, and we discuss its origin in a short subsection later one. Before, however, let us present the results of another simulation corresponding to the case of $\lambda=1.0$. 
In Fig.\ref{fig:10new} we present the plots, at $t=0$, $t=500$ and $t=3000$,  of $\varphi_1$ and $\varphi_2$ obtained in a simulation for $\lambda=1.0$  and also the time dependence of the total energy of the fields at $x\in[-40,40]$.
We note that this time, like in the previous case, after the `outside' solitons have moved out and were absorbed at the boundaries,
the fields $\varphi_1$ and $\varphi_2$ were very similar to each other except that they were nonzero from $0$ to $\pi$ ($\varphi_1$) and $\pi$ to $2\pi$ ($\varphi_2$) and the solitons remained localised around $x\approx0$. In this they resembled what has happened for $\lambda=1.2$ in the period of time between 
$t\sim600$ and $t\sim950$ {\it i.e.} before the breather was emitted.
Looking at it very carefully we note that the bound state  of solitons also did move a little (compare the plots of $\varphi_i$ at $t=500$ and $t=3000$). So there was probably also an emission of a very small breather but this breather was too small to be visible in our plots.

\subsubsection {A possible explanation}
Note that the final state of the simulation for $\lambda=1.0$ has a soliton (in $\varphi_1$) going from 0 to $\pi$ and a soliton (in $\varphi_2$) from $\pi$ to $2\pi$. Both solitons, together, are `locked' and so describe a `bound state'  located around $x=0$. This is also true initially for $\lambda=1.2$ of the previous figure.  Note also, that after the bound state had been formed, the two fields are essentially related, modulo small `breather's and radiation effects' so that $\varphi_2=\varphi_1+\pi$. Thus, $\sin(\varphi_2)=-\sin(\varphi_1)$ and our bound state system of two solitons can be interpreted as one soliton of one field corresponding to $\varphi=\varphi_1=\varphi_2-\pi$ plus some extra radiation.

The energy of such a field is given by
\begin{align}
    E\,=\,\frac{1}{2}{\int_{a}^{b}\,dx(2-\lambda)\Big[(\partial_x\varphi)^2\,+\,\frac{64}{4-\lambda^2}\sin^2(\varphi)\Big]},
\label{extra1}
\end{align}
where $\varphi\equiv\varphi_1=\varphi_2-\pi$ and $\varphi(a)\approx 0$ and $\varphi(b)\approx\pi$.


This expression looks like an energy of an exited one soliton of only one field.
The lowest energy of such a field is determined by the BPS condition for such a field which must satisfy
\begin{align}
\partial_x\varphi\,=\,-\frac{8\sin{\varphi}}{\sqrt{4-\lambda^2}}.
\label{extra2}
\end{align}
 The energy (\ref{extra1})  
 reads $E(\lambda)=8\sqrt{\frac{2-\lambda}{2+\lambda}}$ and so is clearly smaller  than the energy of our field
(see Fig.\ref{fig:10new} and compare with $E(\lambda=1.0)\approx 4.618$) so our field is emitting its excess of energy. The same has been true in all other cases of our simulations.


\subsection{Scattering of fields $\varphi_1$ and $\varphi_2$}

Next we performed some simulations of two soliton systems (corresponding to two kinks in each field), initially located at some distance from each other and sent towards each other with a small velocity (we used $v=0.1$). To do this we followed the same procedure
as before. Given that we were taking solitons far away from each we useed the Lorentz covariance of a separated soliton to give it velocity by setting $\frac{\partial\varphi_i}{\partial t}=v\frac{\partial \varphi_i}{\partial x}$
and correcting $dx$ by the Lorentz factor $\frac{1}{1-v^2}$.
This time we had to use larger grids and we experimented with using more lattice points or using slightly 
larger $dx$. In fact there no numerical differences so we used larger $dx$.

\begin{figure}[h!]
  \centering
 \subfigure[]{\includegraphics[width=0.55\textwidth,height=0.35\textwidth]{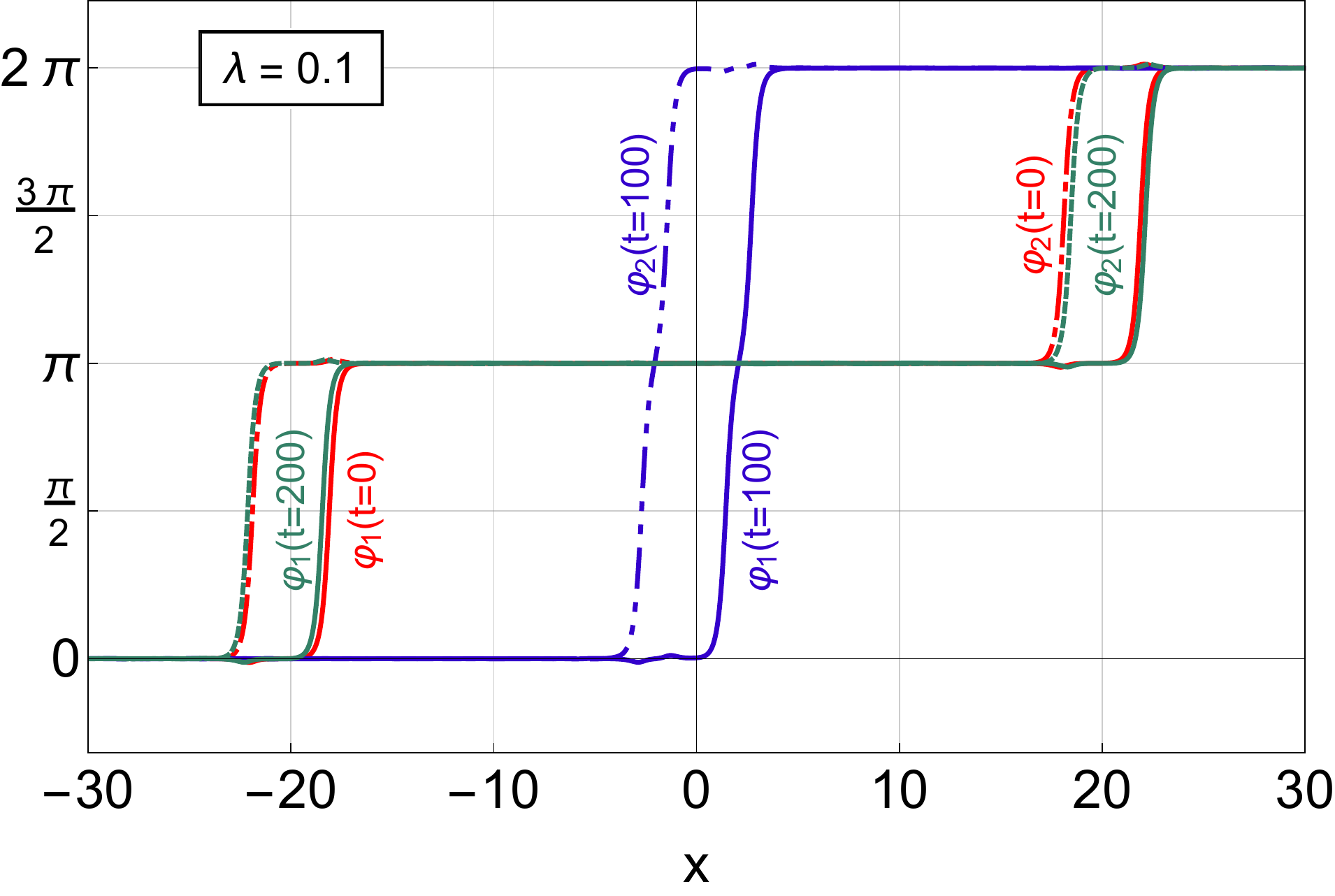}}
 \hskip0.5cm
 \subfigure[]{\includegraphics[width=0.4\textwidth,height=0.35\textwidth]{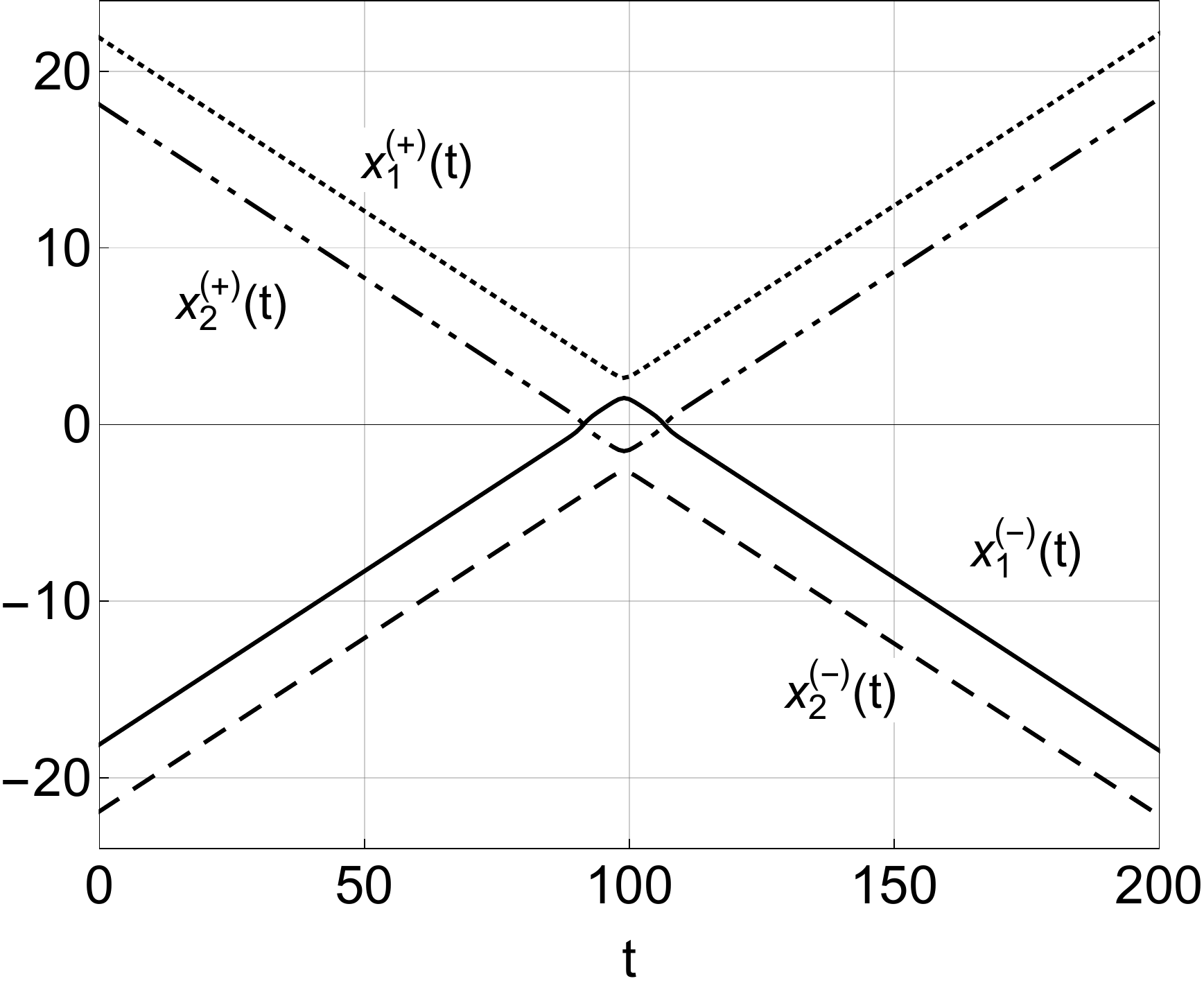}}              
    \caption{(a) Fields $\varphi_1$ and $\varphi_2$ for $\lambda=0.1$ at $t=0$, $t=100$ and $t=200$. (b) The time dependence of 
the trajectories of all solitons determined in this simulation, where $\varphi_a(x_a^{(-)})=\frac{\pi}{2}$ and $\varphi_a(x_a^{(+)})=\frac{3\pi}{2}$ for $a=1,2$.}
  \label{fig:11sol}
\end{figure}

We have performed many such simulations. They were all consistent with what we have observed for the static  systems - modified, of course, by their relative
motion towards each other. Here we present the results of some typical simulations.
\begin{figure}[h!]
  \centering
{\includegraphics[width=1.0\textwidth,height=0.6\textwidth]{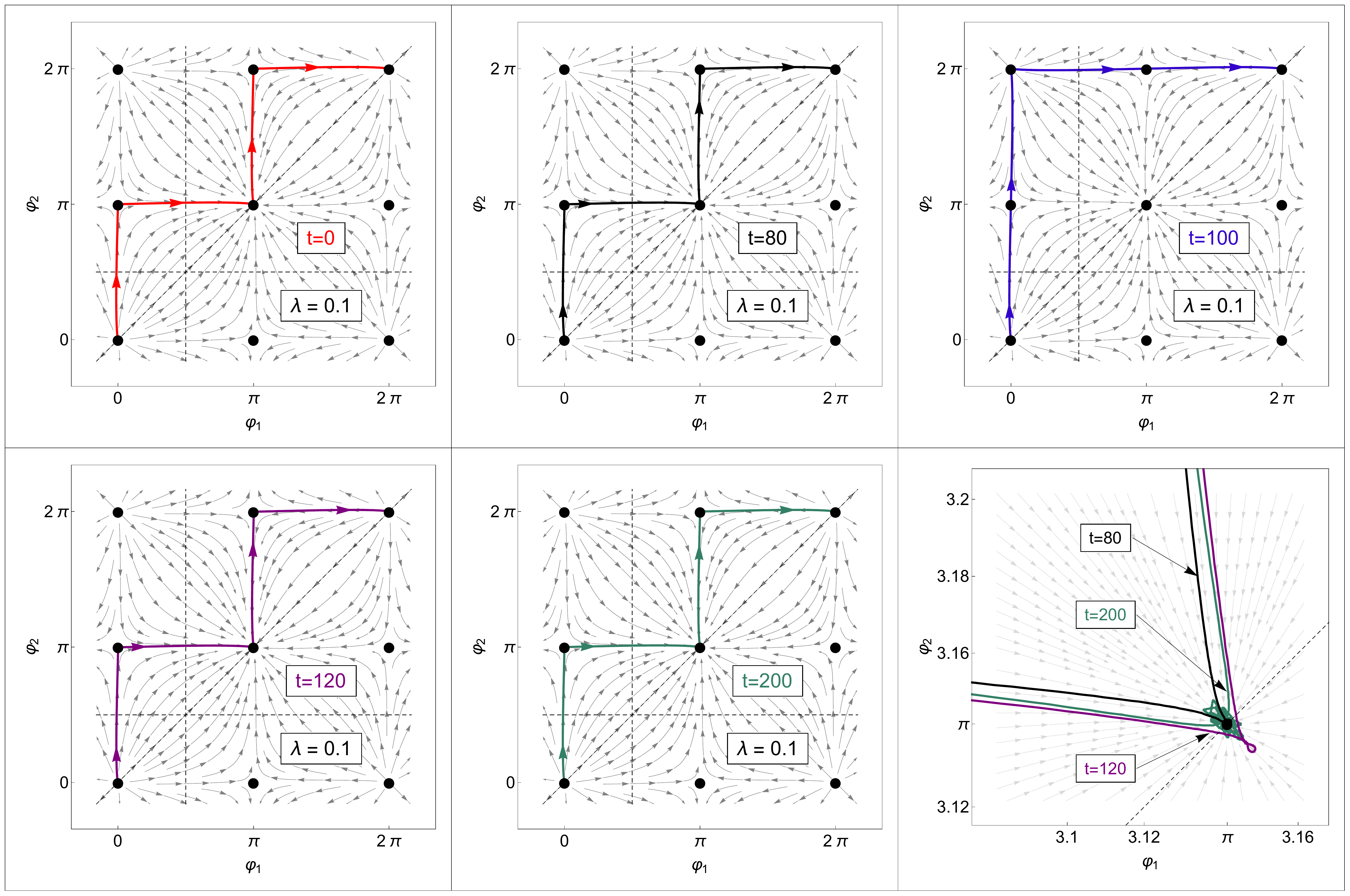}}         
    \caption{The $\lambda=0.1$ case at $t=0$, $t=100$ and $t=200$ on the background provided by flow with $\epsilon=1$.}
  \label{fig:11solc}
\end{figure}

For $\lambda=0$ or small values of $\lambda$ we saw straight  reflection. This is clear from the plots in Fig.\ref{fig:11sol}(a). They show the initial fields $\varphi_1$ and $\varphi_2$ at $t=0$ and later at $t=100$ and $t=200$ for $\lambda=0.1$.
The energy of the system was well conserved. Fig.\ref{fig:11sol}(b) shows the trajectories of the solitons. The plots are very similar to what was seen for $\lambda=0$ or $\lambda=0.2$. { In figures Fig.\ref{fig:11solc} we present the trajectories of the solutions in the space of fields for  5 values of time.}

\begin{figure}[h!]
  \centering
  \subfigure[]{\includegraphics[width=0.55\textwidth,height=0.4\textwidth]{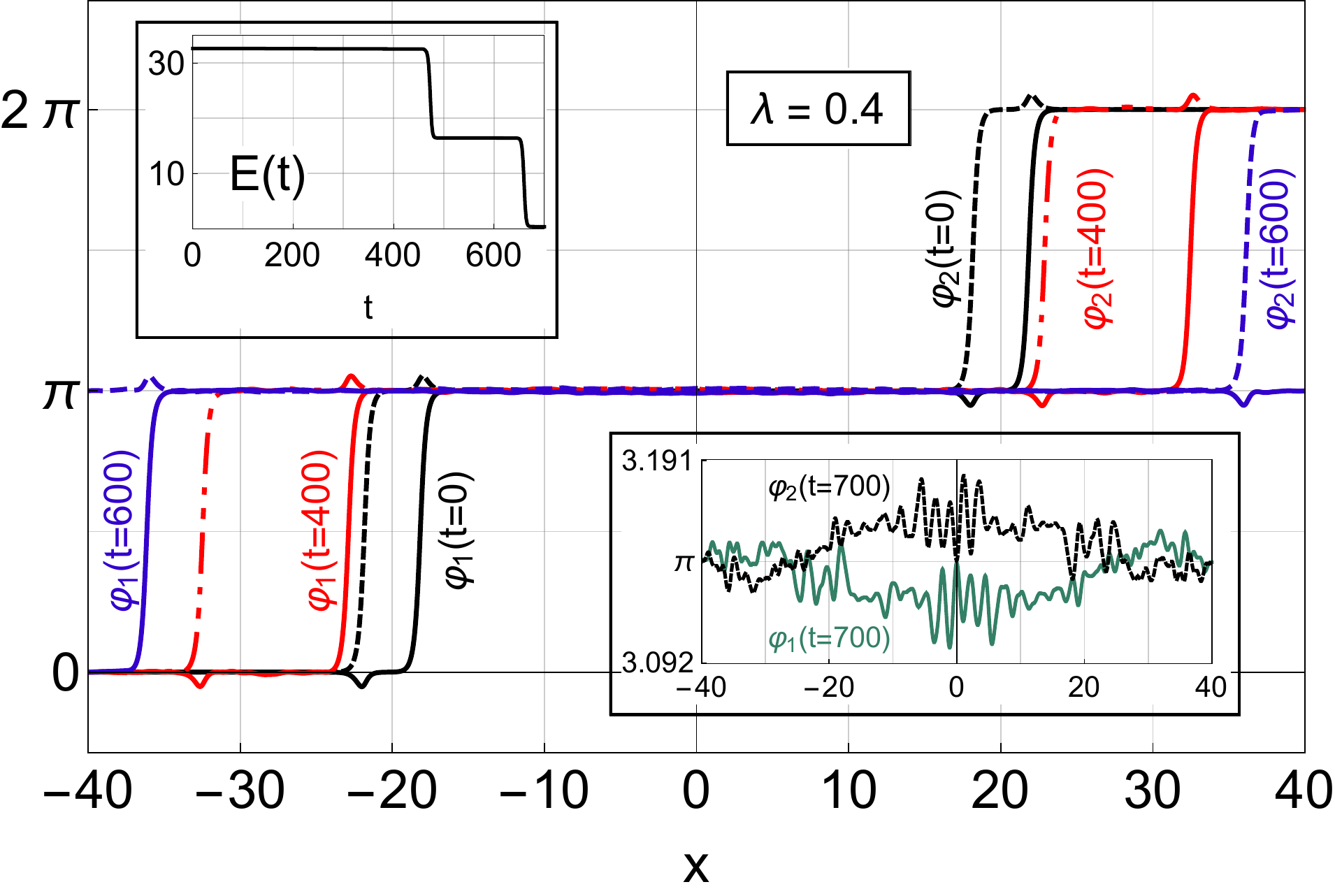}}
  \hskip0.5cm
   \subfigure[]{\includegraphics[width=0.4\textwidth,height=0.4\textwidth]{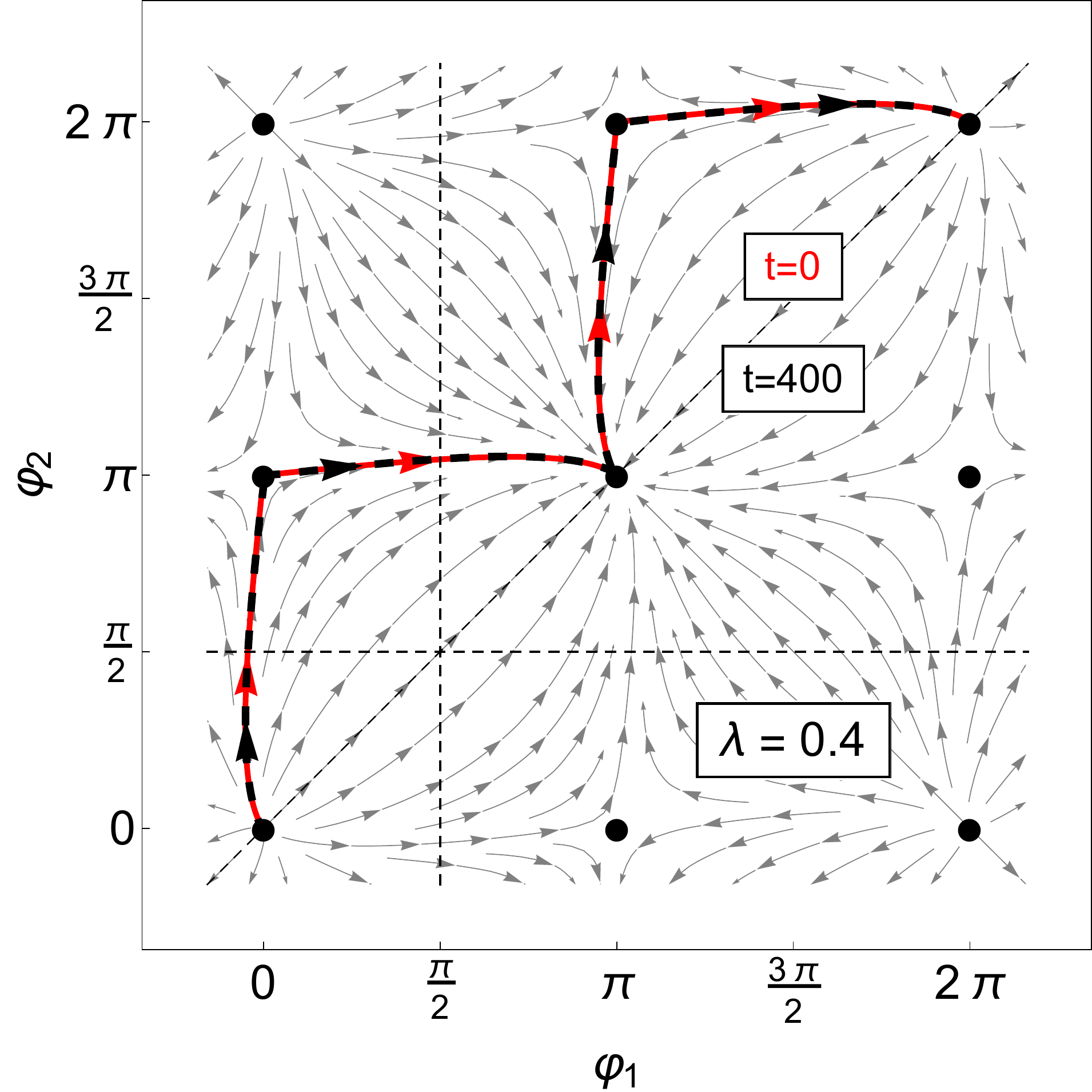}}
   \caption{(a) Fields $\varphi_1$ and $\varphi_2$ for $\lambda=0.4$ and energy of the system seen in this simulation. (b) Evolution in space of fields at $t=0$ and $t=400$. The flow corresponds to $\epsilon=1$.}
  \label{fig:12sol}
\end{figure}
\begin{figure}[h!]
 \centering
  \subfigure[]{\includegraphics[width=0.6\textwidth,height=0.35\textwidth]{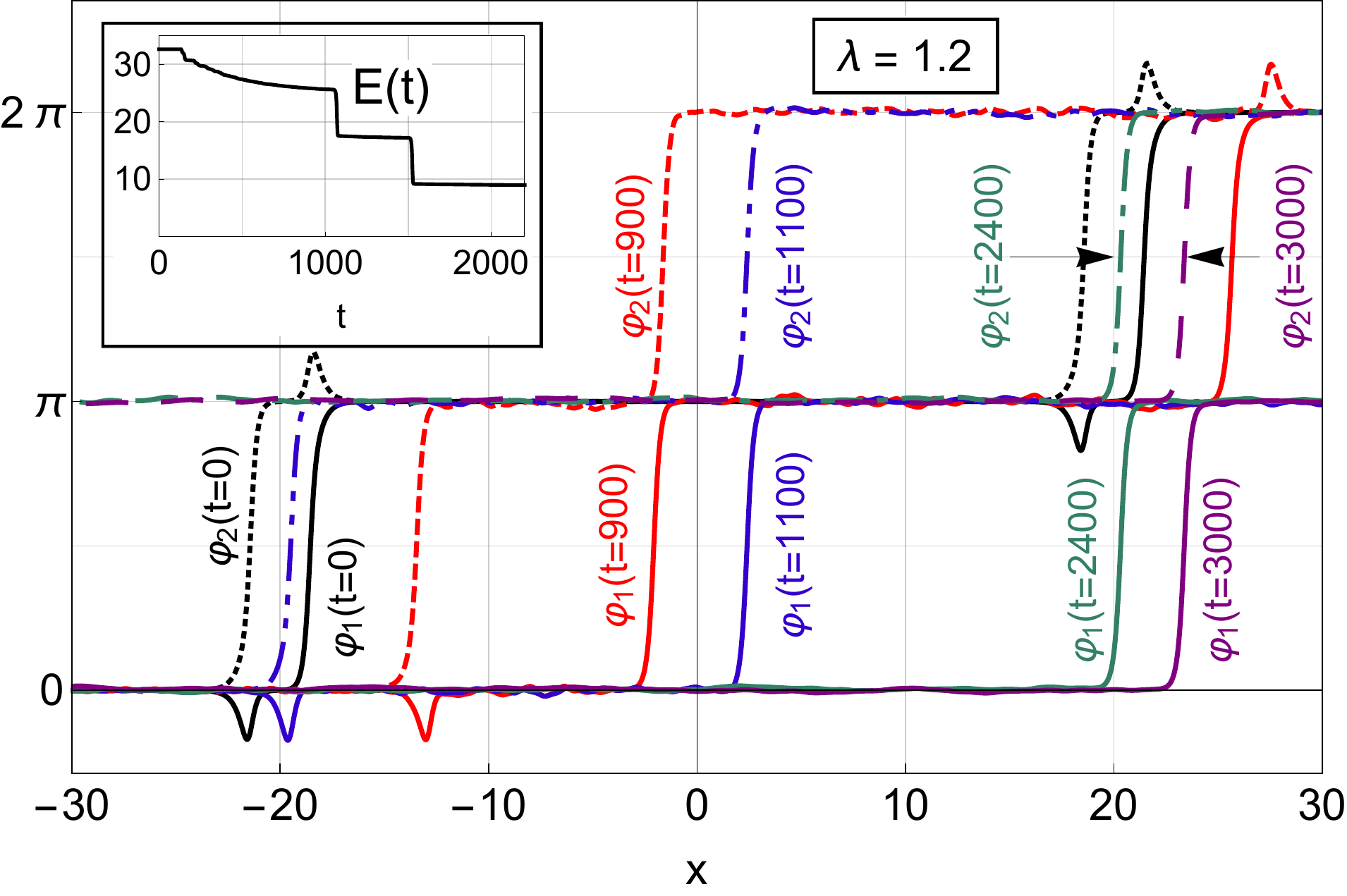}}
  \hskip0.4cm
   \subfigure[]{\includegraphics[width=0.35\textwidth,height=0.35\textwidth]{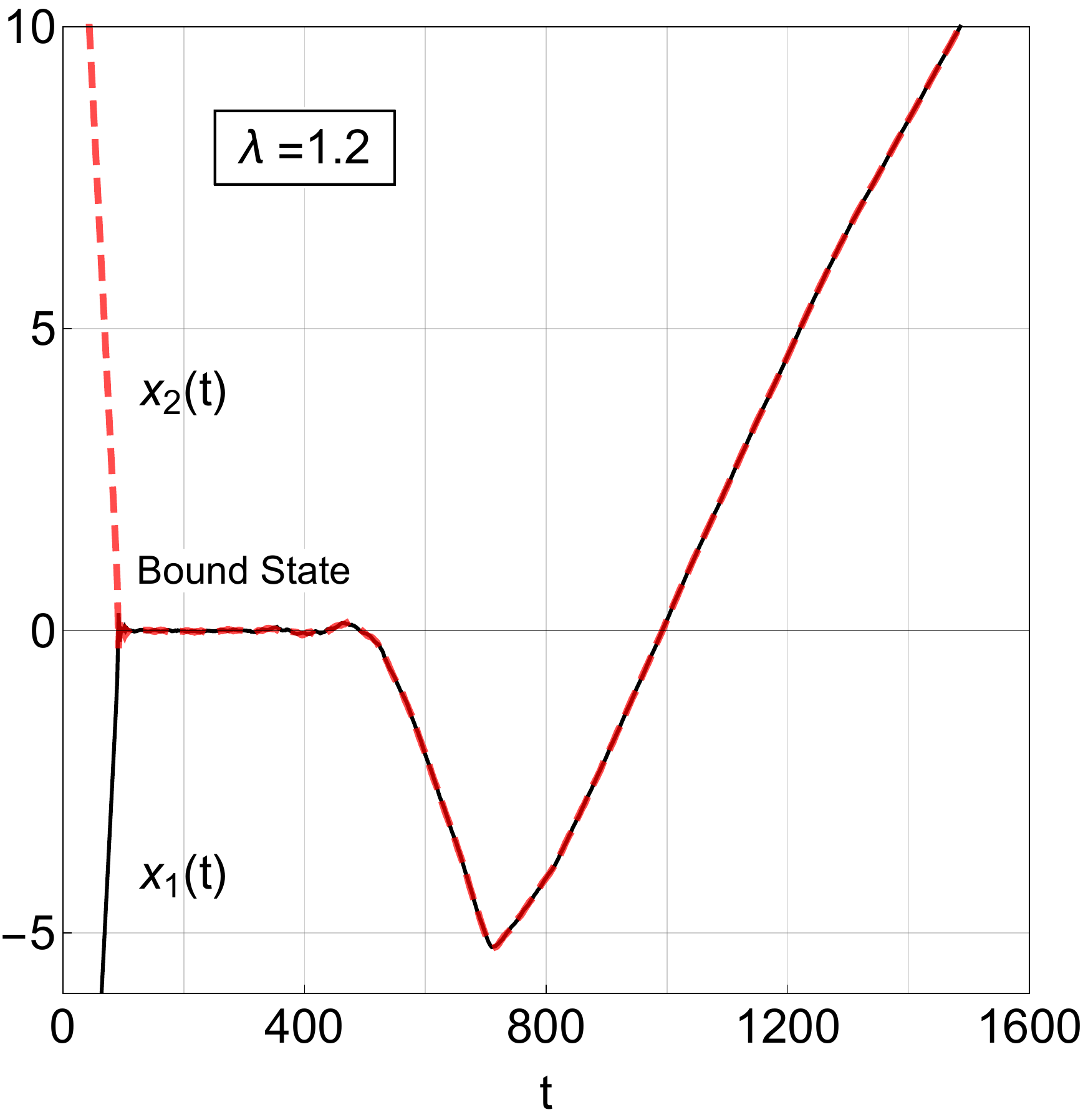}}

    \caption{(a) Evolution of the fields corresponding to $\lambda=1.2$ and the energy of the system, (b) Trajectories of the final soliton as seen from both $\varphi_1$ and $\varphi_2$.}
  \label{fig:13sol}
\end{figure}

For larger values of $\lambda$ we have seen a reflection followed by some absorption at the boundaries. 
In Fig.\ref{fig:12sol} we present some results of the simulation for $\lambda=0.4$. 
Snapshots of fields at $t=0$, $t=400$,  $t=600$ and $t=700$ are shown in figure (a). At $t=600$ one of the two kinks in each field has been already absorbed at the border. An insertion plot shows the noise that is left after the solitons have escaped from the region $x\in[-40,40]$ ({\it i.e.} have already been absorbed at the boundary of our grid). Fig.\ref{fig:12sol}(b) shows the trajectories of solutions at $t=0$ and $t=400$. 
The similarity of the trajectories demonstrates that any further (after reflection) evolution of the system  in the space of fields does not lead to a change of path. This follows from the fact that the kink -`bump' pair moves together in the space of fields $\varphi_1$, $\varphi_2$ in such a way that the path associated with the curve in the space of fields remains the same for all instants of time. 
\begin{figure}[h!]
\centering
{\includegraphics[width=0.4\textwidth,height=0.35\textwidth]{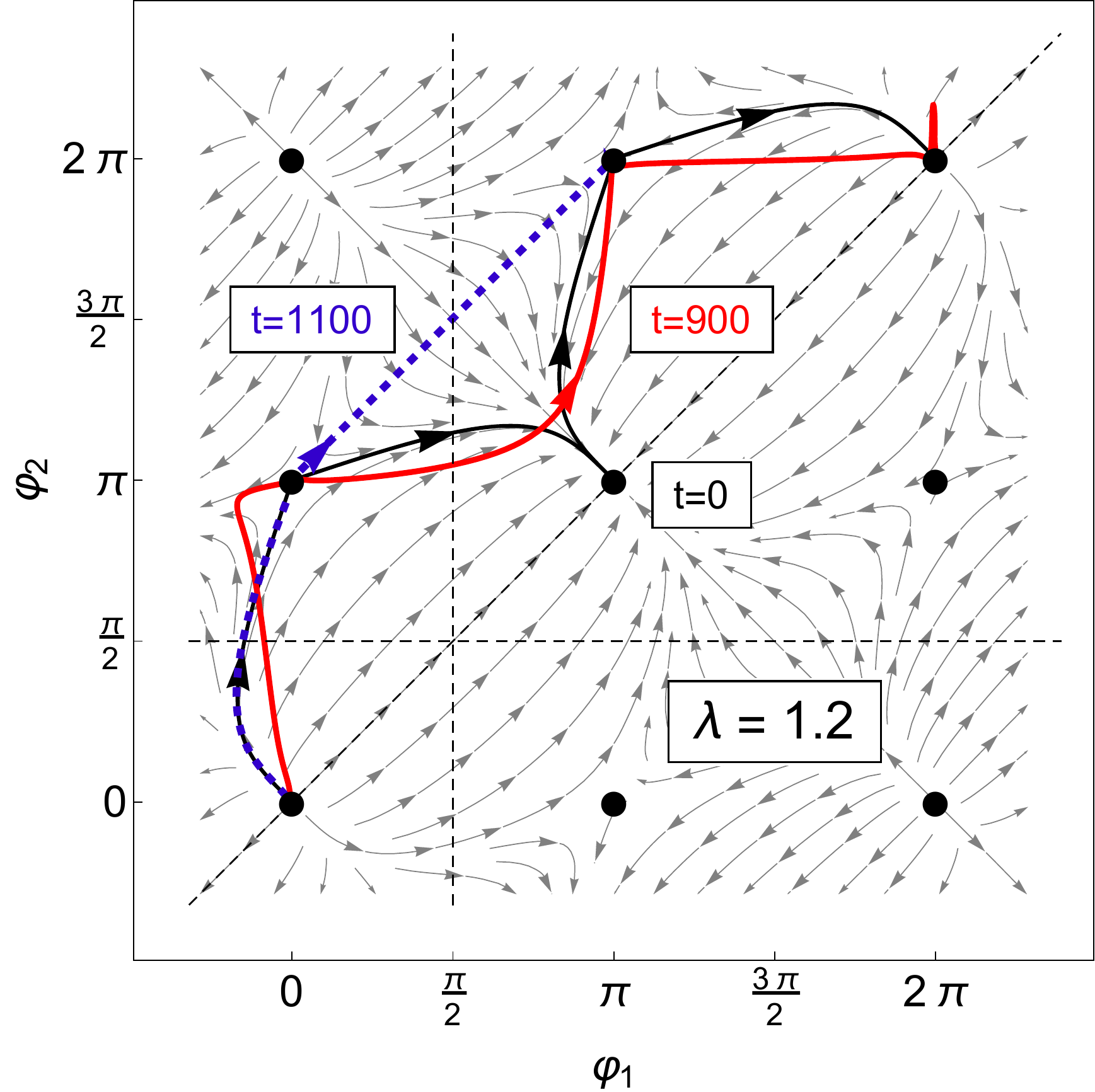}}

    \caption{Evolution of the fields corresponding to $\lambda=1.2$  in the space of fields on background provided by flow with $\epsilon=1$.}
  \label{fig:13solb}
\end{figure}
Obviously,  some small radiation which still exists in the system  causes certain irregularities of the path in the vicinity of the vacua.

For even larger values of $\lambda$ the process is even more complicated and it resembles what was seen for static cases. In Fig.\ref{fig:13sol} we present our plots for the case of $\lambda=1.2$. Note that the functions $\varphi_1$ and $\varphi_2$ at $t=3000$  are very similar in shape. We see very clearly a close relation to what we have seen in the static case. Fig.\ref{fig:13solb} shows evolution of the system in the space of fields. The central part of the curve  between minima $(0,\pi)$ and $(\pi,2\pi)$ has evolved in such a  way that it finally became orthogonal to the modified gradient flow.

\section{Breather-like fields}
Next we had a look at some breather-like configurations. So we started by constructing, for each function $\varphi_1$ and $\varphi_2$, breather like configurations. We have done this as before, {\it i.e.} by taking a superposition 
of a kink and an antikink.  Then we have performed various studies of the evolution of such systems of fields.
Like in the previous sections - first we have looked at the static case (in this case we have to put the kink and the antikink close together) and later we have sent them towards each other.

\subsection{Static case - very close}

Let us first discuss our results of the simulations of the static configurations. In fact, for all values of $\lambda$ the results of our simulations were very similar.
\begin{figure}[h!]
  \centering
 {\includegraphics[width=0.65\textwidth,height=0.4\textwidth]{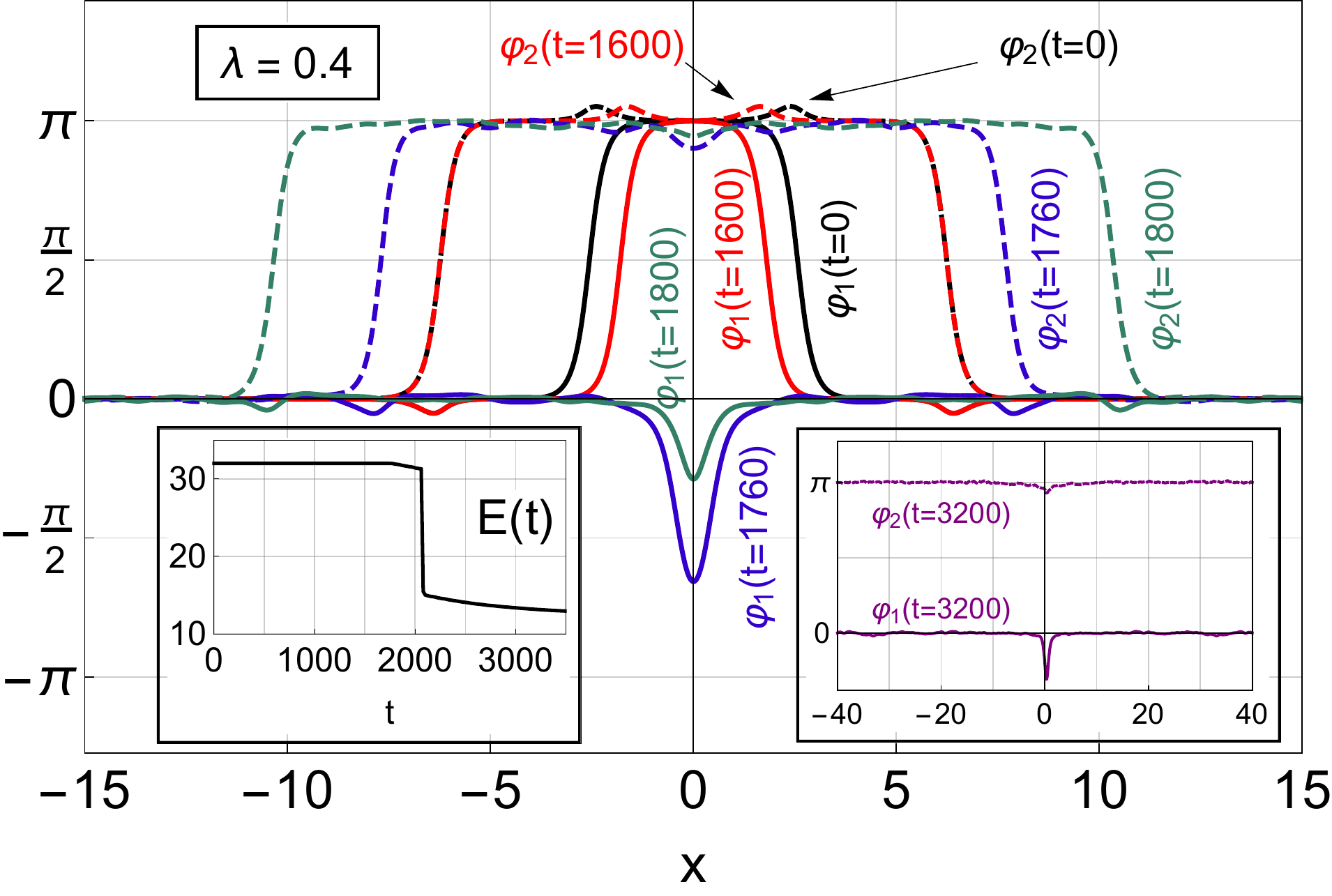}}                
    \caption{Evolution of the fields corresponding to $\lambda=0.4$ and the energy of the system. }
  \label{fig:14sol}
\end{figure}
They have always showed the creation of a `breather-like' structure and the two extra solitons moving out to the boundaries. At the boundaries the two extra solitons were 
absorbed and the `breather-like' structure continued radiating out energy gradually decreasing in magnitude. So the `breather-like' structure was really an oscillon.
The details of the process depended on $\lambda$ and on the initial distance between the solitons. For a given distance, as $\lambda$ increased, the solitons attracted more strongly  
and the oscillon radiated energy faster ({\it i.e.} was less stable).
\begin{figure}[h!]
  \centering               
 {\includegraphics[width=0.8\textwidth,height=0.37\textwidth]{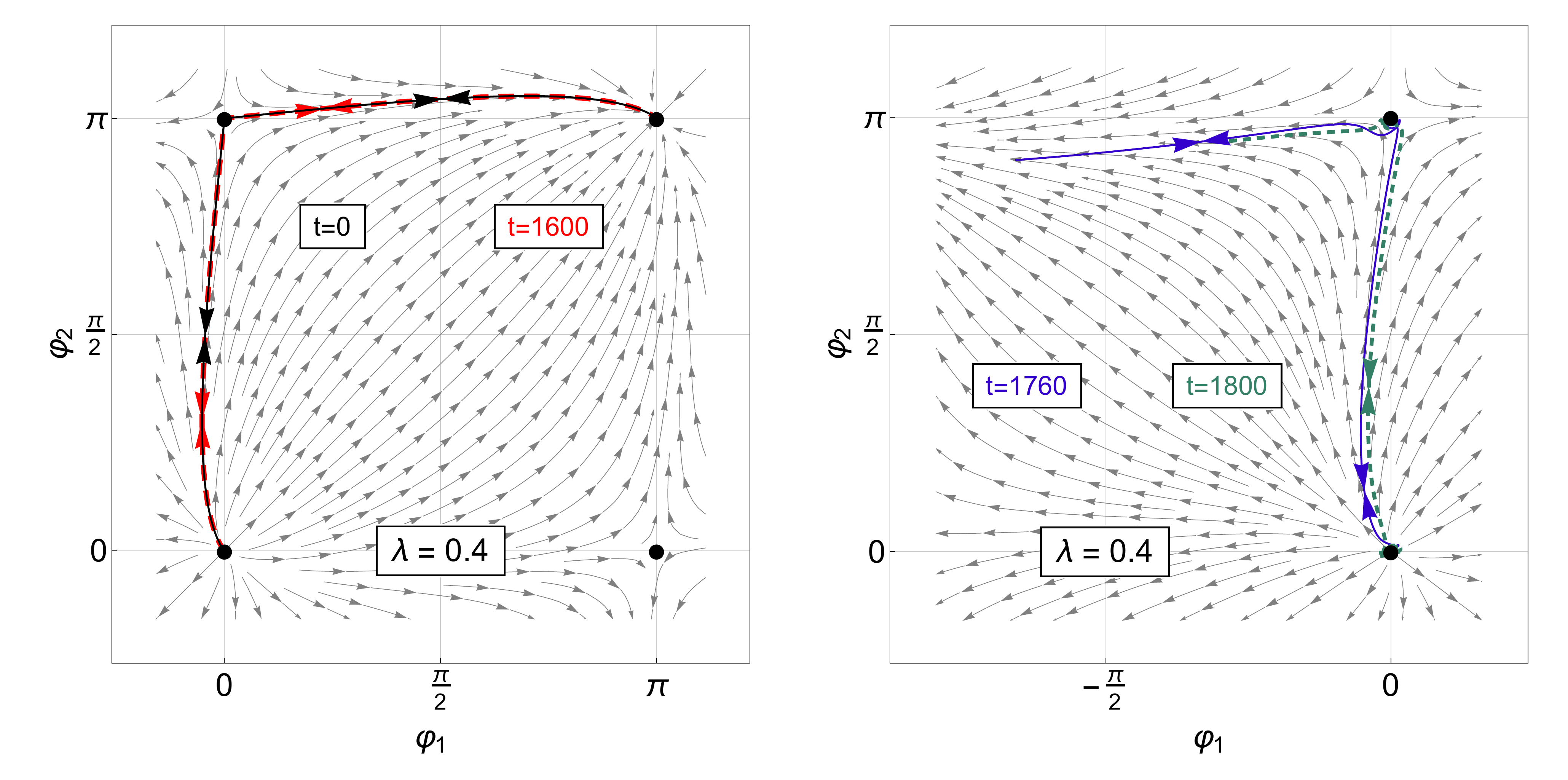}}
    \caption{ Evolution of the fields corresponding to $\lambda=0.4$  in the space of fields: each path corresponds to one of these two curves which differ from each other by their orientation. The flow presented here corresponds to the choice $\epsilon=1$.}
  \label{fig:14solb}
\end{figure}

\begin{figure}[h!]
  \centering
 \includegraphics[width=0.55\textwidth,height=0.35\textwidth]{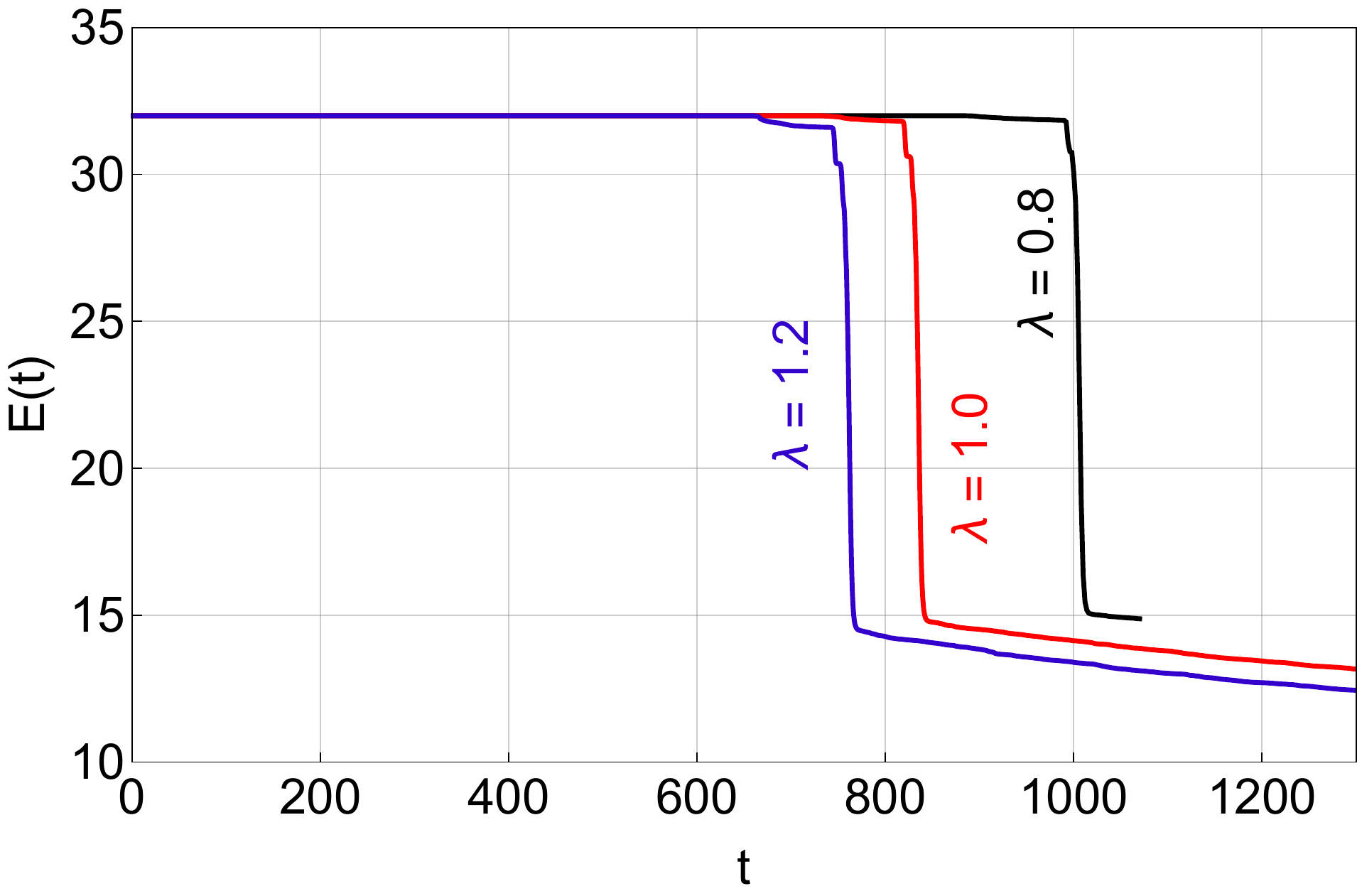}
   \caption{Time dependence of the energy for  $\lambda=0.8$, $\lambda=1.0$ and  $\lambda=1.2$.}
  \label{fig:15sol}
\end{figure}

In Fig.\ref{fig:14sol} we present a few plots of what was seen in a typical simulation. The case discussed here was obtained in a simulation for $\lambda=0.4$. Looking at the evolution of the system in the space of fields shown in Fig.\ref{fig:14solb} we note that the kink structure is associated with the part of the curve which approximately follows the gradient flow. On the other hand, the `breather like' object is represented by a path whose length varies with time. This path does not follow the gradient flow. All paths correspond to these two curves which differ from each other by their orientation. 

In Fig.\ref{fig:15sol} we present the energy plots for other (larger) values of $\lambda$. We clearly see that as we increase $\lambda$ the energy drop starts earlier.

\subsection{Solitons and antisolitons - sent towards each other}
We have also looked at solitons and antisolitons sent towards each other. We used the same approach as in the case of double solitons, 
{\it i.e.} determining initial time derivatives from the spatial derivatives of the fields.
The initial structures were sent to each other, like in the case of double solitons, each with $v=0.1$ towards $x=0$.

We have carried out our studies for various values of $\lambda$. For small values of $\lambda$ we have found a reflection and a flip of each field followed by the absorption at the boundaries.


\begin{figure}[h!]
  \centering
 \subfigure[]{\includegraphics[width=0.55\textwidth,height=0.35\textwidth]{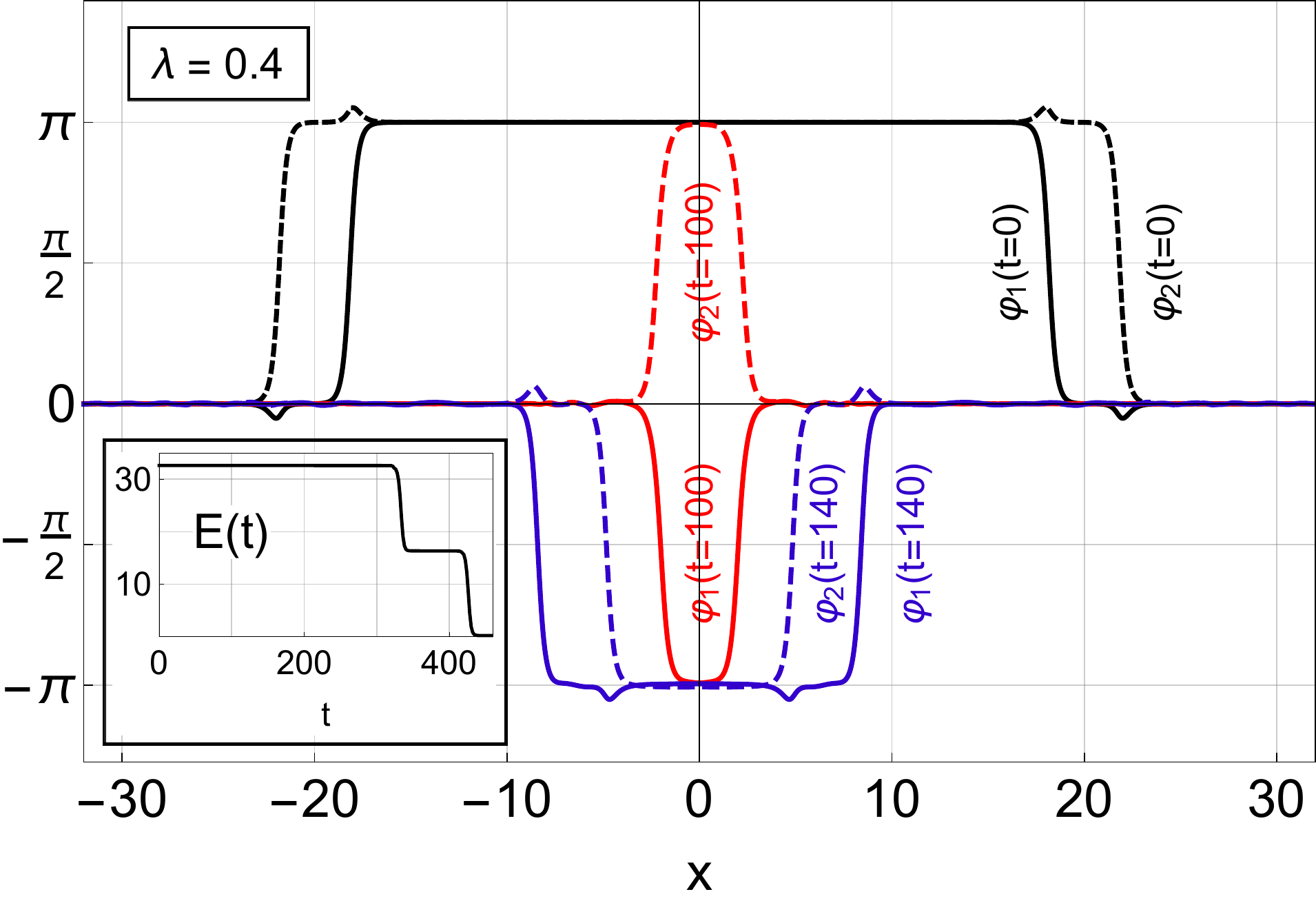}}  
 \hskip 0.5cm              
  \subfigure[]{\includegraphics[width=0.35\textwidth,height=0.35\textwidth]{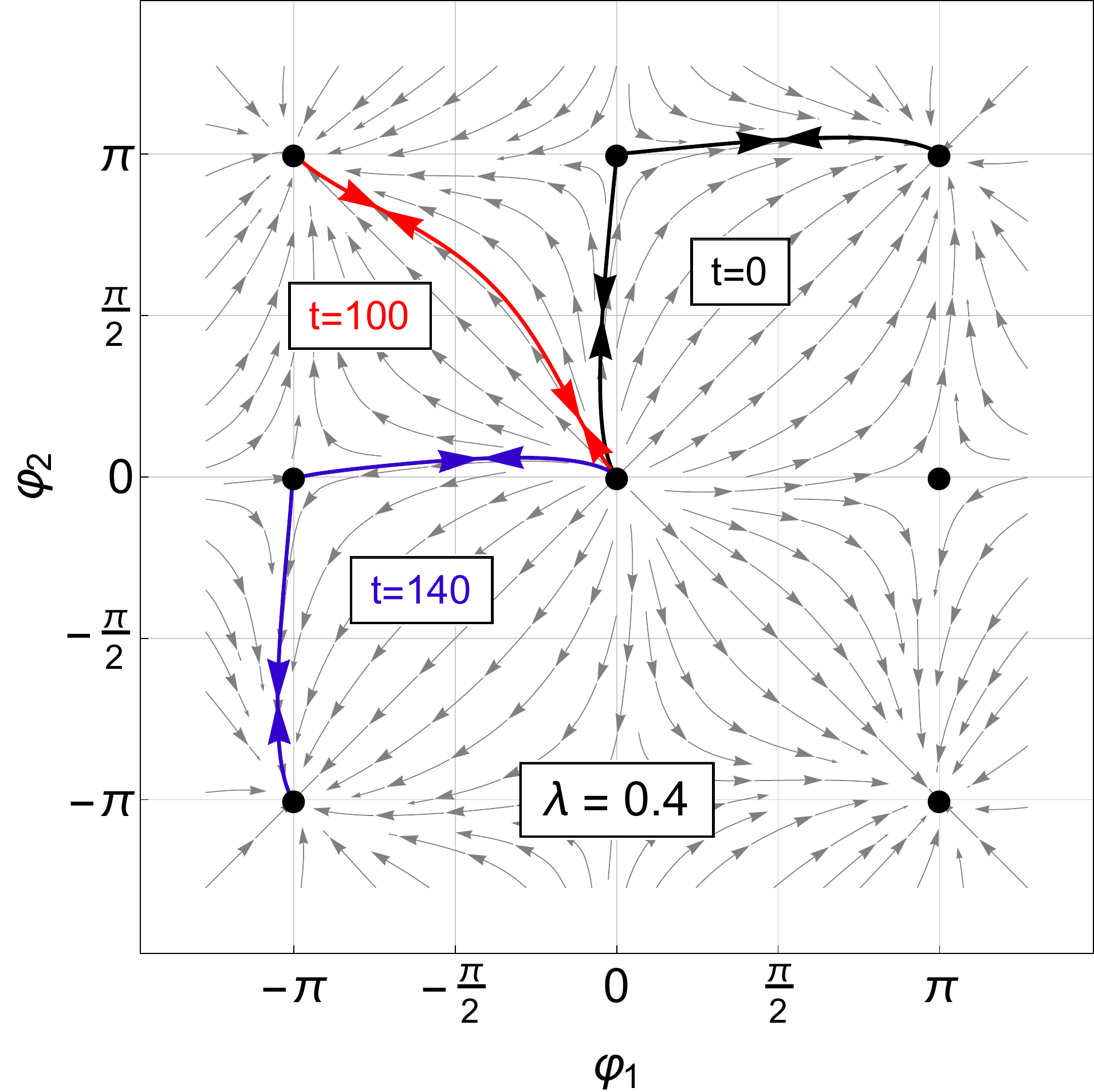}}
   \caption{(a) Plots $\varphi_1$ and $\varphi_2$ for $\lambda=0.4$, at $t=0$,  $t=100$ and  $t=140$. (b) Evolution of the system in the space of fields.}
  \label{fig:16sol}
\end{figure}

In Fig.\ref{fig:16sol}(a)  we have plotted fields $\varphi_1$ and $\varphi_2$, for a few values of $t$, as seen in a simulation for $\lambda=0.4$. We have also plotted the total energy of the simulation and we have noticed that after $t=400$ the reflected kinks were already absorbed at the boundaries.
 Fig.\ref{fig:16sol}(b) shows the solutions in the space of fields. We have performed many simulations for various small values of $\lambda$ and all the results looked very similar for all small values of $\lambda$ including $\lambda=0.0$ so we do not include them here.

\begin{figure}[h!]
  \centering
  \includegraphics[width=0.7\textwidth,height=0.45\textwidth]{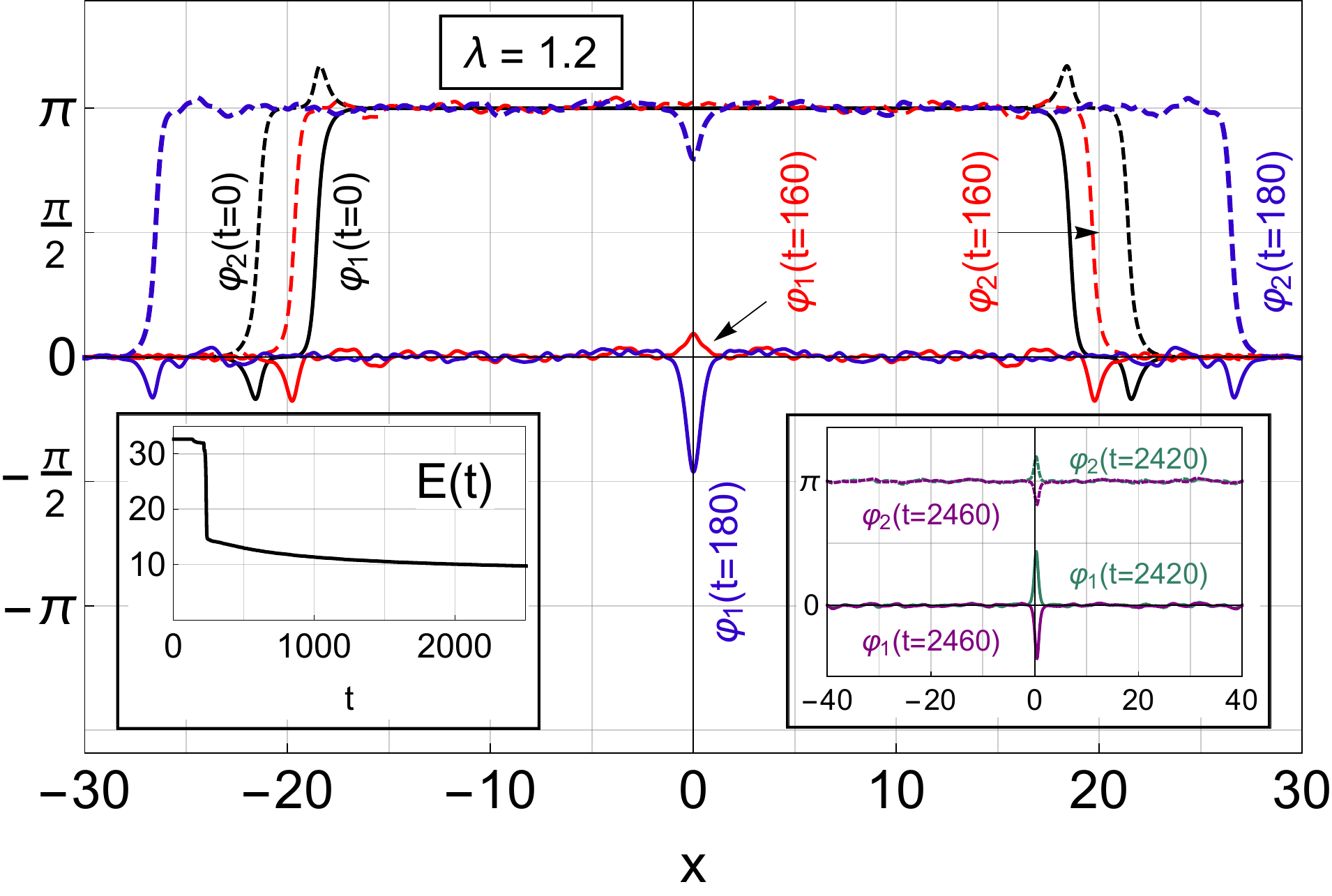}
   \caption{Plots of $\varphi_1$ and $\varphi_2$ for $\lambda=1.2$, at $t=0$, $t=160$, $t=180$,
$t=2420$ and $t=2460$ and  a plot of the time dependence of the total energy.
}
  \label{fig:17sol}
\end{figure}

For larger values of $\lambda$ ($\lambda=0.6$ and larger) the system ended up creating an oscillon.
Again there was very little difference in the way the simulation proceeded, and so here, in Fig.\ref{fig:17sol},
we present our results for one such $\lambda$, namely, $\lambda=1.2$.

The plots show very clearly the appearance of an oscillon which seems be long-lived (it emits some energy
but as the plot of the time dependence of the energy shows this emission seems to be slowing down).
Looking at the fields  $\varphi_1$ and $\varphi_2$ we note that after the oscillon had been formed both fields oscillate 
in phase and both go up and down at the same times. The only difference is that $\varphi_1$ oscillates
around 0 and $\varphi_2$ oscillates around $\pi$.

Can they alternate ({\it i.e.} oscillate our of phase)? To check this we looked at other values of $\lambda$.
At $\lambda=-1.0$ we found such a case.

\begin{figure}[h!]
  \centering
   \includegraphics[width=0.6\textwidth,height=0.4\textwidth]{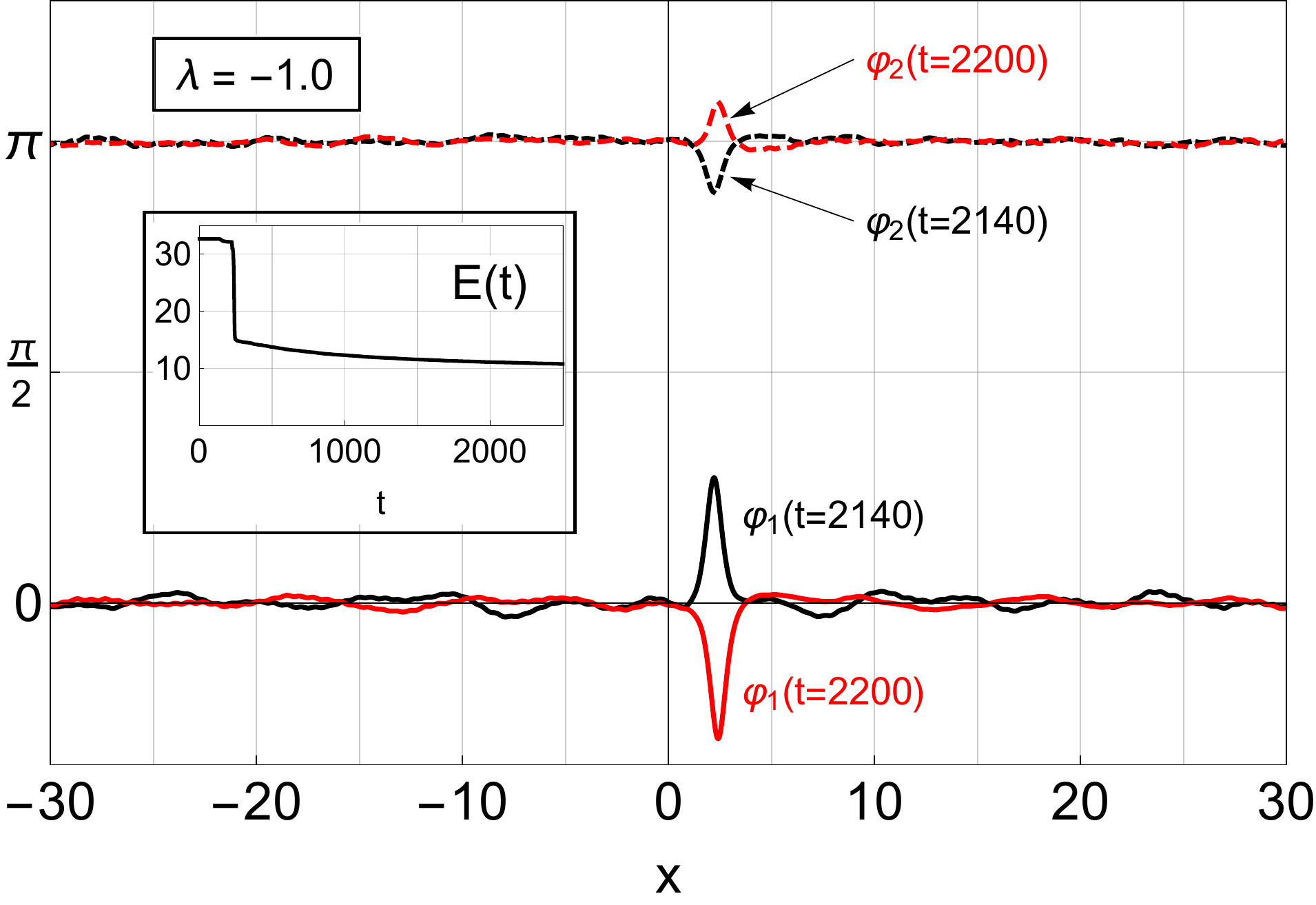}
   \caption{Fields $\varphi_1$ and $\varphi_2$ for $\lambda=-1.0$ at $t=2140$ and $t=2200$. The time dependence of the total energy.}
  \label{fig:18sol}
\end{figure}

In Fig.\ref{fig:18sol} we present the plots of $\varphi_1$ and $\varphi_2$ at two values of $t$ when the oscillon had already been formed
and existed for a while.
They show very clearly that, this time, the oscillations are out of phase.
The plot of the total energy shows that the oscillon is relatively long lived.

\section{Other two breather systems}

In our previous work we considered systems in which for $x<0$ both $\varphi_1$ and $\varphi_2$ varied from 
0 to $\pi$ and for $x>0$ varied from $\pi$ to $0$. Hence we called them $\epsilon=1$ systems.
However, we can also consider the systems in which for $x<0$ $\varphi_1$ varied between 0 and $-\pi$ 
and $\varphi_2$ varied between $-\pi$ and 0 and for $x>0$ $\varphi_1$ varied between $-\pi$ and 0
and $\varphi_2$ between 0 and $-\pi$. This we would call $\epsilon=-1$, taking the definitions from 
the region $x>0$. 

Here we discuss the results of some of such simulations.
All the simulations we described here involved initially static fields.
We have performed several such simulations and here we are describing our results obtained in some of these cases; some involving kink antikink systems with positive $\lambda$ and the others with negative $\lambda$. For $\lambda=0$, we observed small attraction so the kinks and antikinks were, initially, not placed too far away from each other.

Basically, the results of our simulations depended on the sign of $\lambda$. Given our starting values, the kink and antikink of the $\varphi_1$ would often lie on the outside of the kink and antikink of $\varphi_2$.

\subsection{Negative $\lambda$}

We have performed many simulations of systems involving negative $\lambda$. In all our simulations the two solitons for $x<0$ (antikink of $\varphi_1$ and kink of $\varphi_2$) moved towards each other, but as they belonged to different fields they could not annihilate, and so passed each other and then bounced back towards their initial positions. The same happened for $x>0$. So the two solitons (anti-kink of $\varphi_1$ and kink of $\varphi_2$)
for $x<0$ started oscillating around their positions and the same thing happend for the other two solitons (located at $x>0$). Their positions remained quite close at all times but the oscillations are clearly visible in the `flipping' 
of their `bumps' on the fields.

\begin{figure}[h!]
  \centering
   \includegraphics[width=0.48\textwidth,height=0.3\textwidth]{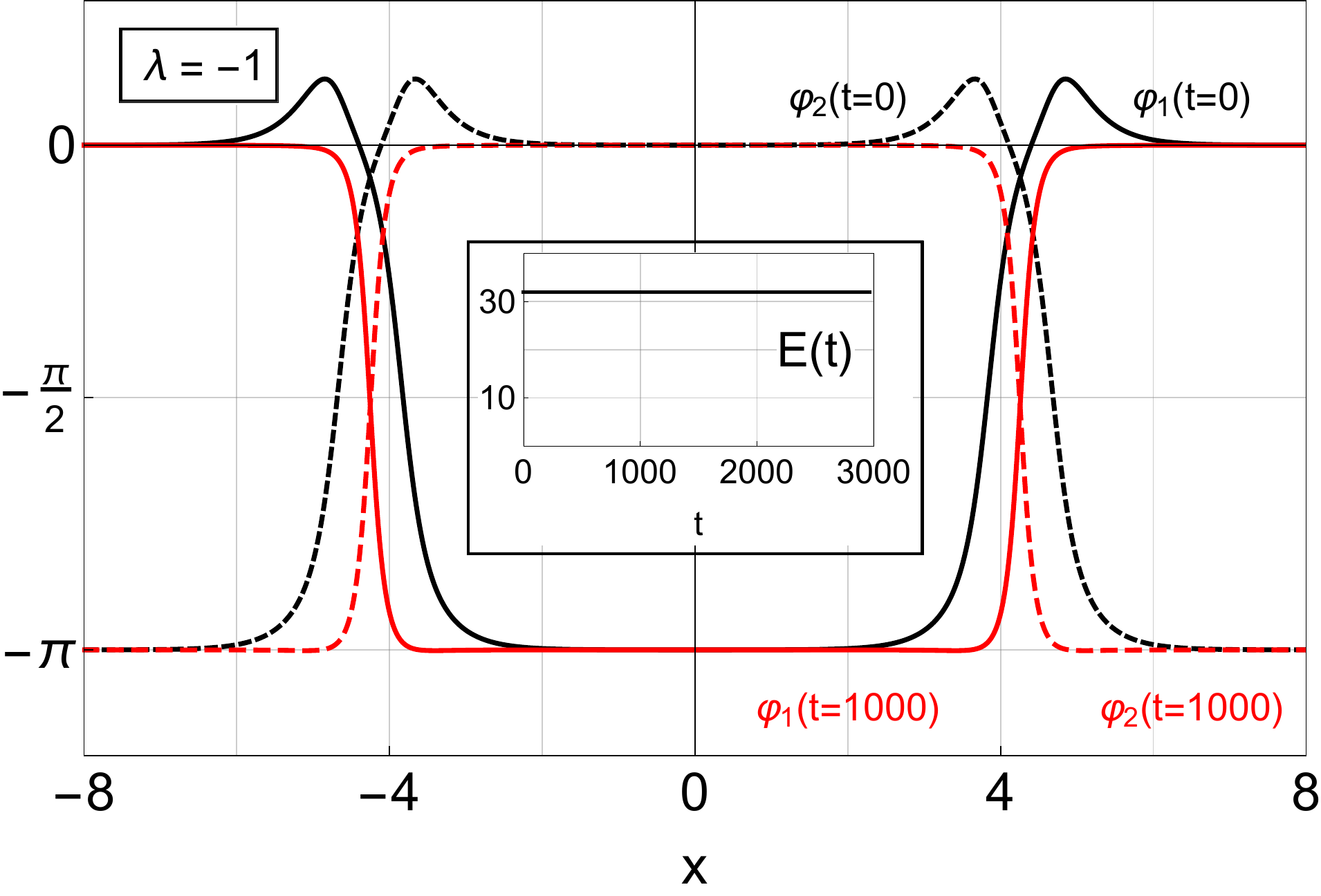}
   \hskip0.5cm
   \includegraphics[width=0.48\textwidth,height=0.3\textwidth]{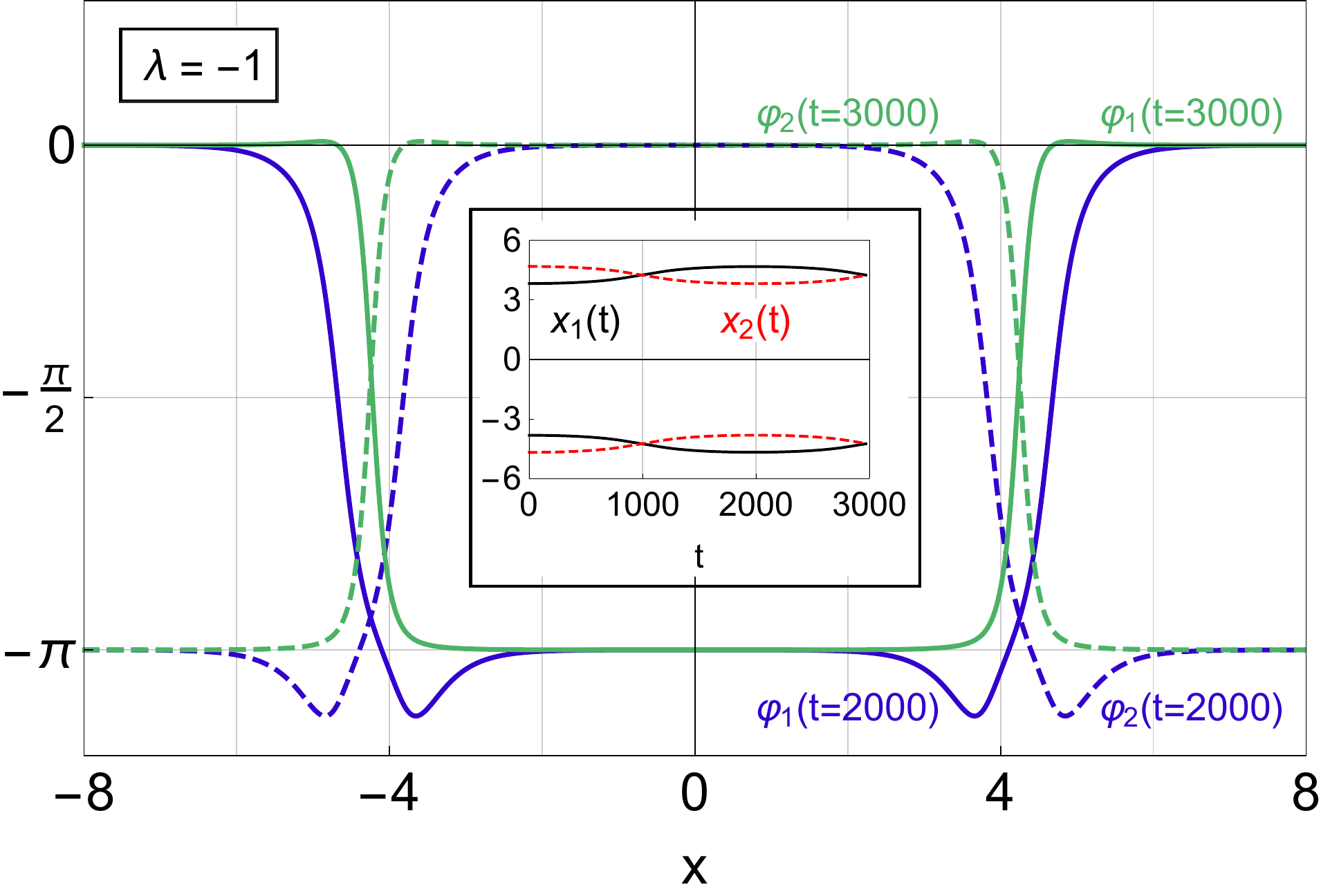}
   \caption{Plots of fields, at various values of time,  obtained in a simulation for $\lambda=-1.0$. The inserts show the total energy and the trajectories of the solitons.}
  \label{fig:21sol}
\end{figure}

In Fig.\ref{fig:21sol} we have put the plots of the fields for $\lambda=-1.0$. The plots show them at four values of time $t=0$, $t=1000$ and $t=2000$ and $t=4000$.  
The continuous curves denote $\varphi_1$ and the dashed ones $\varphi_2$. The curves for $t=0$ are black, $t=1000$ red, $t=2000$ blue and $t=3000$ green. The total energy is fully conserved (up to $10^{-7}$).
The insert in the fig b) shows the trajectories of the solitons seen in this simulation. These trajectories demonstrate a simple and very regular oscillation of the solitons in the left hand part of the plot ({\it i.e.} for $x<0$), and similarly, below, the same for the solitons at $x>0$.

We have also performed similar simulations for other values of negative $\lambda$. All the plots were very similar except that for smaller values of $\vert\lambda\vert$ the oscillations were slower and for larger values of $\vert\lambda\vert$ they were more frequent. So, as the one full oscillation for $\lambda=-1.95$ required $t\sim 82$, for $\lambda=-1.8$ it required $t\sim 345$. For $\lambda=-1.5$ it required already $t=\sim450$
which increased further to $t\sim 3850 $ for $\lambda=-1.0$. For $\lambda=-0.5$ it required $t\sim 10000$, and at smaller values of $\vert \lambda \vert$ the times were even larger.

\begin{figure}[h!]
  \centering
  {\includegraphics[width=0.5\textwidth,height=0.3\textwidth]{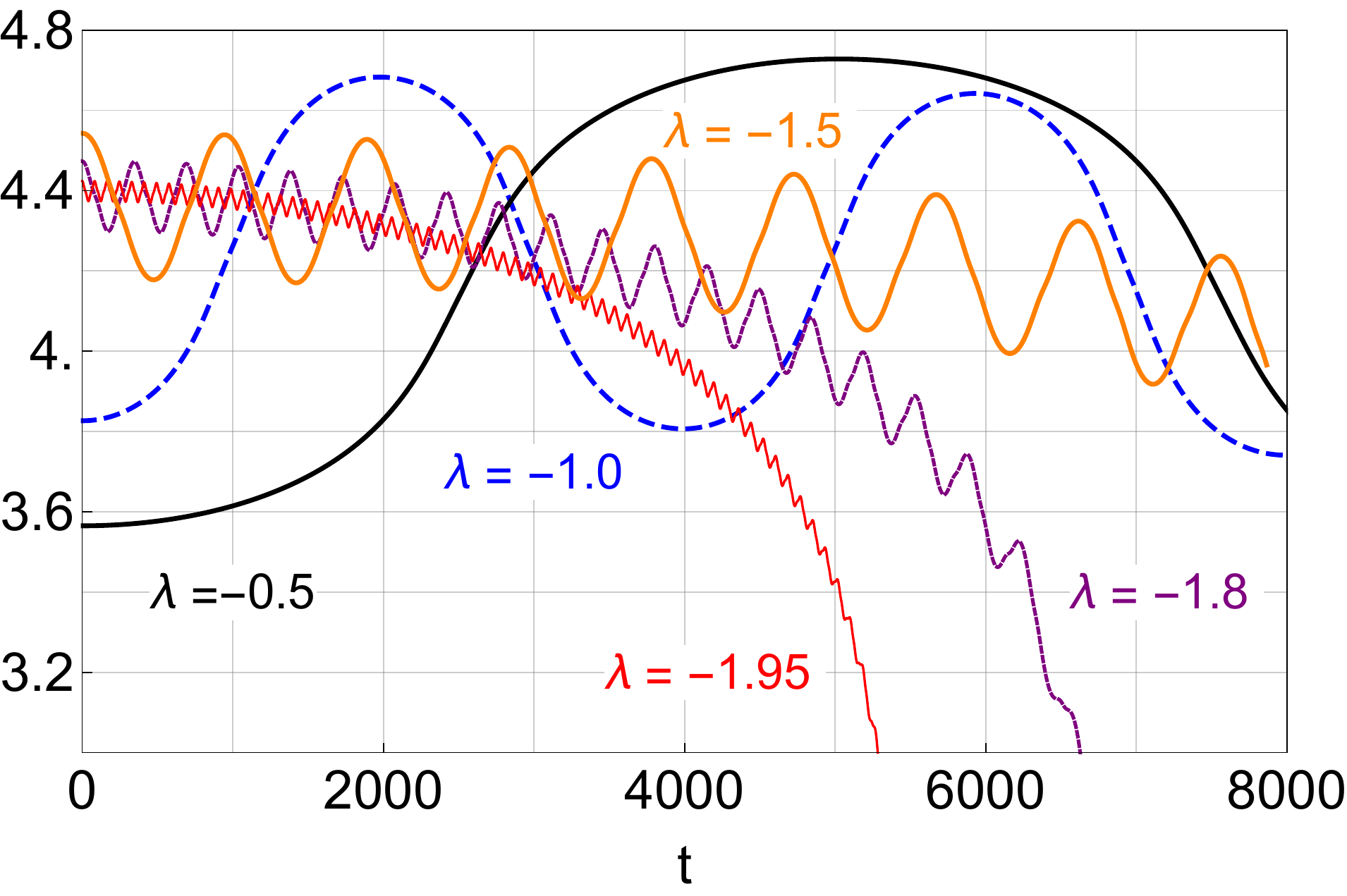}}
      \caption{Plots of trajectories $x_1(t)$, obtained for different values of $\lambda<0$.}
  \label{fig:27sol}
\end{figure}

However, we have continued running our simulations much longer and then we have 
discovered that during these oscillations the whole system of solitons was slowly moving together.
 The speed of this motion also depended on $\lambda$, and for larger values
of $\vert\lambda\vert$ the motion was faster.
In Fig. \ref{fig:27sol} we have plotted trajectories of one of the solitons seen in several simulations
for various values of $\lambda$. 

\begin{figure}[h!]
  \centering
  {\includegraphics[width=0.95\textwidth,height=0.4\textwidth]{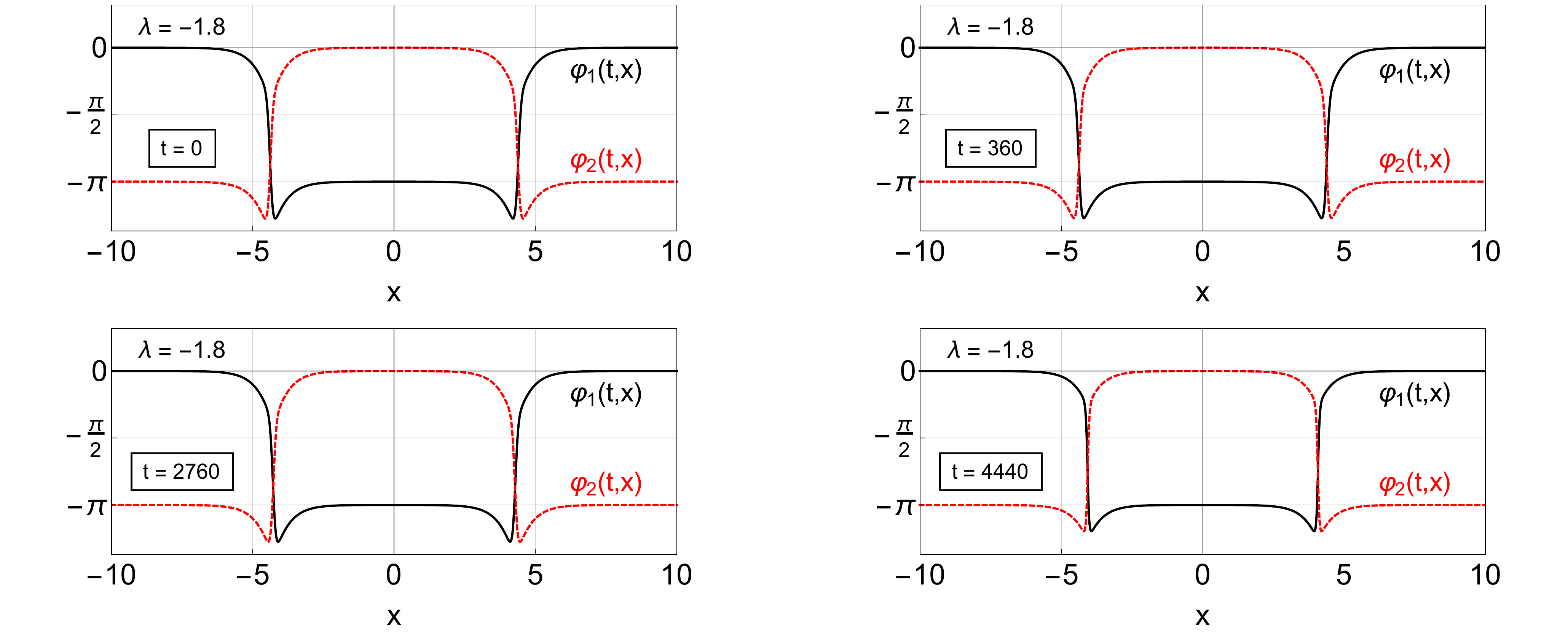}}
      \caption{Plots of fields, at four values of time $t$, seen in the simulation for  $\lambda=-1.8$.}
  \label{fig:25sol}
\end{figure}

\begin{figure}[h!]
  \centering
   \includegraphics[width=0.8 \textwidth,height=0.26\textwidth]{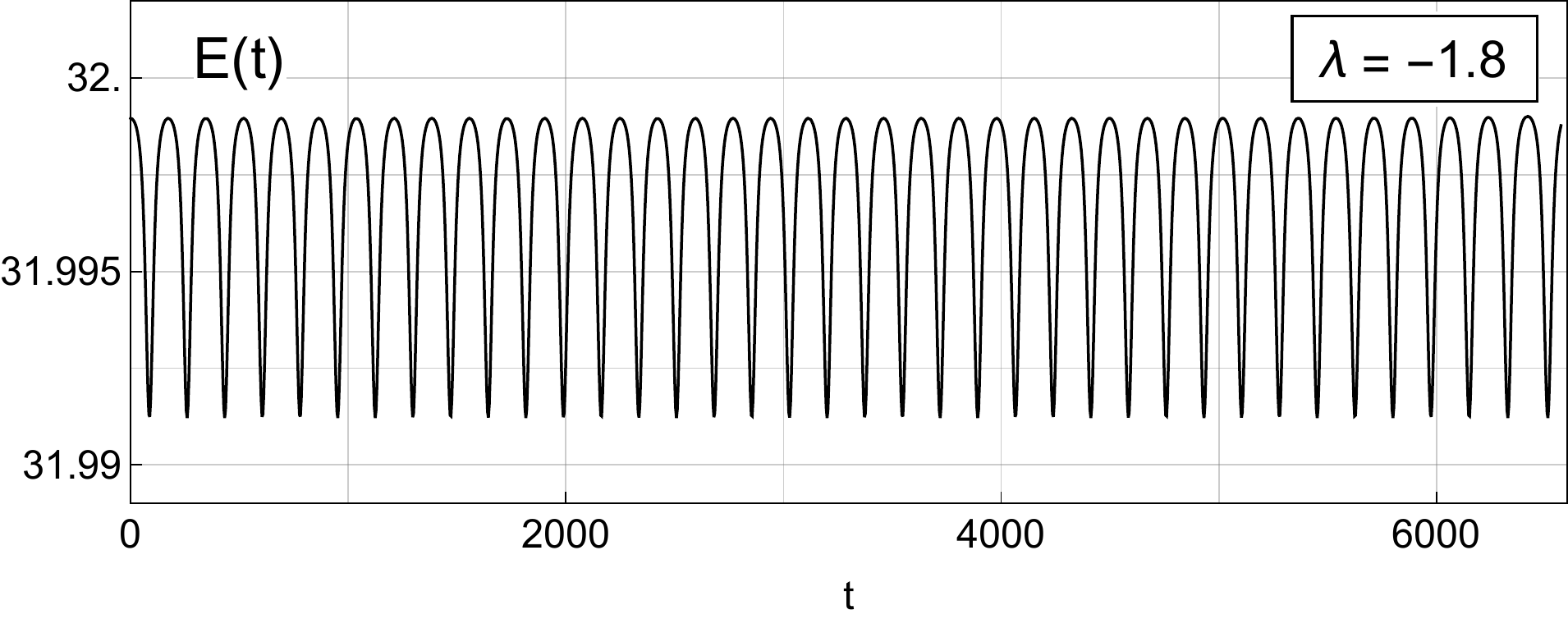}
   \caption{Total energy seen in the simulation for $\lambda=-1.8$}
  \label{fig:21asol}
\end{figure}

In Fig.\ref{fig:25sol} we present the plots of the fields seen in the simulation for $\lambda=-1.8$ for four values of time  ($0$, $360$, $2760$ and $4440$) which show very clearly that, apart from small changes of the `positions' of solitons, the fields look exactly the same. 
During these oscillations the energy was very well conserved as can be see from the plot of the energy shown in Fig.\ref{fig:21asol}. The small oscillations are numerical artifacts in the calculation of the energy (when the solitons of the fields are very close together, to save the computer time, we calculate this energy approximately).  

\begin{figure}[h!]
  \centering
{\includegraphics[width=0.9\textwidth,height=0.3\textwidth]{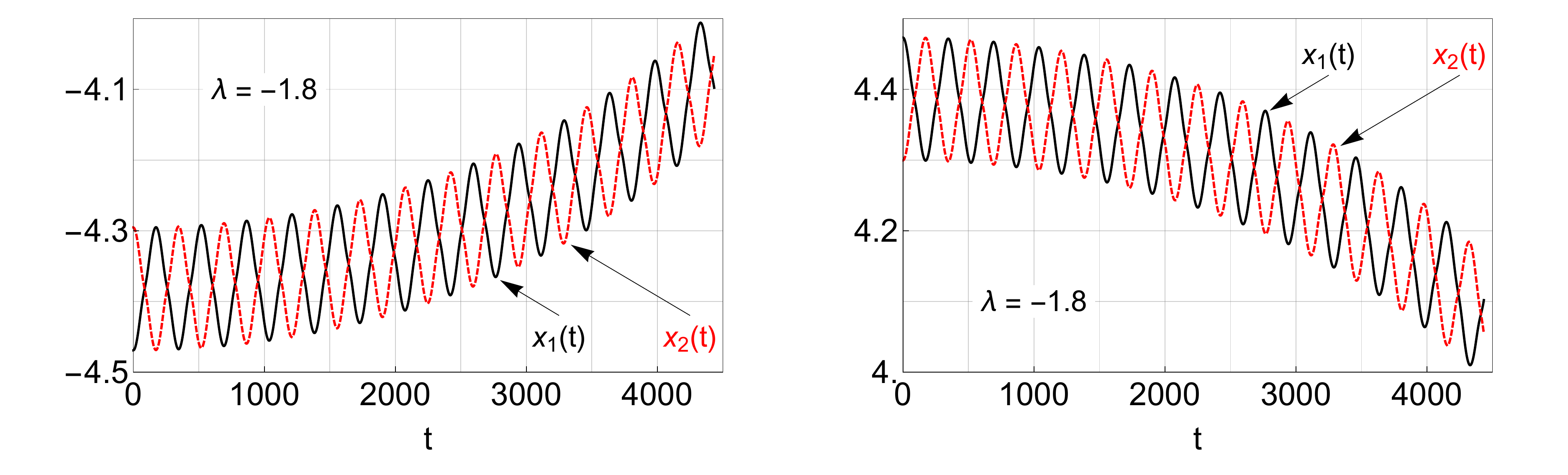}}
        \caption{Plots of trajectories $x_a(t)$, $a=1,2$ for $\lambda=-1.8$ where $\vp_a(t,x_{a}(t))=-\frac{\pi}{2}$.}
  \label{fig:26sol}
\end{figure}

\begin{figure}[h!]
  \centering
  {\includegraphics[width=0.9\textwidth,height=0.3\textwidth]{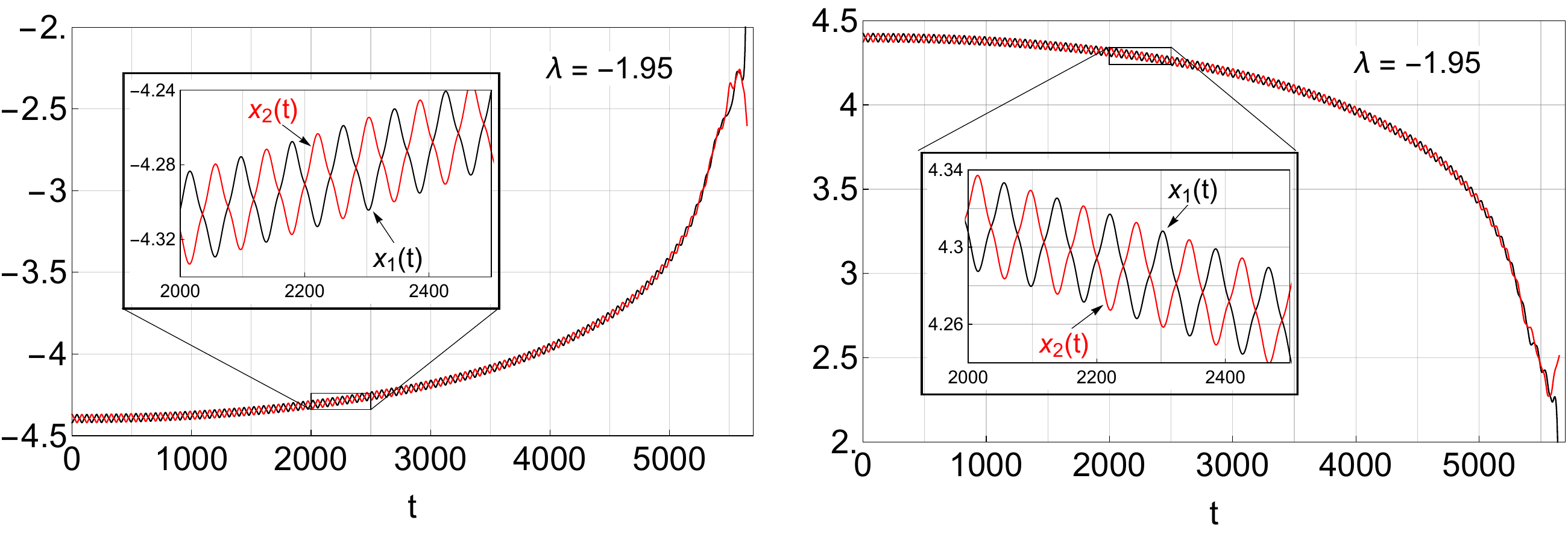}}
      \caption{Plots of trajectories $x_1(t)$ and $x_2(t)$ obtained in the simulation for $\lambda-1.95$.}
  \label{fig:27traj}
\end{figure}

In Fig.\ref{fig:26sol}  we present more detailed plots of the  oscillations of the trajectories 
for the simulation with $\lambda=-1.8$ and in Fig.\ref{fig:27traj} we have put the plots of the trajectories of the solitons 
obtained in the simulation for $\lambda=-1.95$. 


\subsection{Positive $\lambda$}

\begin{figure}[h!]
  \centering
   \includegraphics[width=0.6 \textwidth,height=0.35\textwidth]{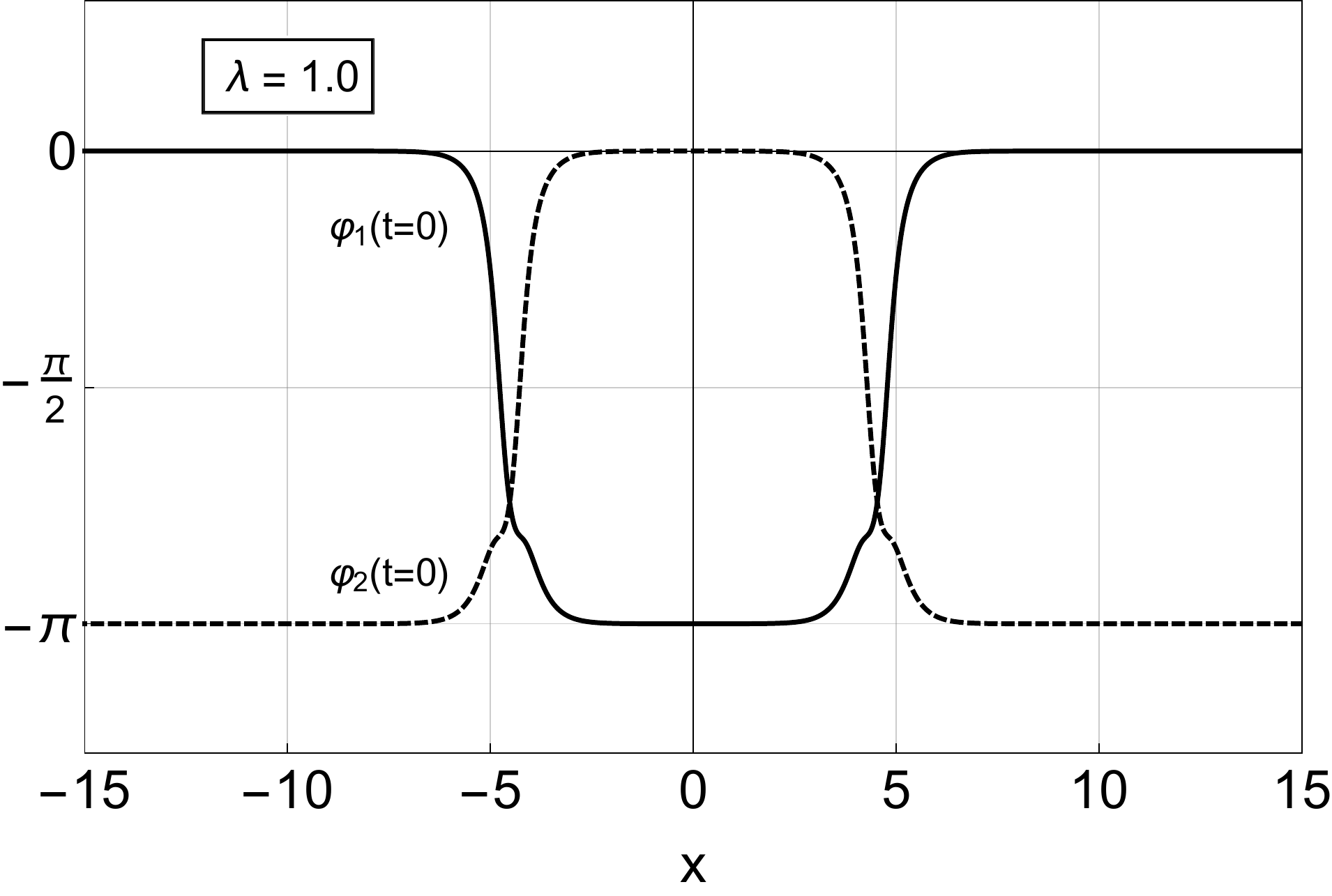}
   \caption{Initial fields $\varphi_1$ and $\varphi_2$ for $\lambda=+1.0$.}
  \label{fig:20sol}
\end{figure}

For $\lambda>0$ we have started the simulations with the initial conditions as shown in Fig.\ref{fig:20sol}.
First we performed the simulation for $\lambda=1.0$. 

As the system started to evolve, the external solitons of the system ({\it i.e.} the kink and anti-kink of the $\varphi_1$ field), started moving towards the boundaries, while 
the kink and the anti-kink of the $\varphi_2$ system started moving towards each other. When they got very close to each other, the solitons annihilated into waves of energy moving out to the boundaries. The moving waves then hit the original solitons and speeded them up in their movement towards the boundaries.

\begin{figure}[h!]
  \centering
   \includegraphics[width=1.0\textwidth,height=0.35\textwidth]{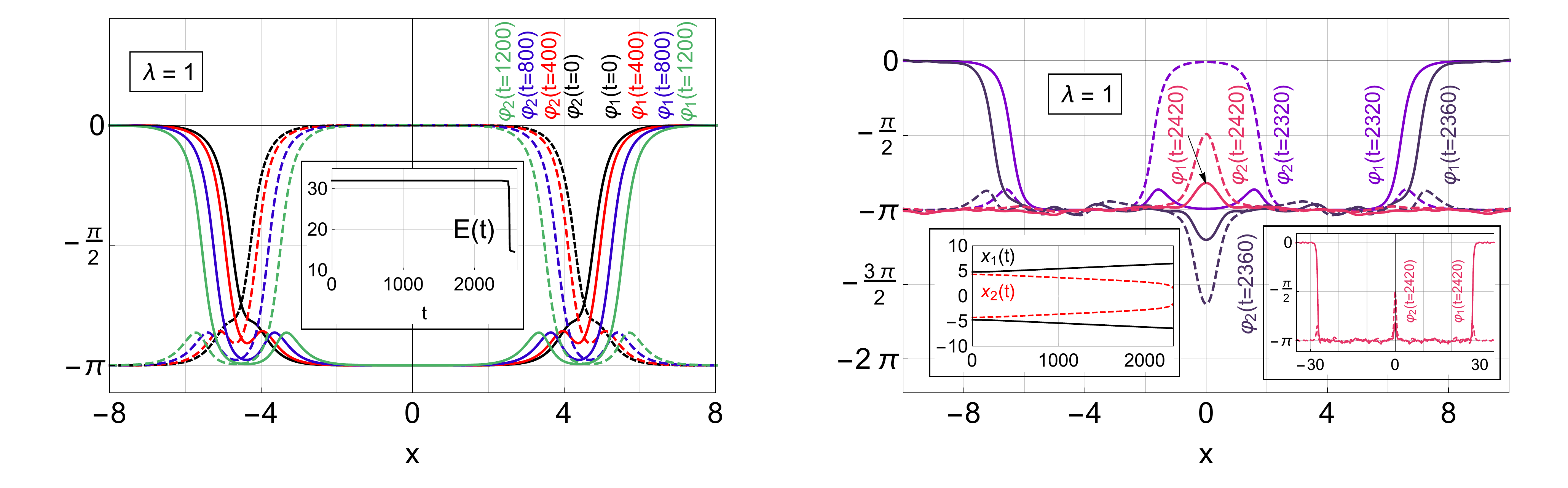}
   \caption{Fields $\varphi_1$ and $\varphi_2$ for $\lambda=1.0$ at various important times. The inserts show the total energy in the plot, and the trajectories of the solitons.}
  \label{fig:22sol}
\end{figure}

In Fig.\ref{fig:22sol} we present the plots of the fields at various times of the simulation. The left hand figure shows the fields before the annihilation
at the centre. In our simulation we have observed that the
radiation generated by the annihilation moved towards the outside solitons and accelerated their motion.  The plot on the right hand side starts with
the fields at $t=2320$ (just before the annihilation) and shows also the fields at $t=2360$ (just after the annihilation).
Hence we can deduce that the annihilation of the solitons took place around $t\sim 2350$. The other plots in the figure on the right show the fields after the annihilation. After the annihilation everything happened very quickly.
The waves of the radiation, resulting from the annihilation,  accelerated the motion of the outside solitons, 
 as is clear from the plots of fields at $t=2420$ (the main figure and subfigure), and resulted in the creation of an 
oscillon.

The inserts on the right hand side show the trajectories of the solitons before the annihilation and the fields just at the annihilation. The insert in the figure on the left shows the total energy; the sudden drop at the end of the curve describes the energy in the region of the simulation and is associated with the outgoing solitons crossing the boundaries of the plot (and so their energy is not included).

For other values of $\lambda>0$ the evolution was very similar. Each time, the two outside solitons started moving away to the boundaries, and the two middle ones started moving together each other. This motion was originally very very slow (dependent on the value of $\lambda$), but then, in all cases, gradually, the solitons accelerated and annihilated each other. 
 
\begin{figure}[h!]
  \centering
  {\includegraphics[width=0.95\textwidth,height=0.6\textwidth]{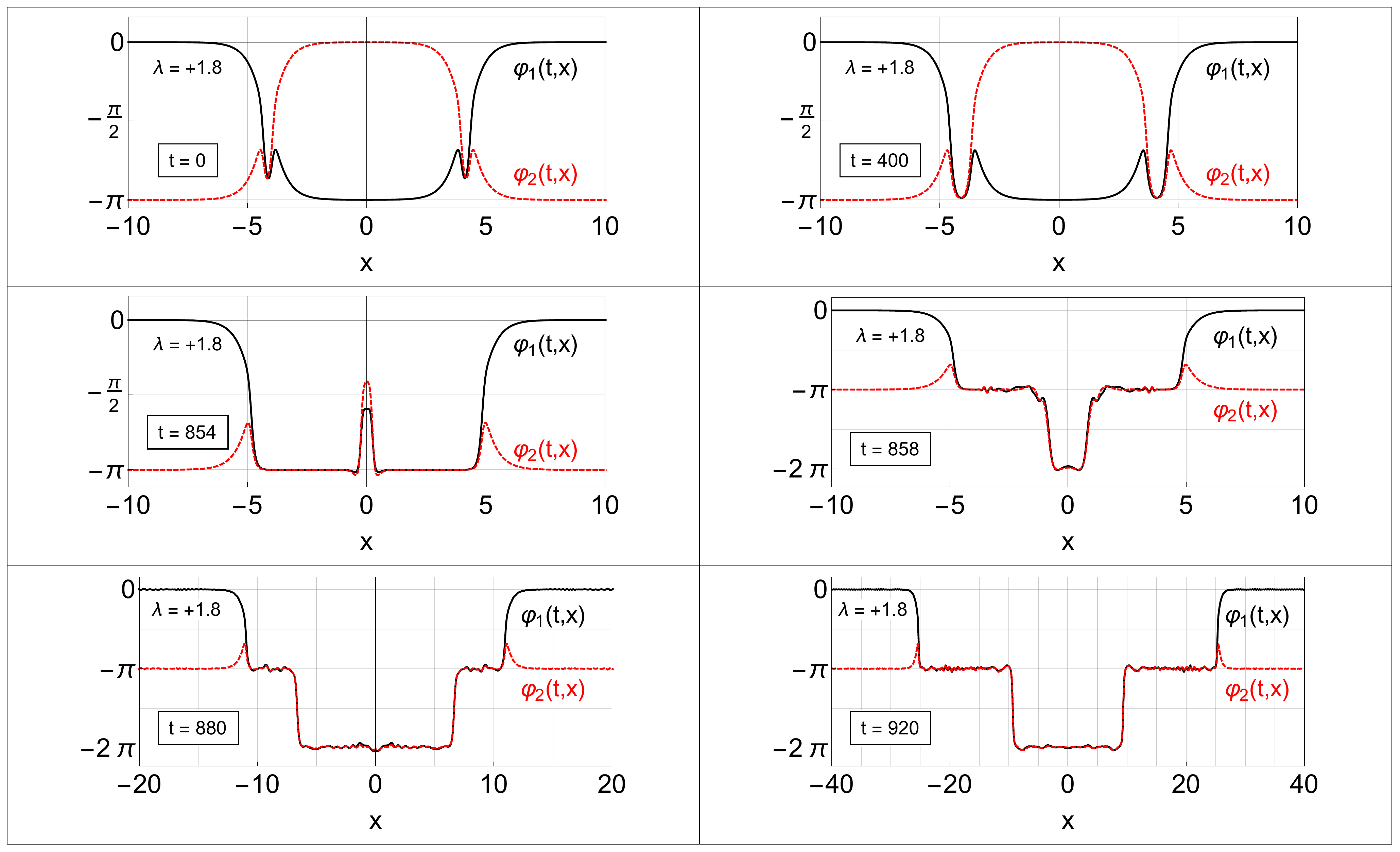}}
      \caption{Plots of the fields, at various important values of time $t$,  for $\lambda=1.8$.}
  \label{fig:28sol}
\end{figure}

\begin{figure}[h!]
  \centering
  {\includegraphics[width=0.4\textwidth,height=0.3\textwidth]{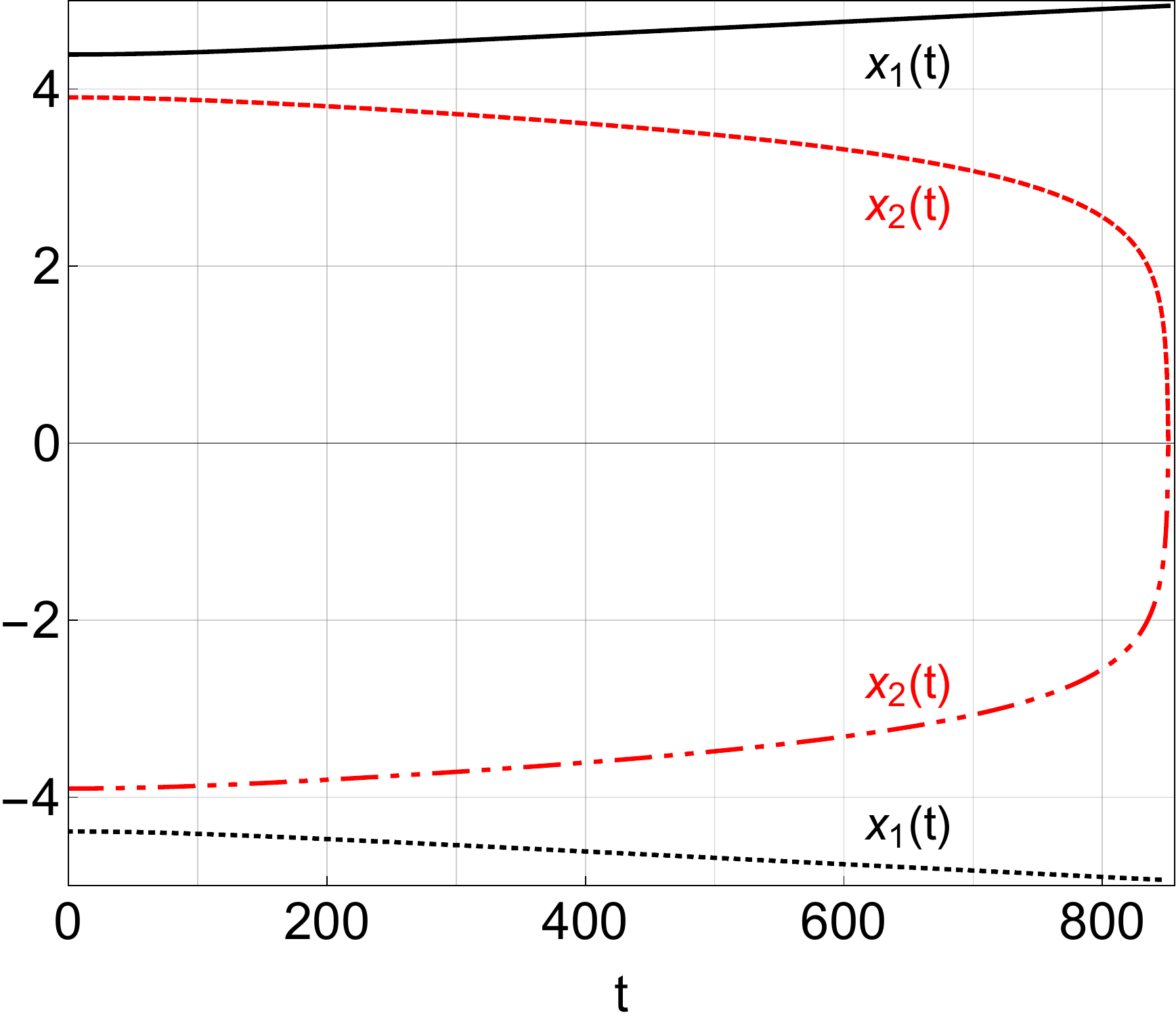}}
      \caption{Plots of the trajectories $x_a(t)$, $a=1,2$ for $\lambda=1.8$ where $\vp_a(t,x_{a}(t))=-\frac{\pi}{2}$.}
  \label{fig:29sol}
\end{figure}

This behaviour was always very similar in its nature except that, for lower values of $\lambda$, the evolution
was slower. Thus for $\lambda=1.$ the annihilation took place at $t\sim 2350$ while for $\lambda=0.5$ it  was delayed till $t\sim 6500$.
For larger values of $\lambda$ - the annihilation happened earlier. This is seen from
Fig.\ref{fig:28sol}, in which we present the plots of the fields 
(at various values of $t$) for $\lambda=1.8$. In the following plot (Fig.\ref{fig:29sol}) we present the trajectories of the solitons seen in that case.
It is clear from this picture that the annihilation took place around $t\sim 860.$ For larger values of $\lambda$ the annihilation took place even earlier.

\section{\bf BPS-systems revisited}

\subsection{General Comments}
We have tried to explain the oscillatory behaviour observed and discussed in the last section but then we realised that in all our cases we had two separate motions; one oscillatory and one involving the overall attraction of the kink antikink in each field.
We have looked at this phenomenon also by altering the distance between the structures in each field and this confirmed our expectations further. Thus the oscillations of the solitons for $x<0$ had no origin in the existence of further solitons for $x>0$; they only modified the overall motion. 
To check this further we looked also at the systems involving 
2 kinks in each field. There also were some motions which suggested possibility of attraction or repulsion (though these modifications were very small). Hence we have decided to look again at the ``static" solutions of the BPS equations.

 Of course, as is well known and easy to check, these static solutions were also solutions of the full equations and so should not evolve at all. However, our BPS fields had to be determined numerically and as such, had small numerical errors. Of course, we tried to reduce them as much as possible and so used the programs written in double precision {\it etc} but some numerical errors, although extremely small, were inavoidable. Thus we decided to rerun our simulations involving BPS fields and this time running the programs for very long times.

\subsection{Kink-antikink systems}

We have performed many such long simulations for various values of $\lambda$ (both positive and negative). We even looked at 
$\lambda=0$. All simulations, at first showed no motion and then, for $\lambda>0$ we saw very slow repulsion and for 
$\lambda<0$ a slow attraction.

In Fig.\ref{fig:30sol} we plot the fields for a few values of time - for the system with $\lambda=-1.8$
In the next figure we plot the results of the same simulation 
using our prepotential $U$. We see the attraction very clearly. 
\begin{figure}[h!]
  \centering
  {\includegraphics[width=0.9\textwidth,height=0.45\textwidth]{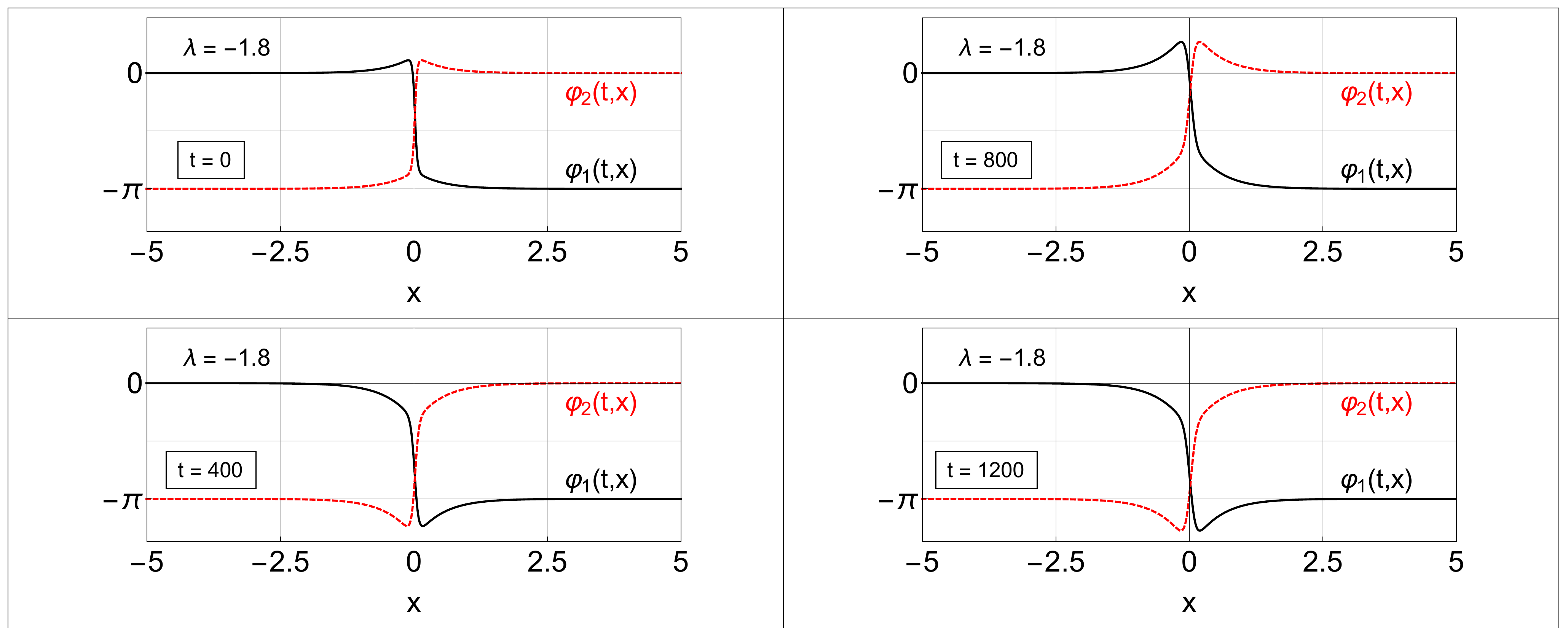}}
      \caption{Plots of fields $\phi_1$ and $\phi_2$ at $t=0$, 400, 800 and 1200 for $\lambda=-1.8$.}
  \label{fig:30sol}
\end{figure}

\begin{figure}[h!]
  \centering
  {\includegraphics[width=0.5\textwidth,height=0.5\textwidth]{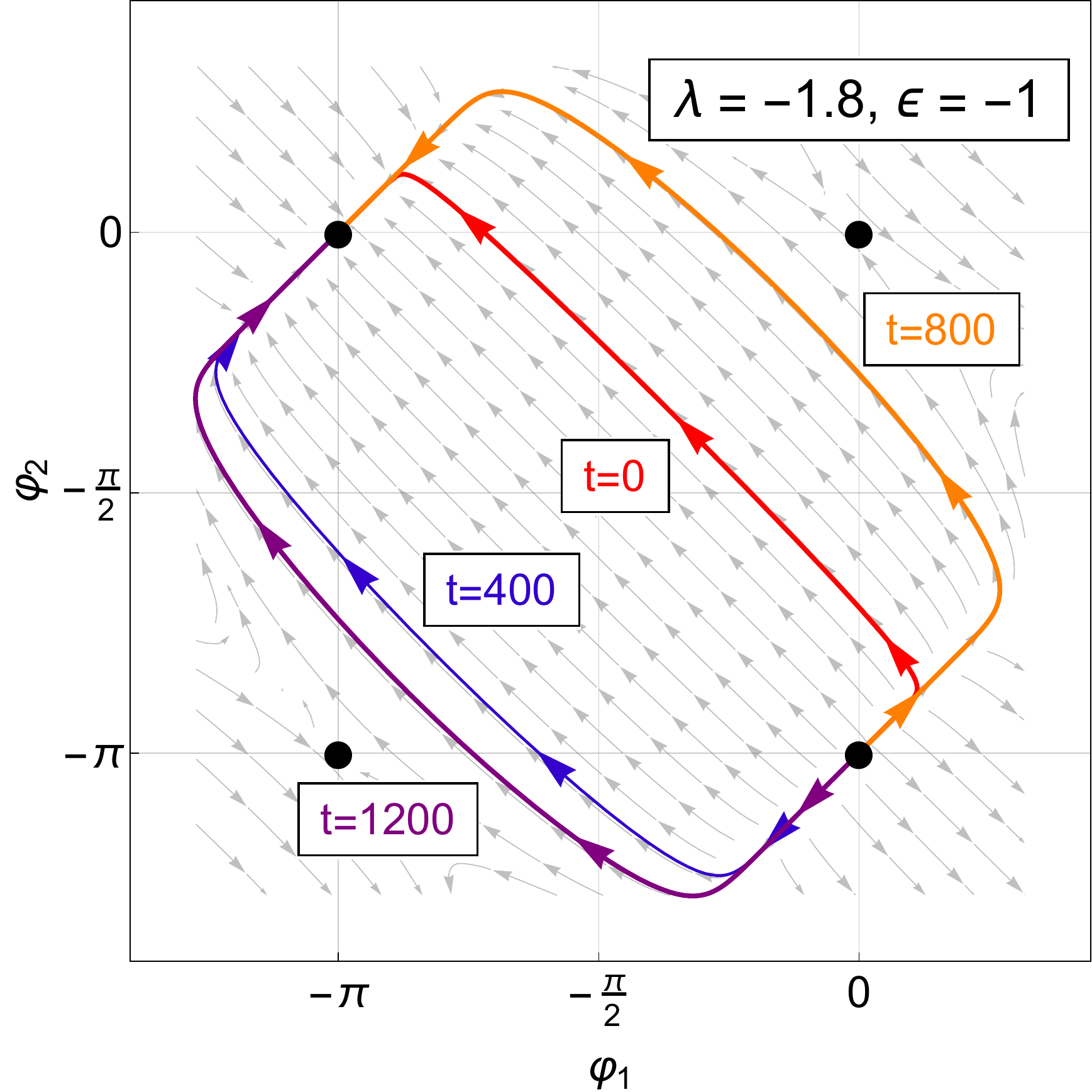}}
      \caption{Evolution in the space of fields for the case $\lambda=-1.8$.}
  \label{fig:31sol}
\end{figure}

\subsection{Origin of these movements}

Thinking about this further we have realised that our two field BPS system has two zero modes. This is clear as the system involves 
two first order equations and so its solutions depend on 2 constants - in our case, values of $\phi_1$ and $\phi_2$ fields at $x=0$.
Of course, this is clear from looking at the curves of the BPS solutions plotted on the background of the prepotential $U$.
One of these zero modes corresponds to the overall translation of both solitons; the other, when $\lambda\ne0$, involves a complicated 
relation between the fields. 

The BPS solutions, for $\lambda\ne0$, had been calculated numerically and so they possess very small numerical errors. Moreover,
the time evolution of the BPS solutions can induce further numerical errors. Of course, we have tried very hard to reduce such errors (writing 
our programs carefully and using double precision {\it etc} but such errors are unavoidable).

For $\lambda=0$, the system reduces to two decoupled Sine-Gordon models and in this case the numerical errors lead to the system sending out only radiation
\cite{radiation}. This is due to the fact that all numerical effects do preserve the overall momentum of the system and as there is no interaction between 
the fields, the position of either soliton cannot change (as this would correspond to nonvanishing momentum).
However, in our system we have an extra possibility of the motion of solitons which preserves the conservation of the overall momentum.
So the two solitons can move away from each other, or move towards each other. And which motion takes place would depend on the perturbation itself.

\subsection{Further comments and comparison with the kink/kink system}

So, we have decided to look at other values of $\lambda$ for the kink-antikink systems and for the kink-kink ones.
First we looked at the $\lambda=1.8$ case for the kink-antikink system.
As expected we have observed  a slow repulsion. 

\begin{figure}[h!]
  \centering
   \includegraphics[width=0.9\textwidth,height=0.45\textwidth]{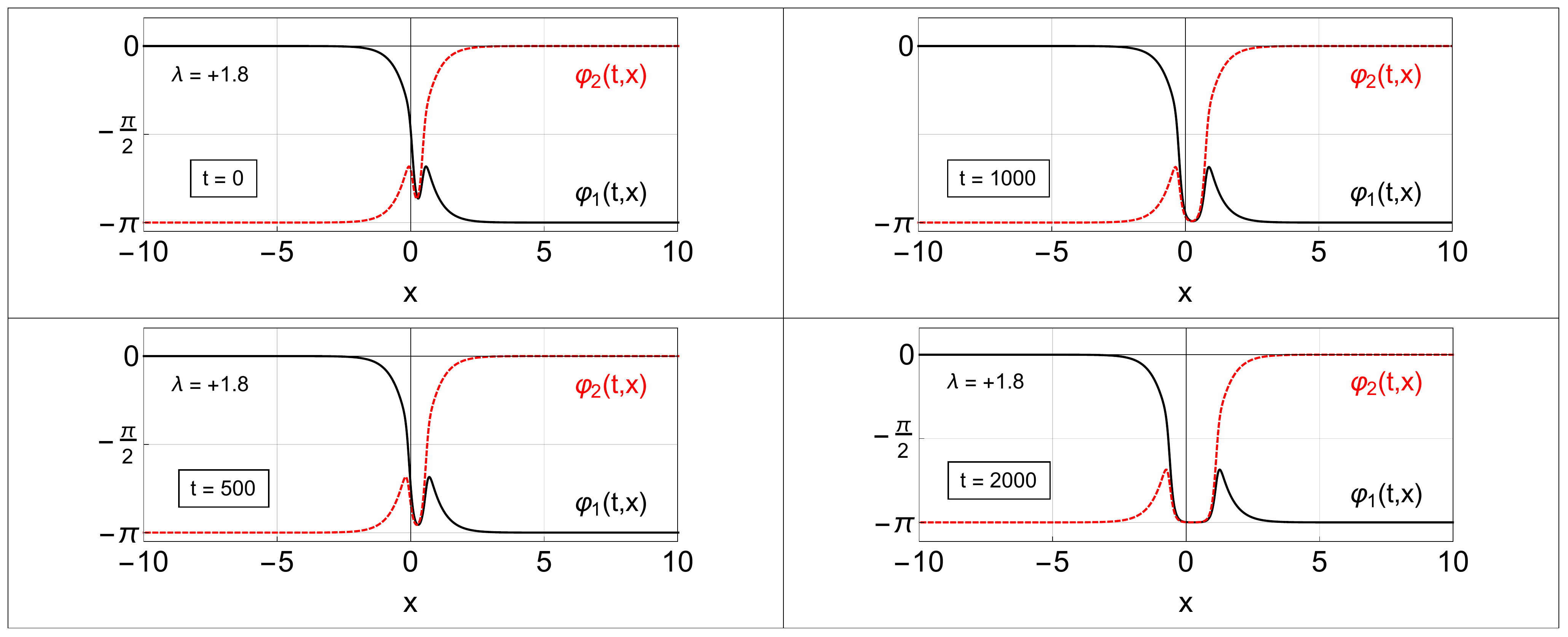}
\caption{Plots of fields $\varphi_1$ and $\varphi_2$ at $t=0$, 500,1000 and 2000 for $\lambda=+1.8$}.
\label{fig:32sol}
\end{figure}

In Fig.\ref{fig:32sol} we plot the fields for a few values of time - for the system with $\lambda=+1.8$.

In the next figure we plot the results of the same simulation using our prepotential $U$, {\it i.e.} as trajectories in the space of fields.
\begin{figure}[h!]
\centering
\subfigure[]{\includegraphics[width=0.3\textwidth, angle =0]{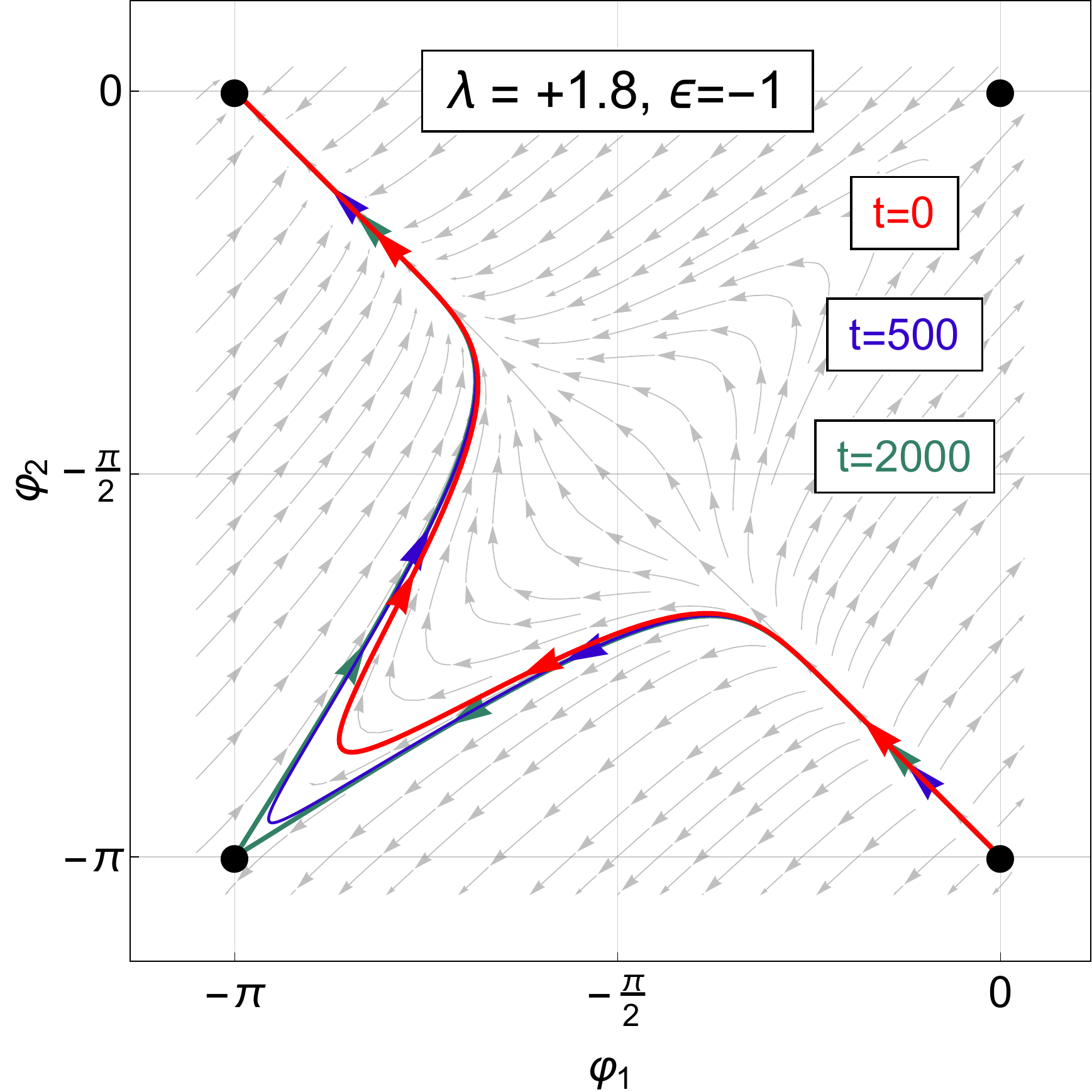}}\hskip0.3cm               
 \subfigure[]{\includegraphics[width=0.3\textwidth, angle =0]{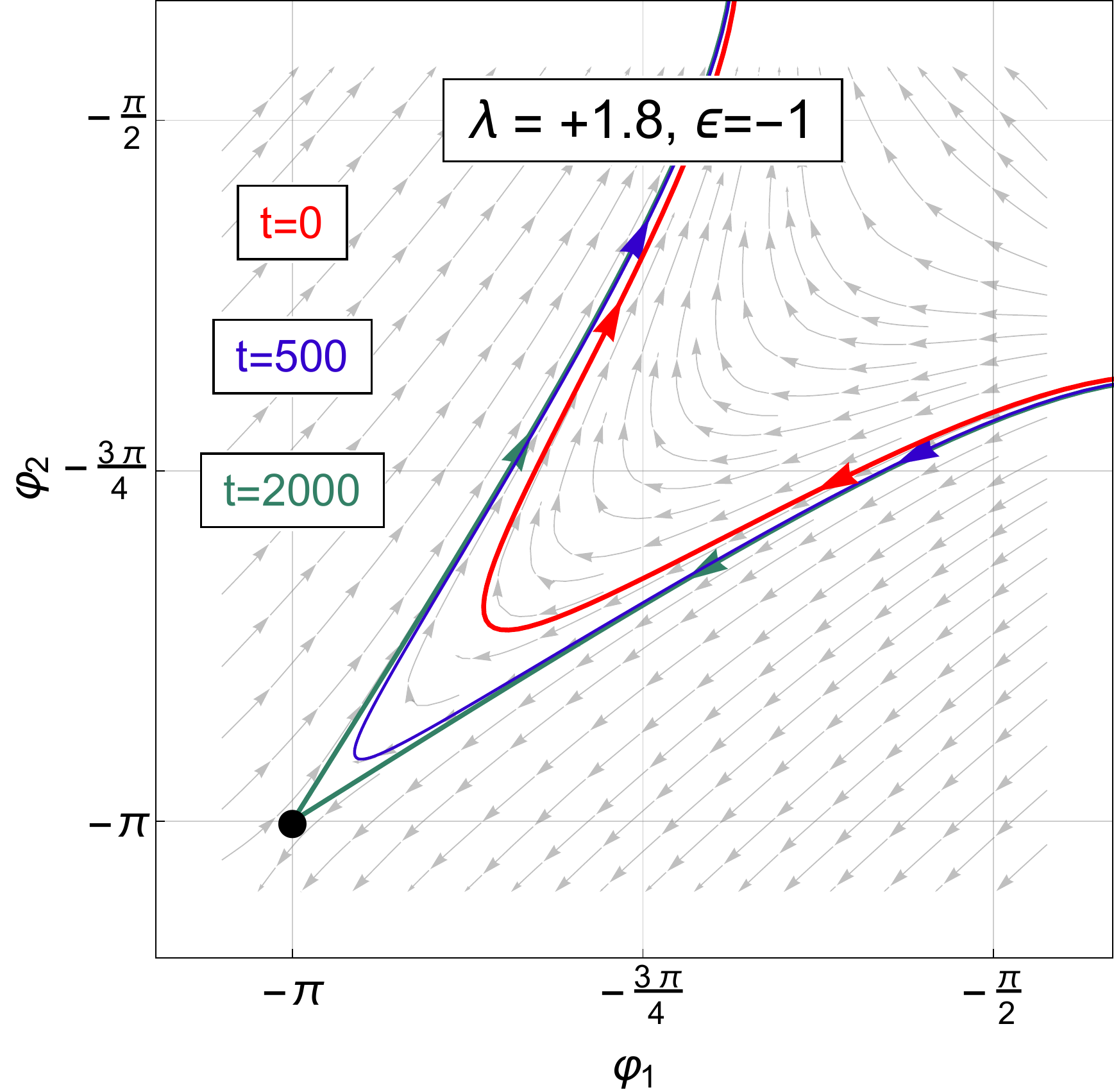}}\hskip0.3cm
 \subfigure[]{\includegraphics[width=0.3\textwidth, angle =0]{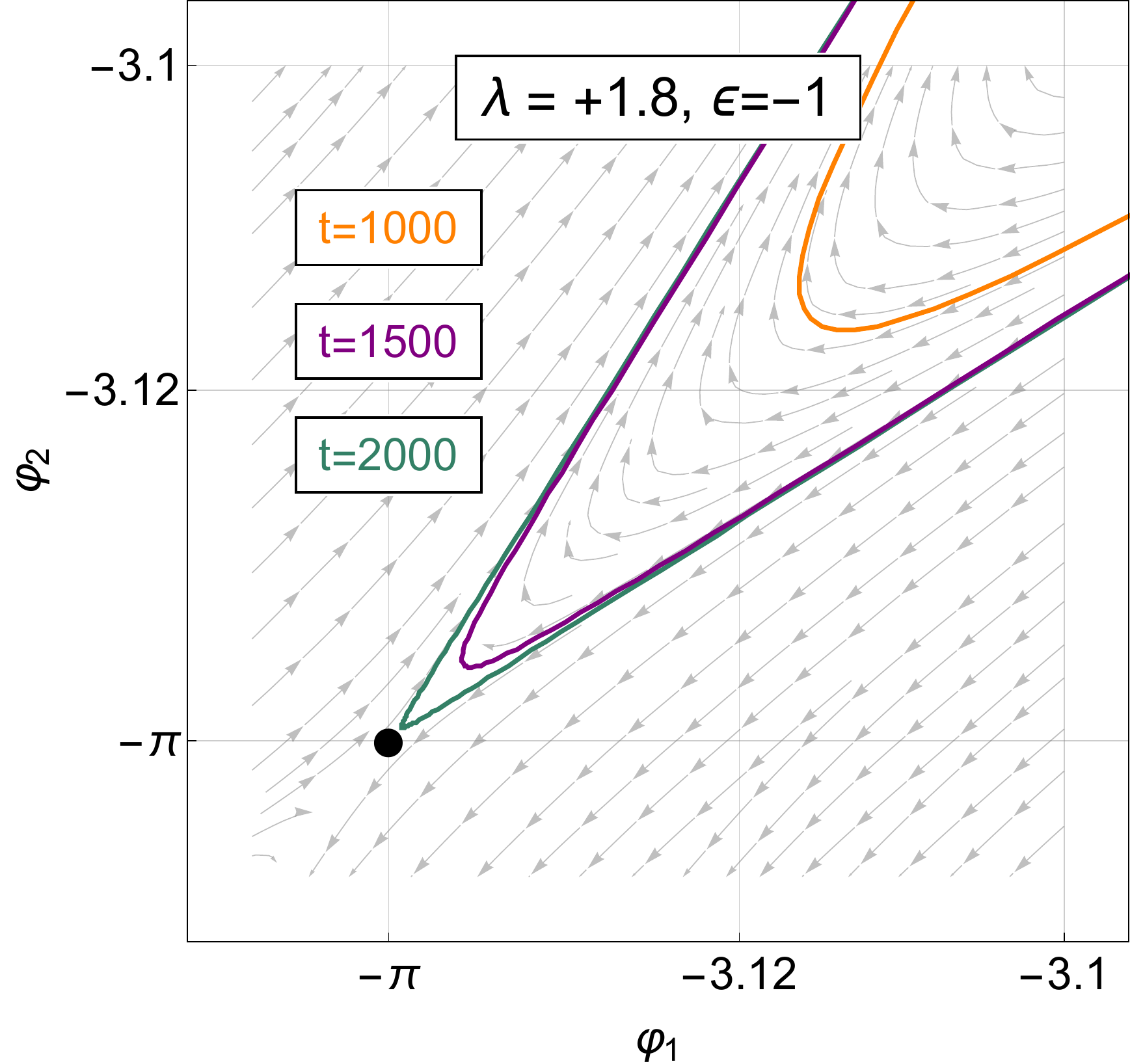}}
\caption{a) Basic figure, b) and c) Close ups of a).}
\label{fig:33sol}
\end{figure}
We see that as $t$ increases the solitons move away from each other ({\it i.e.} they move to $\pm\infty$ which in plots shows
them as moving towards $(-\pi,-\pi)$).
Note that in both cases the energy is {\bf extremely} well conserved.

Next we looked at the same phenomena for systems involving two kinks. We performed several long simulations. This time the motion was significantly 
slower but,  again, we saw attraction and repulsion. This time, however, attraction was observed for $\lambda>0$ and repulsion for $\lambda<0$.

\begin{figure}[h!]
  \centering
\subfigure[]{\includegraphics[width=0.4\textwidth,height=0.3\textwidth, angle =0]{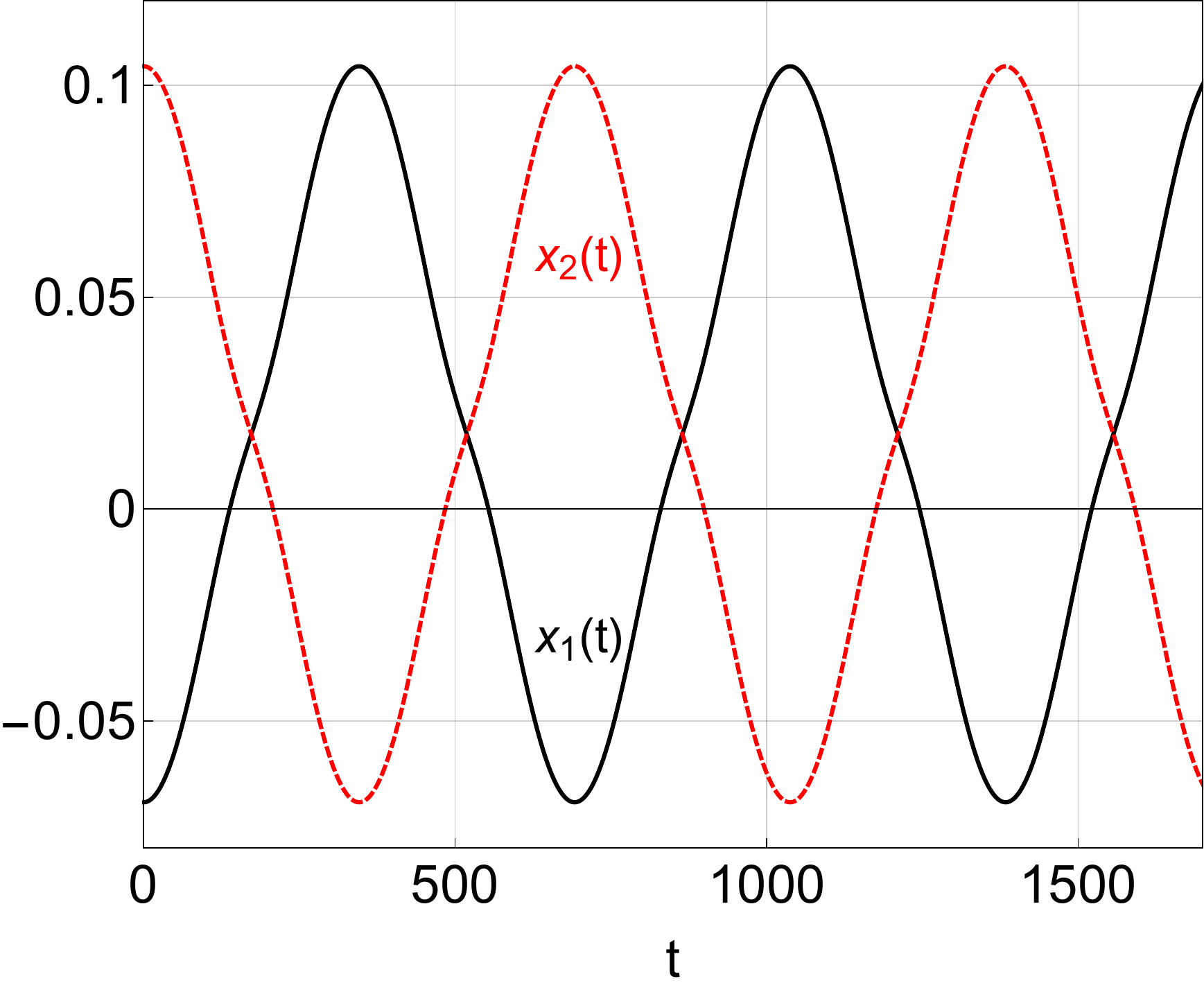}}\hskip0.7cm 
 \subfigure[]{\includegraphics[width=0.4\textwidth,height=0.3\textwidth, angle =0]{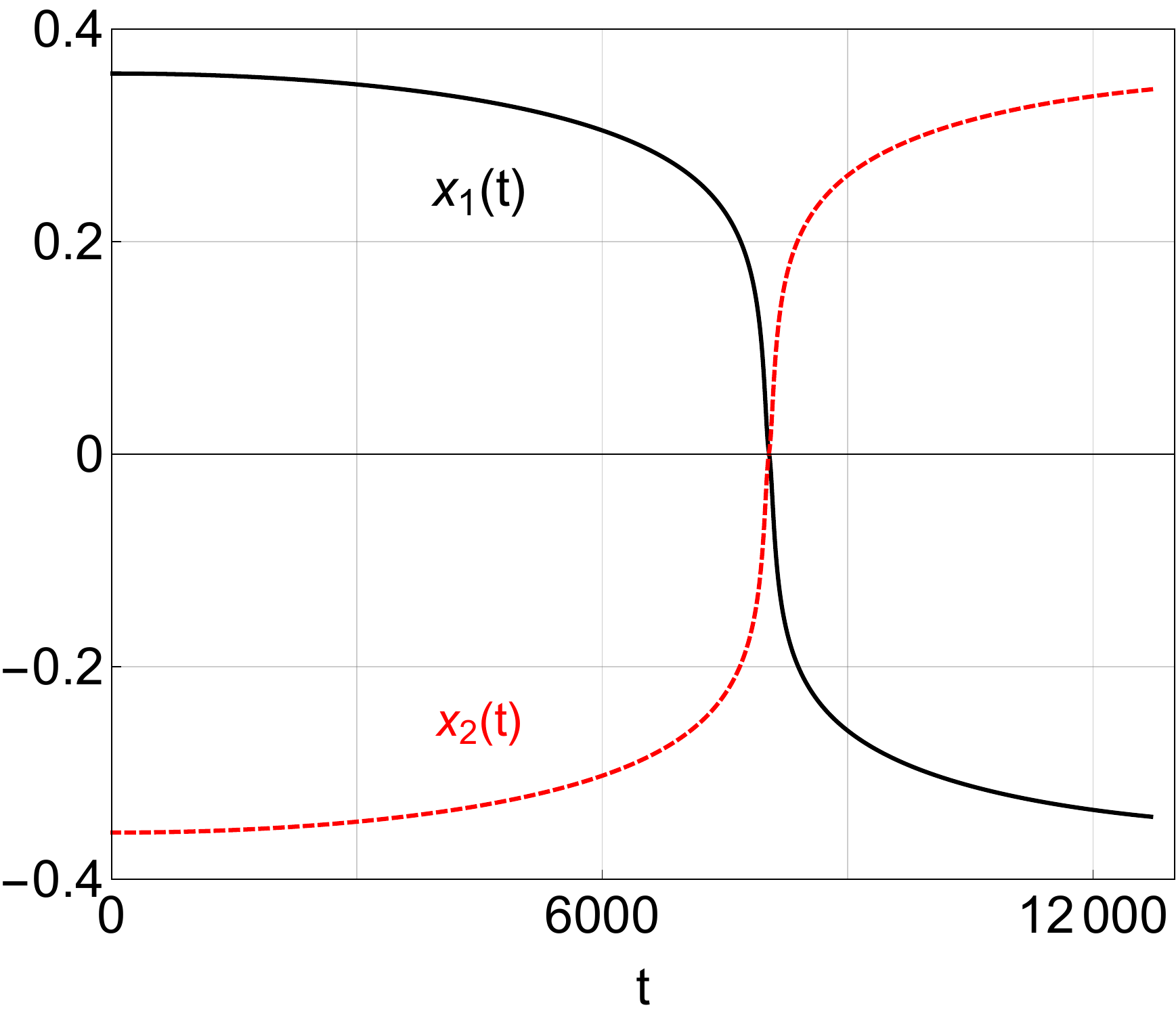}}
\caption{Trajectories in the attractives cases; a) kink-antikink for $\lambda=-1.8$, b) kink-kink for $\lambda=1.95$}
\label{fig:34sol}
\end{figure}

In Fig.\ref{fig:34sol} we present the plots of the trajectories seen in the cases of $\lambda=-1.8$ and $\lambda=1.95$.  

In the next figure (Fig.\ref{fig:35sol}) we plot the trajectories for the kink-antikink system for $\lambda=1.8$, and kink-kink for $\lambda=-1.95$.

\begin{figure}[h!]
\centering
\subfigure[]{\includegraphics[width=0.4\textwidth,height=0.3\textwidth, angle =0]{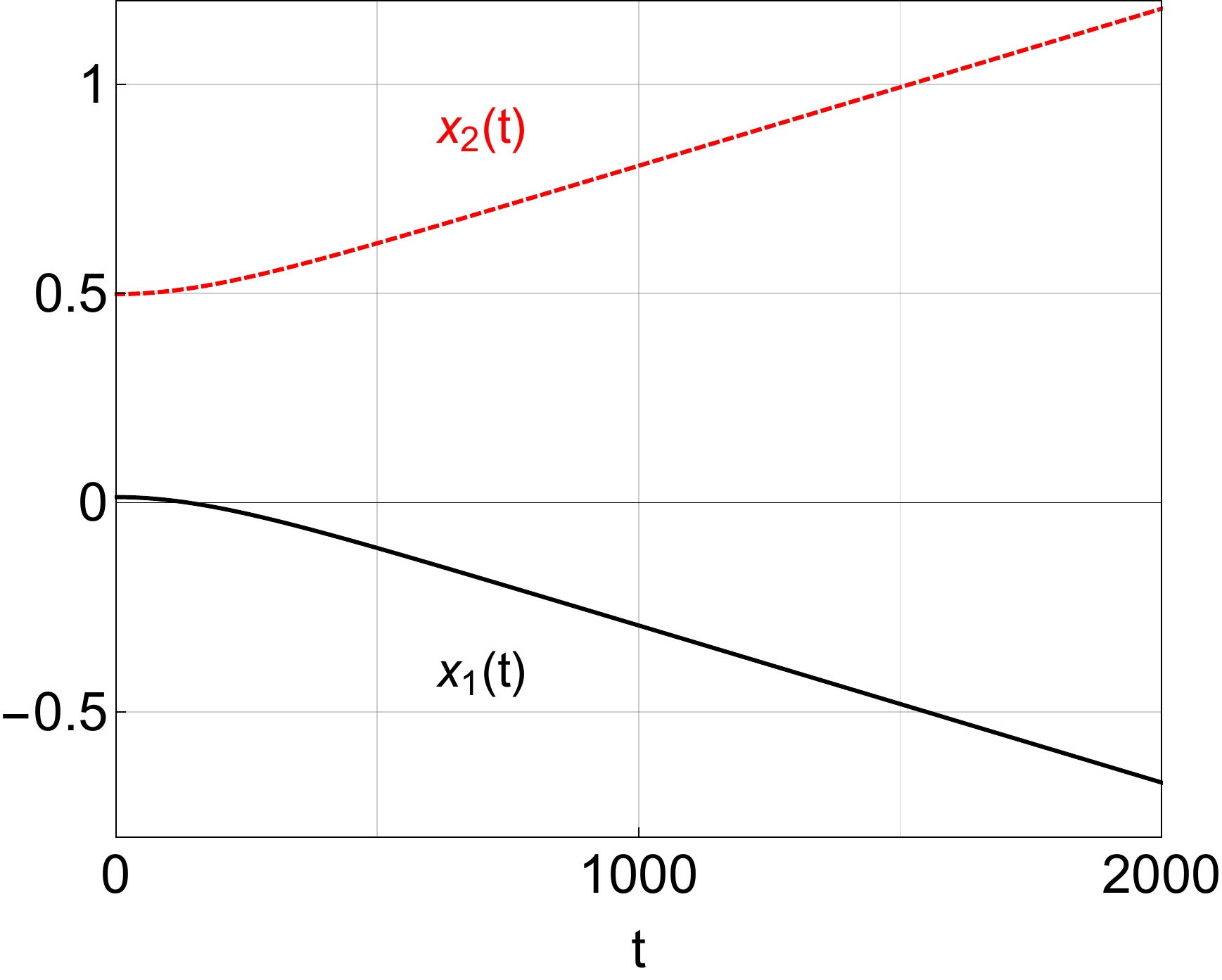}}\hskip0.7cm                
\subfigure[]{\includegraphics[width=0.4\textwidth,height=0.3\textwidth, angle =0]{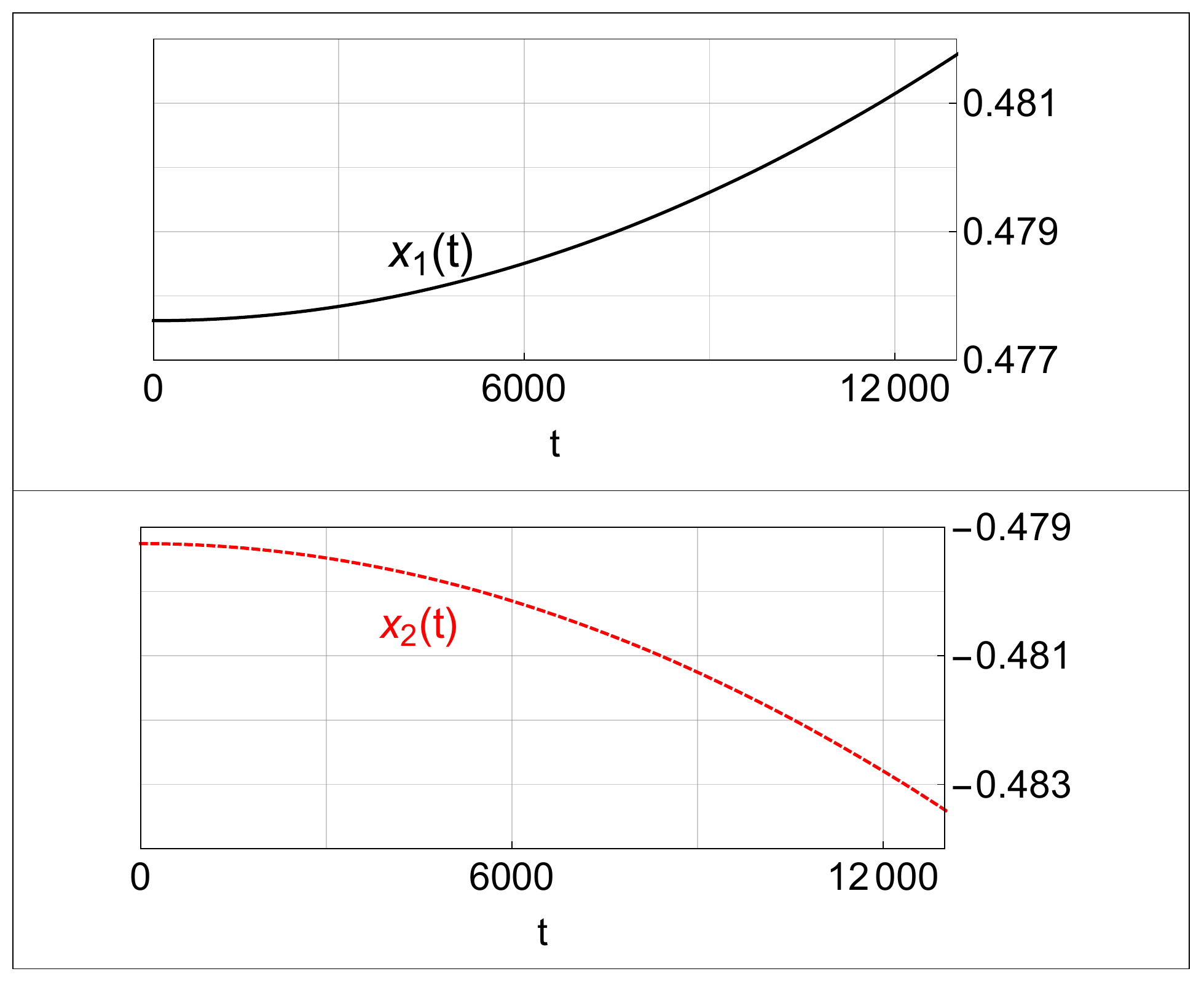}}
\caption{Trajectories in the repulsive cases; a) kink-antikink for $\lambda=1.8$, b) kink-kink for $\lambda=-1.95$}
\label{fig:35sol}
\end{figure}

These plots show very clearly that in the kink-kink cases the motion is much slower (look at the range of $t$ in all cases).
Can we explain this difference?

\begin{figure}[h!]
\centering
\subfigure[]{\includegraphics[width=0.4\textwidth,height=0.25\textwidth, angle =0]{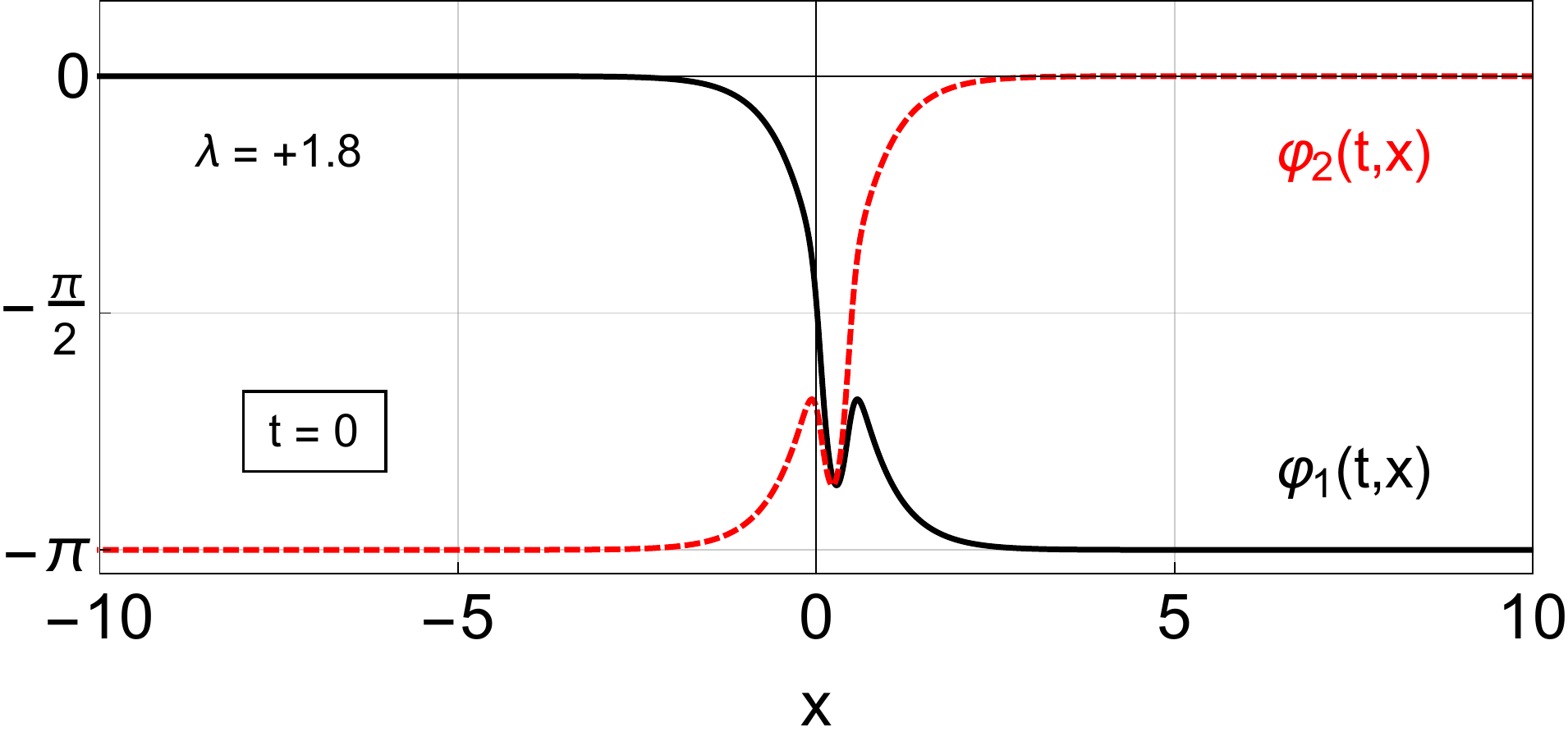}}\hskip0.3cm                
\subfigure[]{\includegraphics[width=0.4\textwidth,height=0.25\textwidth, angle =0]{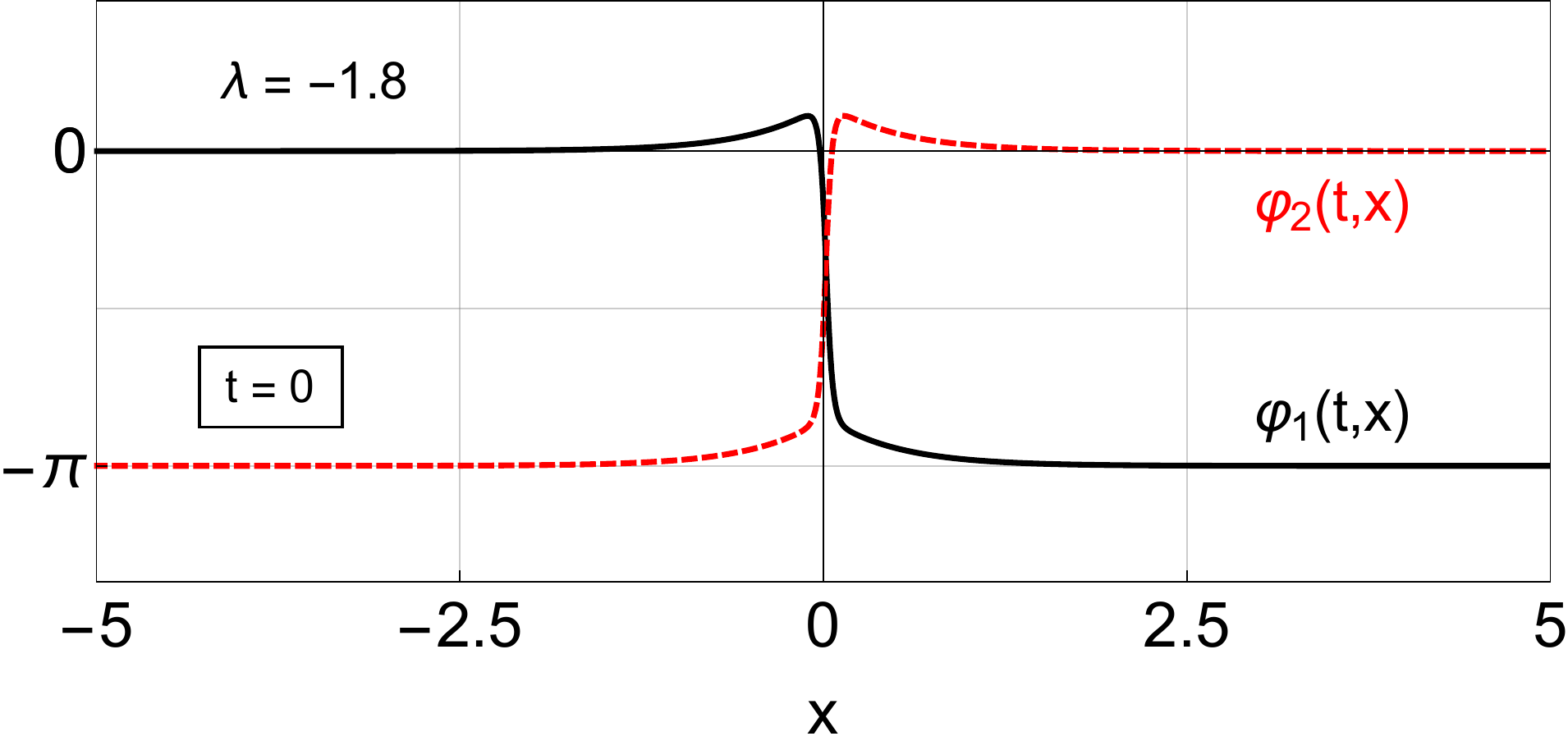}}
\caption{The initial fields of the kink-antikink cases a) for $\lambda=+1.8$, b)  for $\lambda=-1.8$}
\label{fig:36sol}
\end{figure}

This is clearly very difficult. Looking at the field configurations describing the two cases of the initial kink-antikink 
systems ($\lambda=\pm 1.8$) shown in Fig.\ref{fig:36sol}
we see that the solitons at $t=0$ are very close together for the case of $\lambda=-1.8$ and a bit further for $\lambda=1.8$.
The case of $\lambda=-1.8$ is attractive (as we know from plots shown in Fig.\ref{fig:30sol} and Fig.\ref{fig:31sol}) while the case of $\lambda=1.8$ leads to the repulsion (see Fig.\ref{fig:32sol} and Fig.\ref{fig:33sol})). Why is this the case? We have not, unfortunately, 
managed to sort this out. Trying to see what would happen in both cases if the solitons are moving towards each other it is clear that both possibilities are allowed (and the fields would have evolved through $\varphi_1=-\varphi_2-\pi$) when the two solitons lie on top of each other and so their `bumps' have disappeared.

\begin{figure}[h!]
\centering
\subfigure[]{\includegraphics[width=0.4\textwidth,height=0.25\textwidth, angle =0]{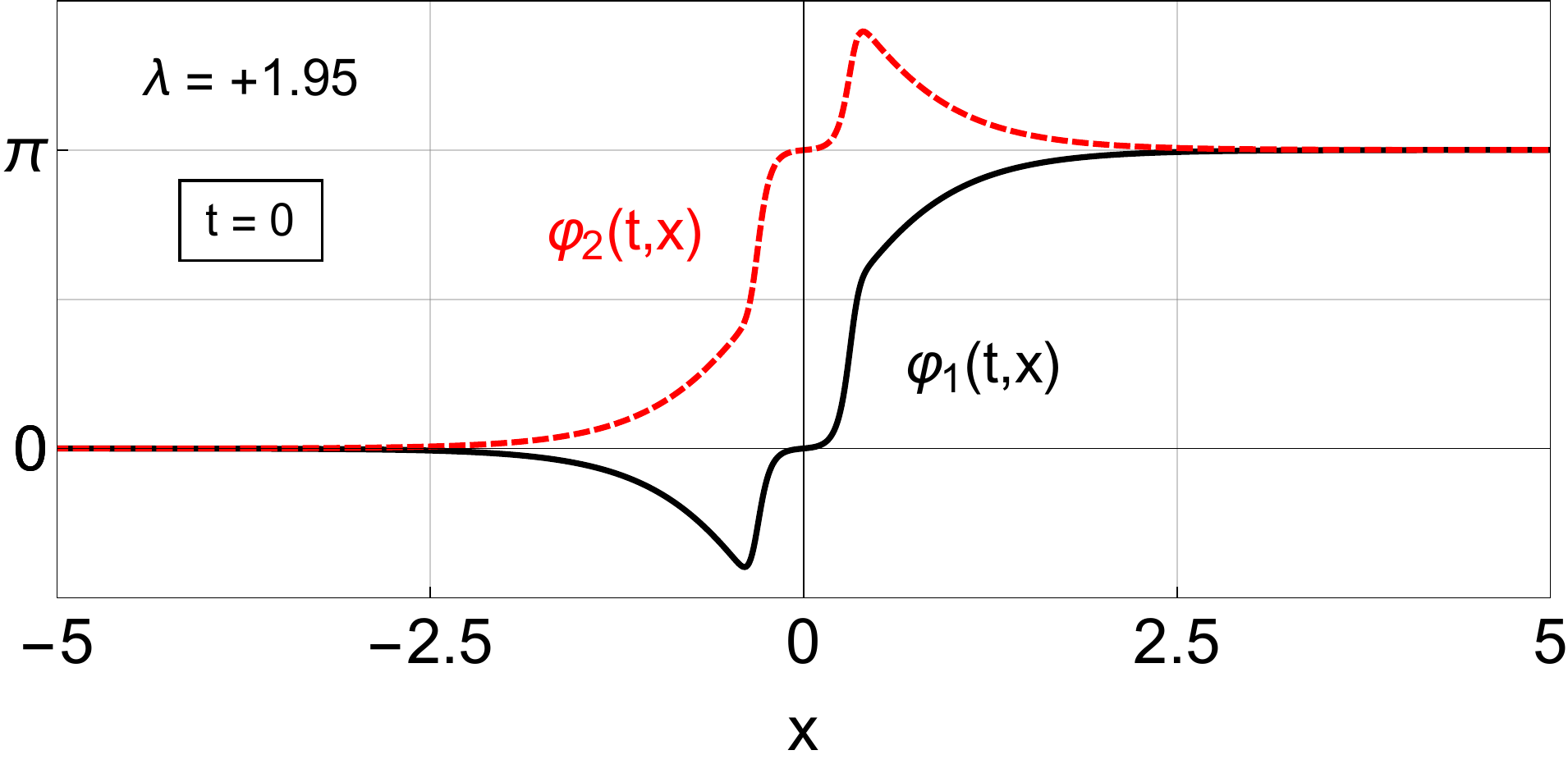}}\hskip0.3cm                
\subfigure[]{\includegraphics[width=0.4\textwidth,height=0.25\textwidth, angle =0]{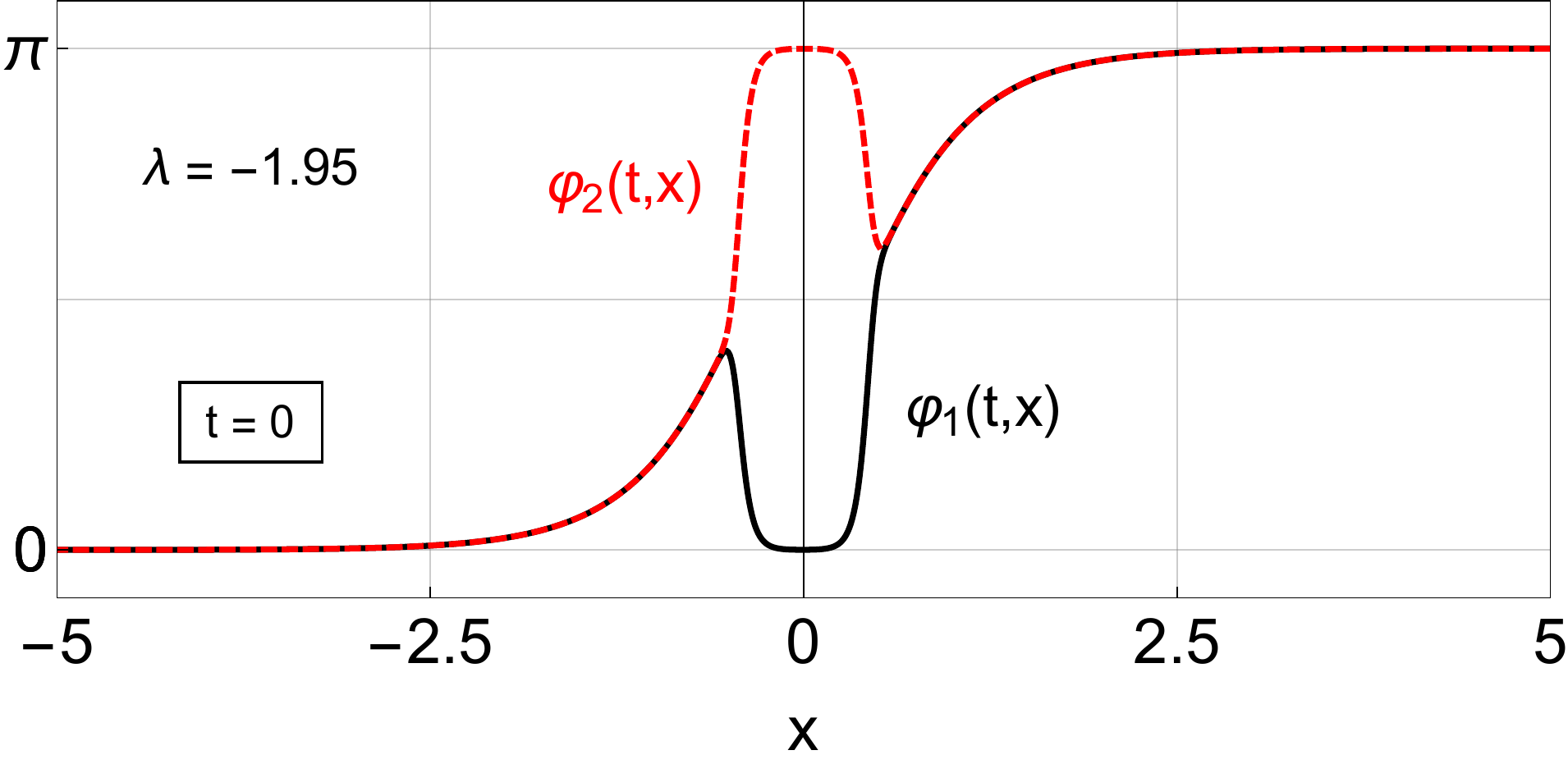}}
\caption{Fields of the kink-kink cases a) for $\lambda=+1.95$, b)  for $\lambda=-1.95$}
\label{fig:37sol}
\end{figure}

In Fig.\ref{fig:37sol} we plot the initial files for $\varphi_1$ and $\varphi_2$ for the cases of  two kinks for $\lambda=\pm 1.95$. This time we see that the initial distance between two solitons is much larger. It is easy to check that, again, both motions are possible and they go through the field
configuration $\varphi_1=\varphi_2$ in which both solitons lie on `top of each other'. Again, we have, so far, not gained any understanding why $\lambda>0$ leads to the attraction and so to the oscillations nor why the $\lambda<0$ leads to the repulsion.

We plan to consider this further in our future work.

\section{Conclusions and Further Comments}

In the work presented in this paper
 we have followed the work reported in \cite{first} and looked in more detail at further properties of solitons in two Sine-Gordon field models which possess BPS properties. In this paper we have looked at systems involving 
two kinks or a system involving a kink and an antikink in each field. Like in a one field case, when we have two solitons per field, the model, in general, looses its BPS properties.
In the case involving one field only, two kinks in such a system repel and when we have a kink and anti-kink they attract. However, when the two Sine-Gordon fields are coupled together the situation is even more complicated. In our case we have a parameter $\lambda$ that controlls the strength of this coupling. To preserve the original BPS property of the system (when we had only one soliton per field)
we have to impose the condition $\vert\lambda\vert < 2$.

First of all we have tested our approach by setting $\lambda=0$ and have confirmed that in cases involving two solitons per field (either both being kinks, or one being a kink, the other an anti-kink), and when the two fields were non-interacting with each other the two kinks in each case repelled, and a kink and an antikink attracted as expected. This is well known from the studies of the one field Sine-Gordon model where it was shown that the energy of two initial static kinks depends on the distance between these kinks and it decreases with the length of this distance. Hence the solitons try to move away from each other. The differences in the energy are incredibly small and so the resultant motion is also very slow. For kinks-anti-kinks the effects are larger and so the motion is faster.  To test our procedure further we pushed the kinks towards each other - they still repelled. First they moved towards each other but then the repulsion won over and they started moving away from each other.

Then we looked at the cases with $\lambda\ne0$. First we considered static configurations. For small values of $\lambda$ we had small repulsion like for $\lambda=0$. This is partly explained by the observation that as the energies of BPS solitons are independent of $\lambda$ 
the energy effects are similar to those for $\lambda=0$. However, these effects grow a little with the growth of $\vert\lambda\vert$.   For larger values of $\vert\lambda\vert$ all solitons and their `bump-like' modifications due to the other fields interacted in  more complicated ways.

We have reported our results for $\lambda=0.6$ and $\lambda=1.2$. In each case the interaction between the bumps and the solitons have resulted in the two outside solitons slowly moving away towards the boundaries and the two more central ones forming a `bound' state close to $x\sim0$ and a lot of energy being sent out towards the two outside solitons. The bound state involved the $\varphi_1$ field which close to $x\sim0$ varied between 0 and $\pi$ and the $\varphi_2$ field
which varied between $\pi$ and $2\pi$. As $\vert\sin(\phi+\pi)\vert=\vert\sin(\phi)\vert$ this bound state can be interpreted 
as a soliton of a one field Sine-Gordon model whose energy
is less than $2\pi$ and so this soliton radiated its excess of energy towards the two outside solitons which, after being hit by this energy,  accelerated towards the boundaries. The bound state could also emit a breather and so start moving away from $x\sim0$. We saw this very clearly for $\lambda=1.2$ and this was also the case for $\lambda=0.6$, but this time the energy of the sent out breather was rather small so the motion of the bound state was also small and the breather itself was hardly visible. 
 
We have also looked at the cases when the solitons were given an initial kick towards each other. In the $\lambda=0$ case this resulted in a delayed repulsion. The same happened for $\lambda\ne0$. There was a small phase-shift as expected and, in each case, after the repulsion the solitons moved towards the boundaries.

We have also looked at two breather systems. In this case our results were very interesting.
When $\lambda=0$, as expected, we had annihilation, sometimes, through the intermediate state of forming a breather. However, this breather was gradually losing energy and so was really an oscillon. 

For small values of $\lambda$ the results were very similar.
Sometimes we had more than one breather as in the case described in  Fig.\ref{fig:14sol}

In some cases we have found that the results depended crucially on the sign of $\lambda$. The magnitude of its value controlled the speed of the observed processes (growing 
with the growth of this value).

For $\lambda>0$ we observed the motion of the two central solitons towards each other, while the two external ones moved towards the boundaries. The annihilation of the solitons at the centre released large amounts of the energy which them moved towards the external solitons and speeded them up. The final system involved external solitons moving towards the boundaries and some radiating oscillons.
For $\lambda<0$ the system behaved differently. The solitons at positive and negative parts of the space oscillated around 
their positions - radiating very little energy (if at all).
The speed of these oscillations grew with the increase of $\vert \lambda\vert$. The system looked very stable and the energy of the system decreased only very little.


We have tried to explain these oscillations and have realised that are driven by the oscillations of the basic BPS solutions
for kink-kink and kink-antikink systems. In principle, such solutions should be stable (as they satisfy the Bogomolnyi bounds). However, as these solutions were found numerically, and the checks of their time dependence have to be carried out 
numerically, they contain minute numerical errors. However, systems of more than one field, have more than one zero mode
and these zero modes can be excited by these small numerical errors. This is what we have seen in our simulations.
The numerical errors were extremely small and so the motions 
of the ostensibly `static' solutions were extremely slow too.
As all numerical errors were extremely small, and the system had no overall momentum these small errors excited a zero mode which controlled the motion of the solitons towards each each other or away from each other. Hence one of the outcomes of our work has been the warning that for numerical studied systems, we can expect such behaviour to occur. Of course, such behaviour can arise only if the system has more than one zero mode, {\it i.e.} it involves more than one field. This is really an important observation and its origin lies in the involvement of interacting fields. In the BPS cases, the `static' fields can easily turn out to be non-static by the involvement of the system's zero modes. So {\bf static} fields can become {\bf nonstatic} and it can complicate the analysis of other phenomena. In our case we have looked at various interactions of more solitons but as such forces are exponential so, if the solitons are too far away from each other, the effects 
of such interactions will be perturbed, by those involving the zero modes. We have looked at other BPS solitonic systems 
and in all cases that we have looked at, this phenomenon was present so we believe this effect is generic. 
Of course, there is another possibility, of both solitons moving in the same direction and the conservation of momentum is achieved by the emitted radiation. Such a case could be generated by the initial numerical errors taking a special form and as such is not very likely to occur.

We hope to be able to say more on this, and other related phenomena, soon.

Finishing, let us add that, although the models that we have studied here, are not integrable, especially for small values of $\lambda$,  the models are not that different
from being integrable. Hence, their properties partially support the ideas of quasi-integrability  \cite{us}. At the same time the property
of the field theory of `being BPS' does not appear to be extremely important. Of course, this is all in (1+1) dimensions.
This may be very different for the field theories in higher dimensions - and, in particular, in models in which multisoliton solutions
can be determined by self-duality ({\it i.e.} like monopoles in (3+1) dimensions or baby skyrmions in (2+1) dimensions).
We plan to look in more detail on such theories in our future work.

\section*{Acknowledgements}
The work reported here was originally started in collaboration with A. Wereszczynski and L.A. Ferreira. The authors would like to thank them for their initial contribution and for their comments.
WJZ wants to thank the Royal Society for 
its grant which supported this research and the University of S\~ao Paulo in S\~ao Carlos where this work was started. The authors would like to thank the Jagiellonian University for 
its hospitality at the Arod\'z retirement meeting in Cracow in October 2018, where this work was discussed.
WJZ's work was supported in part also by the Leverhulme Trust Emeritus Fellowship.

\end{document}